\documentclass[prb,twocolumn,showpacs,preprintnumbers,amsmath,amsfonts,amssymb,floatfix,aps,superscriptaddress]{revtex4-2}

\usepackage{graphicx}
\usepackage{xr}
\usepackage{enumitem}
\usepackage{amssymb}
\usepackage{amsmath}
\usepackage{amsfonts}
\usepackage{bm}
\usepackage{dsfont}
\usepackage{comment}
\usepackage{color}
\usepackage{relsize}
\usepackage{bm}
\usepackage[final]{microtype}
\usepackage[normalem]{ulem}
\usepackage[nodisplayskipstretch]{setspace}
\usepackage{tikz}

\usepackage{hyperref}
\hypersetup{
     colorlinks=true,
     linkcolor=blue,
     filecolor=blue,
     citecolor = blue,
     urlcolor=blue,
   }
\usepackage[export]{adjustbox}

\newcommand{\be}{\begin{equation}}
\newcommand{\ee}{\end{equation}}
\usepackage{units}
\usepackage[left]{lineno}

\graphicspath{{../figures/}}
\begin{document}

\title{First-principles theory of phonon renormalization \\ from nonlinear electron-phonon interactions}

\author{Florian Kluibenschedl}
\email{florian.kluibenschedl@ist.ac.at}
\thanks{These authors contributed equally to this work.}
\affiliation{Institute of Science and Technology Austria (ISTA), 
Am Campus 1, 3400 Klosterneuburg, Austria
}

\author{Matthew Houtput}
\thanks{These authors contributed equally to this work.}
\affiliation{Theory of Quantum Systems and Complex Systems, Universiteit Antwerpen, 
B-2000 Antwerpen, Belgium}

\author{Jacques Tempere}
\affiliation{Theory of Quantum Systems and Complex Systems, Universiteit Antwerpen, 
B-2000 Antwerpen, Belgium}

\author{Cesare Franchini}
\affiliation{Faculty of Physics, Computational Materials Physics, University of Vienna, 
Kolingasse 14-16, Vienna A-1090, Austria}
\affiliation{Department of Physics and Astronomy ``Augusto Righi'', Alma Mater 
Studiorum - Università di Bologna, Bologna, 40127 Italy}

\author{Mikhail Lemeshko}
\affiliation{Institute of Science and Technology Austria (ISTA), 
Am Campus 1, 3400 Klosterneuburg, Austria}

\author{Ragheed Alhyder}
\email{ragheed.alhyder@ist.ac.at}
\affiliation{Institute of Science and Technology Austria (ISTA), 
Am Campus 1, 3400 Klosterneuburg, Austria}

\begin{abstract}
Electron-phonon interactions renormalize phonon frequencies and lifetimes and are central to the dynamical properties of solids. While these effects are usually described within linear electron-phonon coupling, the role of nonlinear electron-phonon interactions for phonon properties remains largely unexplored. In this work, we study phonon renormalization arising from the long-range linear one-electron-one-phonon and the nonlinear one-electron-two-phonon interactions within a diagrammatic framework. We derive the corresponding self-energy diagrams, which depend on the chemical potential and temperature, and evaluate them from first principles for the two polar semiconductors LiF and KTaO$_3$. In both materials, the two interaction channels renormalize the phonon spectrum in qualitatively distinct ways. The linear contribution is sharply localized near the Brillouin-zone center, whereas the nonlinear process couples an incoming phonon to other branches throughout the spectrum. As a result, it renormalizes phonons across the entire Brillouin-zone, with a pronounced temperature dependence governed by the thermal occupation of those branches. This behavior provides a clean experimental signature of the one-electron-two-phonon coupling. While the nonlinear phonon renormalization is small in LiF, it is somewhat larger in KTaO$_3$, which we attribute to its greater number of thermally populated phonon branches at room temperature. Our results establish a general framework to assess nonlinear electron-phonon effects on the phonon properties in materials with stronger lattice fluctuations, including soft semiconductors such as lead-halide perovskites.
\end{abstract}

\maketitle

\section{Introduction}

Phonons, the quantized normal modes of lattice vibrations, are collective low-energy excitations associated with ionic motion in solids~\cite{Mahan2000}. Their spectrum and lifetime encode the dynamical response of the lattice and determine a wide range of thermodynamic and transport properties, including heat capacity, thermal conductivity, and structural stability~\cite{Mahan2000,togoFirstPrinciplesPhonon2015,baroniPhononsRelatedCrystal2001}. 
Theoretically, phonon properties can be systematically described within a field-theoretic framework of lattice dynamics, in which the lattice is coupled to the surrounding electronic system through the electron-phonon interaction~\cite{giustinoElectronphononInteractionsFirst2017}.

Concretely, atomic displacements modify the electronic potential landscape, and the induced change in electronic density feeds back onto the lattice through screening. At the level of a many-body theory, this interplay is captured by electron and phonon Green’s functions and their self-energies, which renormalize the respective propagators and determine both frequency shifts and linewidths of the respective quasiparticles~\cite{schwingerPhononGreensFunction1991,giustinoElectronphononInteractionsFirst2017,bergesPhononSelfEnergyCorrections2023,Kluibenschedl2025}. Standard treatments in the literature usually apply two approximations to the electron-phonon problem. 

The first approximation is that the self-energies are evaluated within lowest order of the electron-phonon interaction. For electrons, this yields the Fan-Migdal and Debye-Waller contributions, while the phonon self-energy is described by electron-hole polarization diagrams \cite{allenEffectsPhononDynamics1974,giustinoElectronphononInteractionsFirst2017}. Already at this level, the electron-hole polarization can significantly alter the phonon properties of materials. For example, metals can exhibit strong screening due to low-energy electronic excitations near the Fermi level, which leads to sizable phonon renormalization effects, including Kohn anomalies and superconductivity-related phenomena~\cite{Kohn1959,piscanecKohnAnomaliesElectronPhonon2004,allenNeutronSpectroscopySuperconductors1972,grimvallELECTRONPHONONINTERACTIONMETALS,lazzeriNonadiabaticKohnAnomaly2006}. By contrast, intrinsic semiconductors and insulators lack low-energy electronic excitations, resulting in weak screening and typically smaller electron-phonon effects on the lattice dynamics~\cite{voglMicroscopicTheoryElectronphonon1976}. Additionally, for metals and semiconductors with large charge carrier densities in the conduction band, arising from doping or photoexcitation, the small ratio of phononic to electronic energy scales typically suppresses vertex corrections to the linear electron-phonon interaction, as formalized by Migdal's theorem~\cite{leeInitioElectrontwophononScattering2020,bauerQuantitativeReliabilityStudy2011,mishraElectronPhononVertex2025}.

The second approximation is that the electron-phonon Hamiltonian itself is usually approximated up to linear order in the ionic displacements. In particular, keeping only the term linear in ionic displacements can break down in materials where strong polar coupling or large-amplitudes soft phonon modes promote nonadiabatic effects~\cite{ponceTemperatureDependenceElectronic2015,zhouElectronPhononScatteringPresence2018,houtputFirstprinciplesTheoryNonlinear2025a}.

\begin{figure*}[t]
    \centering
    \includegraphics[width=0.8\textwidth]{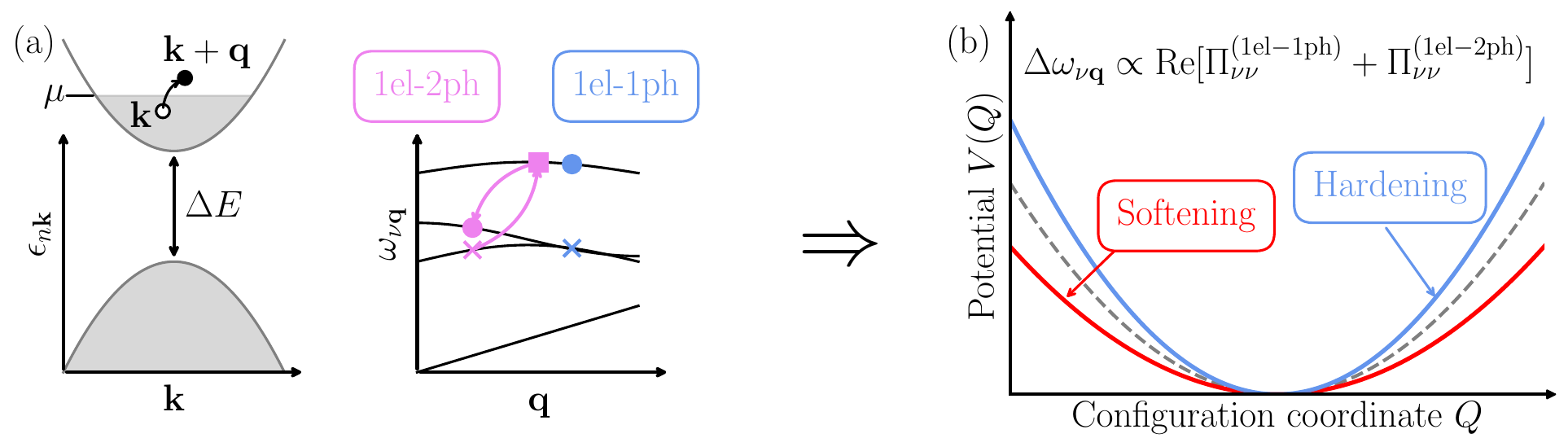}
    \vskip -0.3cm
    \caption{(a) Schematic illustration of the scattering processes between conduction band electrons arising from doping or photoexcitation and phonons within the one-electron-one-phonon (blue) and one-electron-two-phonon (violet) interaction. Crosses indicate annihilated phonons, squares denote intermediate phonon states, and the circle marks the final phonon. In general, the final phonon state can differ in branch from the initial one, but the initial and final momenta must be almost equal. This allows for off-diagonal contributions to the phonon self-energy. (b) Modification of the effective phonon potential due to electron-phonon interactions. Within our diagrammatic framework, this manifests as a frequency shift: a hardening of the phonon corresponds to a shift to higher frequencies (blue shift), while a softening corresponds to a shift to lower frequencies (red shift).}
    \label{fig:fig1}
\end{figure*}

Beyond linear order, the electron-phonon interaction contains nonlinear terms arising from higher-order derivatives of the electron-ion potential with respect to atomic displacements. The leading next order correction is given by the one-electron-two-phonon vertex. On the analytical side, model studies of quadratic electron-phonon coupling have explored its impact on polaron quasiparticle properties~\cite{gogolinQuantumPolaron1991,houtputFrohlichHamiltonianPathintegral2021,ragniPolaronQuadraticElectronphonon2023,kliminAnalyticMethodQuadratic2024,ragniPolaronsArbitraryNonlinear2025}. In particular, these works show that nonlinear coupling can qualitatively modify the polaron problem, leading to effects such as lattice stiffening driven by quantum fluctuations~\cite{gogolinQuantumPolaron1991}. Complementary to these analytical developments, first-principles derivations of the nonlinear electron-phonon interaction have recently been developed for both short- and long-range polar couplings~\cite{Zhenbang2026,houtputFirstprinciplesTheoryNonlinear2025a}. In this framework, the interaction strength is mainly determined by the derivative of the dynamical matrix with respect to the electric field of the electron, and the corresponding diagram contains an additional phonon propagator. This introduces additional scattering channels with a pronounced temperature dependence, in contrast to the lowest-order perturbation theory diagrams arising from the linear electron-phonon interaction. Particularly, in soft semiconductors it was shown that this can lead to significant corrections to electronic properties such as the electron mobility~\cite{Houtput2026}. This naturally raises the question of how the same nonlinear interaction feeds back on the lattice and renormalizes the phonon spectrum itself.

In this work, we therefore investigate the effect of the nonlinear electron-phonon interactions on phonon properties in polar semiconductors as a function of the charge carrier density in the conduction band, arising for example from doping or photoexcitation. A schematic overview of the relevant scattering processes, where free carriers interact with one or two phonons, and their impact on phonon frequencies in terms of blue- or red-shifts is shown in Fig.~\ref{fig:fig1}. We extend the standard diagrammatic expansion of the phonon self-energy by including diagrams up to lowest order in the one-electron-two-phonon vertex, and compare its contributions with the electron-hole polarization bubble. 

As a benchmark, we evaluate both diagrams for LiF, a prototypical polar semiconductor~\cite{antoniusDynamicalAnharmonicEffects2015,sioInitioTheoryPolarons2019,karnaPressuretuningSpatialExtension2025}, and KTaO$_3$, a quantum paraelectric~\cite{singhStabilityPhononsKTaO1996,ranalliTemperatureDependentAnharmonicPhonons2023,essweinFirstprinciplesCalculationElectronphonon2023}, and quantify the resulting corrections to phonon frequencies and lifetimes. Although the nonlinear correction remains small in both materials, it is two to three times larger in KTaO$_3$, which we attribute to more phonon branches being thermally populated at room temperature compared to LiF. The resulting framework provides a systematic route to assess higher-order electron-phonon effects on phonon properties in soft semiconductors, where we anticipate stronger phonon renormalizations, as indicated by previous works on the electron self-energy~\cite{Houtput2026}. 

\section{Hamiltonian}
We consider a crystalline material with $N$ atoms in the primitive unit cell where the electrons interact with the phonons of the lattice. The Hamiltonian for the crystal is given by, $\hat{H} = \hat{H}_{\rm el} + \hat{H}_{\rm ph} + \hat{H}_{\rm el-ph}$. Here, $\hat{H}_{\rm el}$ and $\hat{H}_{\rm ph}$ are the bare electronic and phononic Hamiltonians respectively, and $\hat{H}_{\rm el-ph}$ describes the interaction between electrons and phonons. The equilibrium structure and bare phonon modes are obtained within the usual Born–Oppenheimer framework. The non-interacting electron and phonon system is described by
\begin{align}
\hat{H}_{\rm el} =& \sum_{n,\mathbf{k}} \epsilon_{n,\mathbf{k}} \hat{c}^\dagger_{n,\mathbf{k}} \hat{c}_{n,\mathbf{k}}, \label{eq:electron_hamiltonian} \\
\hat{H}_{\rm ph} =& \sum_{\nu,\mathbf{q}} \hbar \omega_{\nu, \mathbf{q}} \left( \hat{a}^\dagger_{\nu,\mathbf{q}} \hat{a}_{\nu,\mathbf{q}} + \frac{1}{2} \right). \label{eq:phonon_hamiltonian}
\end{align}
Here, $\epsilon_{n,\mathbf{k}}$ are electron energies for an electron in band~$n$ with momentum $\mathbf{k}$. They can be computed from the Kohn-Sham eigenvalues from first-principles calculations, neglecting electron-electron interactions. The fermionic creation (annihilation) operators $\hat{c}^\dagger_{n,\mathbf{k}}$ ($\hat{c}_{n,\mathbf{k}}$), create (annihilate) an electron in the Bloch state $\psi_{n,\mathbf{k}}$. Throughout this work, we assume that the electronic spectrum is spin degenerate and that the electron-phonon interaction is spin independent and spin conserving. We therefore suppress the spin index of the electronic operators, and spin enters only through a degeneracy factor. In materials with spin-split or spin-textured electronic bands, the spin dependence of the electronic states and electron-phonon matrix elements must instead be retained explicitly.

Conversely, energies for a phonon with momentum $\mathbf{q}$ and band index $\nu$ are characterized by the angular phonon frequency $\omega_{\nu, \mathbf{q}}$. All the bands constitute the phononic band structure, with band indices running over the $3N$ vibrational modes and are obtained by diagonalizing the dynamical matrix. The bosonic creation and annihilation operators $\hat{a}^\dagger_{\nu,\mathbf{q}}$ and $\hat{a}_{\nu,\mathbf{q}}$ create and annihilate a phonon with momentum $\mathbf{q}$ in branch $\nu$. 
While several approaches go beyond the harmonic approximation underlying $\hat{H}_{\rm ph}$ in Eq.~\eqref{eq:phonon_hamiltonian}, including the stochastic self-consistent harmonic approximation~\cite{erreaAnharmonicFreeEnergies2014}, the temperature-dependent effective potential method~\cite{hellmanTemperatureDependentEffective2013}, or the special displacement method~\cite{zachariasAnharmonicLatticeDynamics2023}, we focus here on phonon renormalization only induced by linear and nonlinear electron-phonon interactions.

To this end, we expand the electron-phonon interaction perturbatively in the phonon field up to  one-electron-two-phonon processes, such that $\hat{H}_{\rm el-ph} = \hat{H}_{\rm 1el-1ph} + \hat{H}_{\rm 1el-2ph}$, with the linear one-electron-one-phonon vertex~\cite{verdiFrohlichElectronPhononVertex2015}
\begin{equation}
\begin{split}
\hat{H}_{\rm 1el-1ph} =& 
\sqrt{\frac{\Omega_0}{\Omega}} \sum_{\mathbf{k},\mathbf{q},n,m,\nu} g_{mn\nu}(\mathbf{k}, \mathbf{q}) \\
\qquad & \times \left( \hat{a}_{\nu,\mathbf{q}} + \hat{a}^\dagger_{\nu,-\mathbf{q}} \right)
\hat{c}^\dagger_{m,\mathbf{k}+\mathbf{q}} \hat{c}_{n,\mathbf{k}}, \label{eq:1el-1ph_hamiltonian}
\end{split}
\end{equation}
and the nonlinear one-electron-two-phonon vertex~\cite{houtputFirstprinciplesTheoryNonlinear2025a}
\begin{equation}
\begin{split}
\hat{H}_{\rm 1el-2ph} =& 
\frac{\Omega_0}{\Omega} \sum_{\mathbf{k},\mathbf{q}_1,\mathbf{q}_2} \sum_{m,n,\nu_1,\nu_2}
g_{mn\nu_1\nu_2}(\mathbf{k}, \mathbf{q}_1, \mathbf{q}_2) \\
\qquad & \times \left( \hat{a}_{\nu_1,\mathbf{q}_1} + \hat{a}^\dagger_{\nu_1,-\mathbf{q}_1} \right)
\left( \hat{a}_{\nu_2,\mathbf{q}_2} + \hat{a}^\dagger_{\nu_2,-\mathbf{q}_2} \right) \\
\qquad & \times \hat{c}^\dagger_{m,\mathbf{k}+\mathbf{q}_1+\mathbf{q}_2} \hat{c}_{n,\mathbf{k}}. \label{eq:1el-2ph_hamiltonian}
\end{split}
\end{equation}
Here, $\Omega_0$ is the unit cell volume, $\Omega$ is the total system volume, and $N_0 = \Omega/\Omega_0$ describes the number of unit cells in the system. The matrix element $g_{mn\nu}(\mathbf{k}, \mathbf{q})$ describes the scattering of an electron in the state $|\psi_{\mathbf{k},n}\rangle$ into the state $|\psi_{\mathbf{k} + \mathbf{q},m}\rangle$ by absorbing a phonon with momentum $\mathbf{q}$ in branch $\nu$. Since the Hamiltonian is Hermitian, the complex conjugate of the scattering matrix element satisfies $g_{mn\nu}^* \left(\mathbf{k},\mathbf{q}\right) = g_{nm\nu} \left(\mathbf{k} + \mathbf{q}, - \mathbf{q}\right)$. We will refer to this process as a one-electron-one-phonon interaction. 

\begin{figure}[t]
    \centering
    \includegraphics[width=\columnwidth]{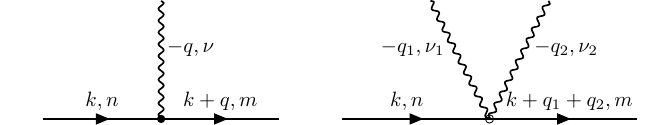}
    \caption{Feynman diagrams for the two electron-phonon interactions from the Hamiltonian in Eqs.~\eqref{eq:1el-1ph_hamiltonian}-\eqref{eq:1el-2ph_hamiltonian}. Solid lines denote electron propagators and wavy lines phonon propagators. (left) The one-electron-one-phonon interaction, $g_{mn\nu} \left(\mathbf{k},\mathbf{q}\right)$, with a filled circle. (right) The nonlinear one-electron-two-phonon interaction, $g_{mn\nu_1\nu_2} \left(\mathbf{k},\mathbf{q}_1,\mathbf{q}_2\right)$, with an open circle. The labels $k$ are four-momenta, e.g. $k=\left(\mathbf{k},\omega\right)$, while $n$, $\nu$ denote electron and phonon branches respectively.}
    \label{fig:vertex_factors}
\end{figure}

The interaction in $\hat{H}_{\rm 1el-2ph}$ occurs when an electron with momentum $\mathbf{k}$ in branch $n$ scatters into an electron state with momentum $\mathbf{k} + \mathbf{q}_1 + \mathbf{q}_2$ in branch $m$ by absorbing two phonons with momentum $\mathbf{q}_1$ and $\mathbf{q}_2$ in branches $\nu_1$ and $\nu_2$. We denote the matrix element of this interaction as $g_{mn\nu_1 \nu_2}(\mathbf{k}, \mathbf{q}_1, \mathbf{q}_2)$, and since the Hamiltonian must be Hermitian, this matrix element satisfies $g_{mn\nu_1\nu_2}^* \left(\mathbf{k}, \mathbf{q}_1, \mathbf{q}_2\right) = g_{nm\nu_1\nu_2}\left(\mathbf{k} + \mathbf{q}_1 + \mathbf{q}_2, -\mathbf{q}_1, -\mathbf{q}_2 \right)$. The vertices for both scattering processes are sketched in Fig.~\ref{fig:vertex_factors}.

So far, the combined electron-phonon Hamiltonian in Eqs.~\eqref{eq:electron_hamiltonian}-\eqref{eq:1el-2ph_hamiltonian} is very general. In particular, the interaction terms can include both short-range and long-range interactions. Subsequently, we focus on the long-range limits of the vertex factors $g_{mn\nu}$ and $g_{mn\nu_1\nu_2}$, which can be parameterized from first-principles calculations~\cite{houtputFirstprinciplesTheoryNonlinear2025a}. 

In the long-range approximation, the ionic polarization field induced by the charge carriers is assumed to be approximately constant over the size of a supercell. Then the Coulomb-like interactions have large contributions near the Brillouin zone center, i.e.~the total momentum~$\mathbf Q=\sum_i\mathbf q_i$ entering the respective polar vertex is small compared with a reciprocal lattice scale, $|\mathbf Q|\ll \pi/a$. In this limit, it was shown that the long-range electron-phonon vertices acquire universal forms governed by Born effective charges and the high-frequency dielectric tensor, leading to a scaling behavior as $g(\mathbf{Q}) \propto 1/|\mathbf{Q}|$, cf.~\cite{verdiFrohlichElectronPhononVertex2015,houtputFirstprinciplesTheoryNonlinear2025a}. Therefore, in the limit of $\mathbf{Q} + \mathbf{G}\to 0$, where $\mathbf{G}$ is a reciprocal lattice vector, the polar long-range vertex is nonanalytic and exhibits the characteristic Fr{\" o}hlich enhancement, whereas the short-range contribution is analytic and remains finite. At low carrier densities, the available electronic phase space restricts the relevant momentum transfers to values of order $k_{\rm F}$. Since \(k_F\ll\pi/a\), the dominant contributions to the relevant diagrams naturally arise from the small-momentum region where the long-range terms are enhanced. 

In materials with large charge carrier densities, coming for example from doping, screening can in principle modify the long-range polar electron-phonon interaction~\cite{giustinoElectronphononInteractionsFirst2017,bergesPhononSelfEnergyCorrections2023}. In particular, when the electronic plasma frequency becomes comparable to the relevant phonon frequencies, dynamical screening and plasmon-phonon hybridization may have to be treated explicitly~\cite{lihmPlasmonPhononHybridizationDoped2024}. In the present work, we do not include screening corrections to the electron-phonon vertices, to focus on the phonon renormalization arising from the linear and nonlinear electron–phonon vertices considered here. We expect this approximation to capture the leading qualitative trends, while a full dynamically screened treatment is left for future work.

The long-range part of the one-electron-one-phonon interaction is~\cite{verdiFrohlichElectronPhononVertex2015},
\begin{equation}
\begin{split}
g^{(\text{long})}_{mn\nu}(\mathbf{k}, \mathbf{q}) 
=& 
\sum_{\mathbf{G} \neq -\mathbf{q}} g_{\nu} (\mathbf{q}, \mathbf{G}) \langle \psi_{\mathbf{k} + \mathbf{q}, m} | 
e^{i (\mathbf{q} + \mathbf{G}) \cdot (\mathbf{r} - \boldsymbol{\tau}_\kappa)} 
| \psi_{\mathbf{k}, n} \rangle, 
\end{split} 
\label{eq:simple_1el-1ph_vertex_long_range}
\end{equation}
with $g_{\nu} (\mathbf{q}, \mathbf{G}) = \mathbf{V}_{\nu} \left(\mathbf{q} + \mathbf{G}\right) \cdot \mathbf{p}_{\nu} \left(\mathbf{q}\right)$. Here, the zero-point amplitudes of the phonon displacement and the dielectric constant are included in the definition of $\mathbf{V}_{\nu} \left(\mathbf{Q}\right)$,
\begin{equation}
\mathbf{V}_{\nu} \left(\mathbf{Q}\right) = \frac{ie^2}{\varepsilon_0 \Omega_0} \sqrt{\frac{\hbar}{2 \omega_{\nu,\mathbf{Q}}}} \frac{\mathbf{Q}}{\mathbf{Q}\cdot \boldsymbol{\varepsilon}_{\infty} \cdot \mathbf{Q}}, \label{eq:def_mod_vertex_1el-1ph}
\end{equation}
whereas the mode polarity is given by a sum over all Born effective charges,
\begin{equation}
\mathbf{p}_{\nu} \left(\mathbf{q}\right) =  \sum_{\kappa} 
\frac{1}{\sqrt{m_{\kappa}}}
\mathbf{Z}_{\kappa} \cdot \mathbf{e}_{\kappa, \nu} \left(\mathbf{q}\right),
\end{equation}
where $\mathbf{e}_{\kappa, \nu} \left(\mathbf{q}\right)$ is the phonon eigenvector associated to atom $\kappa$ in the branch $\nu$. Consequently, the material parameters determining the one-electron-one-phonon matrix element are the Born effective charge tensor $\mathbf{Z}_{\kappa}$ and the dielectric tensor at high frequencies, $\boldsymbol{\varepsilon}_{\infty}$, which are both accessible from first-principles calculations. Moreover, the vertex in Eq.~\eqref{eq:simple_1el-1ph_vertex_long_range} decays as $1/|\mathbf{q} + \mathbf{G}|$, characteristic of long-range polar interactions, and extends the Fr{\" o}hlich vertex by taking into account couplings to multiple phonon branches.

The long-range contribution of the nonlinear one-electron-two-phonon vertex was recently derived in~\cite{houtputFirstprinciplesTheoryNonlinear2025a},
\begin{equation}
\begin{split}
g^{(\text{long})}_{mn\nu_1\nu_2} & (\mathbf{k}, \mathbf{q}_1, \mathbf{q}_2) 
= 
\sum_{\mathbf{G} \neq -\mathbf{q}_1 - \mathbf{q}_2} 
g_{\nu_1\nu_2} \left(\mathbf{q}_2,\mathbf{q}_1 + \mathbf{q}_2 + \mathbf{G}\right) \\
& \times \left\langle \psi_{\mathbf{k} + \mathbf{q}_1 + \mathbf{q}_2, m} \left|
e^{i (\mathbf{q}_1 + \mathbf{q}_2 + \mathbf{G}) \cdot \hat{\mathbf{r}}}
\right| \psi_{\mathbf{k}, n} \right\rangle 
\end{split}
\label{eq:1-el-2-ph_coupling_element}
\end{equation}
with the shorthand notation
\begin{equation}
g_{\nu'\nu} \left(\mathbf{q},\mathbf{Q}\right) 
= 
\frac{e^2}{2\varepsilon_0 \Omega_0} 
\frac{\mathbf{Q}\cdot \mathbf{Y}_{\nu'\nu} \left(\mathbf{q}\right)} {\mathbf{Q}\cdot \mathbf{\varepsilon_{\infty}} \cdot \mathbf{Q}}.
\end{equation}
Besides the $1 / |\mathbf{q}_1 + \mathbf{q}_2 + \mathbf{G}|$ divergence, the coupling strength is governed by the zero-point amplitudes of the involved phonon modes and by the derivative of the dynamical matrix with respect to the internal or external electric field $\boldsymbol{\mathcal{E}}$,
\begin{equation}
\begin{split}
\mathbf{Y}_{\nu_1 \nu_2}(\mathbf{q}) 
=&
\frac{1}{i e} 
\sqrt{ \frac{\hbar}{2 \omega_{\mathbf{q}, \nu_1}} } 
\sqrt{ \frac{\hbar}{2 \omega_{\mathbf{q}, \nu_2}} } 
\sum_{\kappa \alpha} \sum_{\kappa' \beta} \\ 
\times & \mathbf{e}^*_{\kappa \alpha, \nu_1}(\mathbf{q}) 
\frac{\partial \mathcal{D}_{\kappa \alpha, \kappa' \beta}(\mathbf{q})}{\partial \boldsymbol{\mathcal{E}}} 
\mathbf{e}_{\kappa' \beta, \nu_2}(\mathbf{q}).
\end{split}
\label{eq:def_Y_nu1nu2_tensor}
\end{equation}
Here, the indices $\alpha, \beta \in \lbrace x,y,z\rbrace$ denote the Cartesian components of the respective phonon eigenvectors. For a discussion of the symmetry and transformation properties of $g^{(\text{long})}_{mn\nu_1\nu_2} (\mathbf{k}, \mathbf{q}_1, \mathbf{q}_2)$ and $\mathbf{Y}_{\nu_1 \nu_2}(\mathbf{q})$ that are relevant for this work, we refer to App.~\ref{sec:symmetry_transformations}.

\section{Phonon Green's function}

\subsection{Interacting phonon propagator and Dyson equation}

We can describe the lattice dynamics of a material with the retarded displacement-displacement correlation function $D_{\nu\mu}(\mathbf{q},t) = - \langle \mathcal{T}_t \, \hat A_{\nu,\mathbf{q}}(t) \hat A_{\nu,-\mathbf{q}} (0) \rangle$, also known as the phonon Green's function~\cite{giustinoElectronphononInteractionsFirst2017}. Here, $\mathcal{T}_t$ is the time-ordering operator and we define the symmetric phonon operator $\hat{A}_{\nu,\mathbf{q}} = \hat{a}_{\nu,\mathbf{q}} + \hat{a}_{\nu,-\mathbf{q}}^{\dagger}$. For non-interacting phonons within the harmonic approximation, the phonon Green's function is diagonal in branch indices and has poles at the phonon frequencies. In the Matsubara formalism, it reads
\begin{equation}
D^{\left(0\right)}_{\nu\mu}  \left(\mathbf{q},i\omega_n\right) 
=
\delta_{\nu\mu}
\frac{2 \omega_{\nu,\mathbf{q}}}{(i\omega_n)^2 - \omega_{\nu,\mathbf{q}}^2},
\end{equation}
where $\omega_n = 2 \pi n/\beta$ represents the bosonic Matsubara frequencies, with inverse temperature $\beta$ and $n\in\mathbb{Z}$~\cite{Mahan2000}. 

The interaction of phonons with electrons renormalizes the adiabatic phonon spectrum of the noninteracting system, as it is captured by the phonon self-energy. Formally, these effects are incorporated through the interacting phonon Green’s function, which satisfies the Dyson equation,
\begin{equation}
\left(D \left(\mathbf{q},i\omega_n\right)\right)^{-1}_{\nu\mu} = \left(D^{\left(0\right)} \left(\mathbf{q},i\omega_n\right)\right)^{-1}_{\nu\mu} - \Pi_{\nu\mu} \left(\mathbf{q},i\omega_n\right), \label{eq:formal_inverse_of_dyson_equation}
\end{equation}
where $\Pi_{\nu\mu} \left(\mathbf{q},i\omega_n\right)$ is the phonon self energy. For finite temperature calculations, the retarded interacting phonon Green's function, $D^{\rm R} \left(\mathbf{q},\omega\right)$, can be obtained by analytic continuation, $i\omega_n \to \omega + i \delta$ and $\delta > 0$. Then, the phonon quasiparticle energies are the solutions in $\omega$ of the secular equation $\det ((D^{\rm R} \left(\mathbf{q},\omega\right))_{\nu\mu}^{-1} ) = 0$.

To solve the secular equation, it is convenient to define the rescaled inverse phonon propagator
\begin{equation}
(\boldsymbol{\tilde{D}}^{\rm R} (\mathbf{q},\omega))^{-1} = S (\boldsymbol{D}^{\rm R} (\mathbf{q},\omega))^{-1} S,
\end{equation}
with the diagonal transformation matrix 
\begin{equation}
S = \mathrm{diag} (\sqrt{2 \omega_{1,\mathbf{q}}}, \dots, \sqrt{2 \omega_{3N,\mathbf{q}}}).    
\end{equation}
The dimension of the matrix is $3N$: in the results of our next section, $3N=6$ for lithium fluoride and $3N=15$ for potassium tantalate. Diagonalizing the full matrix~$(\boldsymbol{\tilde{D}}^{\rm R} (\mathbf{q},\omega))^{-1}$ self-consistently amounts to solving an eigenvalue problem whose solutions yield the renormalized phonon frequencies $\tilde{\omega}_{\nu\mathbf{q}}$. In this work, we reduce the computational workload by evaluating this matrix on the unperturbed frequencies $\omega = \omega_{\nu,\mathbf{q}}$, and subsequently diagonalize it, which is equivalent to Rayleigh-Schr{\" o}dinger perturbation theory. In this formulation, off-diagonal components of the phonon self-energy are fully retained and can mix different phonon branches. For more details on the calculation see~\cite{Hermes2013,Hermes20132} and App.~\ref{subsec:diagonalizing_dyson_equation_with_off_diagonal_terms}. 

In many situations, however, the dominant contribution to the phonon renormalization arises from the diagonal part of the self-energy, and off-diagonal terms can be treated perturbatively or neglected. Assuming that both the frequency shift and linewidth are small, i.e.~$|\tilde{\omega}_{\nu,\mathbf{q}} - \omega_{\nu,\mathbf{q}}| \ll \omega_{\nu,\mathbf{q}}$ and $|\gamma_{\nu,\mathbf{q}}| \ll \omega_{\nu,\mathbf{q}}$ hold, the secular equation simplifies and one obtains 
\begin{equation}
\begin{split}
\tilde{\omega}_{\nu,\mathbf{q}} 
&\approx 
\omega_{\nu,\mathbf{q}} + \mathrm{Re} \Pi^{\rm R}_{\nu\nu} \left(\mathbf{q},\omega_{\nu,\mathbf{q}}\right),  \\
\gamma_{\nu,\mathbf{q}} 
&\approx 
- \mathrm{Im} \Pi^{\rm R}_{\nu\nu} \left(\mathbf{q},\omega_{\nu,\mathbf{q}}\right). \label{eq:summary_diagonal_deg_approximate_renormalization}
\end{split}
\end{equation}
Note that in a degenerate subspace, we diagonalize the respective matrix in the subspace to get the renormalized phonon frequencies. Here, $\gamma_{\nu,\mathbf{q}}$ denotes the phonon linewidth due to electron-phonon interaction, with lifetime $\tau_{\nu,\mathbf{q}}^{\rm ph} = 1 / (2 \gamma_{\nu,\mathbf{q}})$.

Generally, the resulting renormalized frequencies can be directly compared to experiments that probe the phonon spectrum such as inelastic neutron or x-ray scattering and Terahertz spectroscopy~\cite{liPhononSelfEnergyOrigin2014,burkelInelasticScatteringXRays1991,jepsenTerahertzSpectroscopyImaging2011}.

\subsection{Perturbative expansion up to lowest order}

\begin{figure}[t]
    \centering
    \includegraphics[width=0.9\columnwidth]{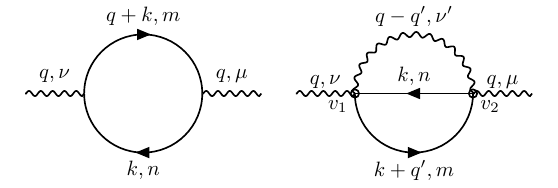}
    \caption{Feynman diagrams contributing to the phonon self-energy at  lowest order in the electron-phonon interaction. (left) Electron-hole polarization bubble coming from the one-electron-one-phonon interaction, $g_{mn\nu} \left(\mathbf{k},\mathbf{q}\right)$. (right) Contribution from the one-electron-two-phonon self-energy. The labels $k$ are four-momenta, e.g. $k=\left(\mathbf{k},\omega\right)$, while the indices $n$ and $\nu$ denote electron and phonon branches, respectively.}
    \label{fig:self_energy_feynman_diagrams}
\end{figure}

In this work, we work in the weak coupling regime and compute self-energy contributions up to second order in the coupling vertices, which are the two diagrams shown in Fig.~\ref{fig:self_energy_feynman_diagrams}.
The diagram on the left corresponds to the standard electron-hole polarization process, in which a phonon decays into an electron-hole pair around the Fermi level. The diagram on the right originates from the nonlinear one-electron-two-phonon coupling and represents the main contribution from the beyond-linear coupling channels to lowest order. In contrast to the polarization bubble, which is governed by electron-hole excitations near the Fermi surface, this contribution involves an additional internal phonon line and therefore depends on both electronic and phononic phase spaces. 

As a consequence, the two contributions exhibit qualitatively different momentum dependences. In the linear polarization bubble, the external phonon momentum $\mathbf q$ is transferred directly to the electron-hole excitation, which strongly restricts the response to the electronic phase space near the Fermi surface. In the nonlinear diagram, by contrast, the electron-hole pair carries an internal momentum $\mathbf q'$ in the Feynman diagram of Fig.~\ref{fig:self_energy_feynman_diagrams}, while the additional phonon carries momentum $\mathbf q-\mathbf q'$. The nonlinear long-range vertex depends on $\mathbf q'+\mathbf G$ and is therefore enhanced when $|\mathbf q'+\mathbf G|$ is small, independently of the external phonon momentum $\mathbf q$. Consequently, $\mathbf q'$ can remain in the long-wavelength electronic regime while the external phonon spans the entire Brillouin zone. The additional internal phonon also introduces Bose-Einstein occupation factors, giving the nonlinear contribution a qualitatively stronger temperature dependence compared to the linear correction. In the following, we will discuss each diagram in more detail. 


\subsubsection{The one-electron-one-phonon diagram}
\paragraph{Derivation.} The left diagram in Fig.~\ref{fig:self_energy_feynman_diagrams} represents the electron polarization bubble, where a phonon scatters by creating an electron-hole pair, which is again absorbed by another phonon. It is the simplest self-energy correction and is routinely discussed in the literature \cite{giustinoElectronphononInteractionsFirst2017, Chaudhary2024}. Calculations of the one-electron-one-phonon self-energy show that a dynamical electronic response can qualitatively modify phonon properties in low-dimensional and metallic systems, among others. For example, in doped graphene and metallic carbon nanotubes, the frequency dependence of the electronic polarization entering the phonon self-energy produces nonadiabatic Kohn anomalies, leading to strong, doping-dependent phonon frequency shifts and linewidth variations that are absent within the adiabatic Born-Oppenheimer approximation~\cite{lazzeriNonadiabaticKohnAnomaly2006,caudalKohnAnomaliesNonadiabaticity2007}. On the other hand, in bulk metals such as MgB$_2$ and CaC$_6$, the self-energy significantly reshapes phonon dispersions over extended regions of the Brillouin zone~\cite{calandraAdiabaticNonadiabaticPhonon2010}. Applying the Feynman rules to the one-electron-one-phonon self-energy diagram in Fig.~\ref{fig:self_energy_feynman_diagrams} gives
\begin{equation}
\begin{split}
\Pi_{\nu\mu}^{\left(\rm 1el-1ph\right)} & \left(\mathbf{q},i\omega_n \right)
= 
- \frac{2}{\hbar^2} \frac{\Omega_0}{\Omega} 
\frac{1}{\beta}
\sum_{\mathbf{k}} \sum_{m,n} \sum_{i\omega_s} \\
& g_{mn\nu} \left(\mathbf{k},\mathbf{q}\right) g_{mn\mu}^* \left(\mathbf{k},\mathbf{q}\right) \\ 
& G_{nn}^{\left(0\right)} \left(\mathbf{k},i\omega_s\right) G_{mm}^{\left(0\right)} \left(\mathbf{q}+\mathbf{k},i\omega_n + i\omega_s\right), 
\end{split} 
\label{eq:def_1_electron_1_phonon_self_energy}
\end{equation}
with fermionic Matsubara frequencies $\omega_s = \pi (2s+1)/\beta$. As motivated in more detail below, we restrict our attention to the case in which the dominant contributions to the one-electron-one-phonon self-energy arise from the long-range limit of the vertex $g_{mn\nu} \left(\mathbf{k},\mathbf{q}\right)$. In this regime, the full vertex can be replaced by $g_{mn\nu}^{\rm (long)} \left(\mathbf{k},\mathbf{q}\right)$, as given in Eq.~\eqref{eq:simple_1el-1ph_vertex_long_range}. The long-range vertex relates the two electronic states $n$ and $m$ through a plane-wave with momentum $\mathbf{q} + \mathbf{G}$ and the noninteracting electron Green's function $G_{nn}^{\left(0\right)} \left(\mathbf{k},i\omega_s\right)$ depends on the band index $n$ only through the dispersion $\epsilon_{n,\mathbf{k}}$. We can therefore rewrite the double sum over the electronic bands as a sum over one band index and a reciprocal lattice vector, i.e.~$\sum_{m,n} \to \sum_{\mathbf{G},n}$.~A detailed proof of this statement is presented in App.~\ref{subsec:summation_over_bloch_states}. The resulting expression is
\begin{equation}
\begin{split}
&\Pi_{\nu\mu}^{\left(\rm 1el-1ph\right)}\left(\mathbf{q},i\omega_n\right) 
=- 
\frac{2 \Omega_0}{\hbar^2} \chi \left(\mathbf{q},i\omega_n\right) \sum_{\mathbf{G}}
g_{\nu} (\mathbf{q}, \mathbf{G}) g_{\mu}^* (\mathbf{q}, \mathbf{G}),
\end{split}
\label{eq:1_electron_1_phonon_self_energy_after_summation_over_to_G_conversion}
\end{equation}
where we have identified the Lindhard polarization function
\begin{equation}
\begin{split}
\chi \left(\mathbf{q},i\omega_n\right) 
=&
\sum_n 
\frac{1}{\beta} \sum_{i\omega_s}
\underset{\rm 1BZ}{\int} \frac{\mathrm{d}^3 \mathbf{k}}{\left(2\pi\right)^3} \\
\times & G_{nn}^{\left(0\right)} \left(\mathbf{k},i\omega_s\right) G_{nn}^{\left(0\right)} \left(\mathbf{q}+\mathbf{k},i\omega_n + i\omega_s\right). 
\end{split} 
\label{eq:Lindhard_definition}
\end{equation}
Performing the sum over Matsubara frequencies and subsequent analytic continuation, $i\omega_n \to \omega + i\delta$, gives the finite temperature expression for the retarded polarization function,
\begin{equation}
\chi^{\rm R} \left(\mathbf{q},\omega\right) = \sum_n \underset{\rm 1. BZ}{\int} \frac{\mathrm{d}^3 \mathbf{k}}{\left(2\pi\right)^3} \frac{n_F \left(\epsilon_{n,\mathbf{k}}\right) - n_F \left(\epsilon_{n,\mathbf{k} + \mathbf{q}}\right)}{\frac{\epsilon_{n,\mathbf{k}}}{\hbar} - \frac{\epsilon_{n,\mathbf{k} + \mathbf{q}}}{\hbar} + \omega + i \delta} \label{eq:retarded_Lindhard_final}
\end{equation}

When evaluating $\chi^{\rm R} \left(\mathbf{q},\omega\right)$, filled bands will not contribute in the sum over $n$. For example, in semiconductors with a gap $\Delta E$ larger than temperature $T$, $\Delta E \gg k_B T$, all bands up to the valence band will be filled, with an exponentially small fraction of electrons being in the conduction band. Consequently, the Lindhard polarization function will be close to zero. However, if there are electrons in the conduction band and we evaluate $\chi^{\rm R} \left(\mathbf{q},\omega\right)$ at the unperturbed phonon frequencies, the most important contributions will come from transitions between occupied and empty states separated by an energy of the order of typical phonon energies, i.e. from low-energy electrons near the Fermi level. 

\paragraph{Long-range limit of $\chi^{\rm R} \left(\mathbf{q},\omega\right)$.} The polarization function $\chi \left(\mathbf{q},\omega\right)$ typically decays rapidly for momenta larger than the Fermi momentum, $\mathbf{q} \gg \mathbf{k}_{\rm F}$ \cite{giustinoElectronphononInteractionsFirst2017}. At small excitation densities, the Fermi momentum is much smaller than the edge of the Brillouin zone, $k_{\rm F} \ll \pi/a$. Consequently, the relevant momentum transfers are restricted to the long-wavelength regime, which justifies using the long-range approximation for the polar coupling vertex. By the same argument, we neglect the $\mathbf{G}\neq 0$ contributions of the sum in Eq.~\eqref{eq:1_electron_1_phonon_self_energy_after_summation_over_to_G_conversion} since the momentum integral converges already in the first Brillouin zone. Thus, keeping only the $\mathbf{G}=0$ term, we obtain
\begin{equation}
\Pi_{\nu\mu}^{\left(\rm 1el-1ph\right)} \left(\mathbf{q},\omega\right)
\approx -
\frac{2\Omega_0}{\hbar^2} g_{\nu} (\mathbf{q}, 0) g_{\mu}^* (\mathbf{q}, 0) \chi \left(\mathbf{q},\omega\right).
\label{eq:1-electron-1-phonon_self_energy_final}
\end{equation}
To keep a compact notation, we drop the superscript $\mathrm{R}$ on retarded functions, and we note that in the remaining part of the manuscript, real-frequency functions always indicate retarded expressions, while imaginary-frequency functions always indicate the Matsubara component. \\

\subsubsection{The one-electron-two-phonon diagram}
\label{subsubsec:one_electron_two_phonon_diagram}

\paragraph{Derivation.} The phonon self-energy diagram on the right in Fig.~\ref{fig:self_energy_feynman_diagrams} arises from the one-electron-two-phonon interaction in Eq. \eqref{eq:1el-2ph_hamiltonian}. Analogously to the electron-hole polarization bubble, the Feynman rules from App.~\ref{appsec:feynman_rules} yield the following mathematical expression for this self-energy diagram,
\begin{widetext}
\begin{equation}
\begin{split}
\Pi_{\nu\mu}^{\rm (1el-2ph)} \left(\mathbf{q}, i\omega_n\right) 
=& 
- \frac{4}{\hbar^2} 
\frac{\Omega_0^2}{\Omega^2} 
\frac{1}{\beta^2}
\sum_{\mathbf{k},m,n} 
\sum_{\mathbf{q}', \nu'} 
\sum_{i\omega_s,i\omega_{\ell}}
g_{mn\nu'\nu} \left(\mathbf{k},-\mathbf{q}+\mathbf{q}',\mathbf{q}\right) 
g_{mn\nu'\mu}^* \left(\mathbf{k},-\mathbf{q}+\mathbf{q}',\mathbf{q}\right) \\
\times & 
G_{nn}^{\left(0\right)} \left(\mathbf{k}, i\omega_{\ell}\right) 
G_{mm}^{\left(0\right)} \left(\mathbf{k} + \mathbf{q}', i\omega_{\ell} + i\omega_s\right) 
D^{\left(0\right)}_{\nu'\nu'} \left(\mathbf{q}-\mathbf{q}', i\omega_n - i\omega_s\right) \label{eq:1electron_2phonon_self_energy_first_expr}
\end{split}
\end{equation}
Again using the long-range approximation, we replace the full one-electron-two-phonon vertex by its long-range form,~$g_{mn\nu'\nu} \to g_{mn\nu'\nu}^{\rm (long)}$, and exactly rewrite the sum over two electronic bands as a sum over one band index and a reciprocal lattice vector, cf.~App.~\ref{subsec:summation_over_bloch_states}. We then obtain
\begin{equation}
\begin{split}
\Pi_{\nu\mu}^{\rm (1el-2ph)} \left(\mathbf{q}, i\omega_n\right) 
=& 
-\frac{4\Omega_0^2}{\hbar^2} 
\frac{1}{\beta^2}
\sum_{n,\nu'} 
\underset{\rm 1. BZ}{\int} \frac{\mathrm{d}^3 \mathbf{k}}{\left(2\pi\right)^3} \underset{\rm 1. BZ}{\int} \frac{\mathrm{d}^3 \mathbf{q}'}{\left(2\pi\right)^3} 
\sum_{i\omega_s,i\omega_{\ell}} 
\sum_{\mathbf{G}} 
g_{\nu'\nu} \left(\mathbf{q},\mathbf{q}' + \mathbf{G}\right) 
g_{\nu'\mu}^* \left(\mathbf{q},\mathbf{q}' + \mathbf{G}\right) \\
\times &
G_{nn}^{\left(0\right)} \left(\mathbf{k}, i\omega_{\ell}\right) 
G_{nn}^{\left(0\right)} \left(\mathbf{k} + \mathbf{q}', i\omega_{\ell} + i\omega_s\right) 
D^{\left(0\right)}_{\nu'\nu'} \left(\mathbf{q}-\mathbf{q}', i\omega_n - i\omega_s\right)
\end{split}
\label{eq:1el-2ph_expression_for_self_energy_after_sum_rewriting}
\end{equation}
In the $\mathbf{k}$ and $i\omega_{\ell}$ integrals over the electronic Green's functions, we recognize Eq.~\eqref{eq:Lindhard_definition} for the Lindhard polarization bubble $\chi \left(\mathbf{q}',i\omega_s\right)$. The remaining $i\omega_s$ Matsubara sum over $\chi \left(\mathbf{q}',i\omega_s\right)$ and $D^{\left(0\right)}_{\nu',\nu'} \left(\mathbf{q}-\mathbf{q}', i\omega_n - i\omega_s\right)$, can then be evaluated independently. Consequently, we define the convolution of polarization bubble and phonon propagator,
\begin{equation}
\Xi_{\nu'}\left(\mathbf{q}',\mathbf{q},i\omega_n\right) 
:=
-\frac{1}{\beta}\sum_{i\omega_s} 
\chi \left(\mathbf{q}',i\omega_s\right) 
D^{\left(0\right)}_{\nu'\nu'} \left(\mathbf{q}-\mathbf{q}', i\omega_n - i\omega_s\right). \label{eq:convolution_Lindhard_phonon}
\end{equation}
The retarded expression for $\Xi_{\nu'}\left(\mathbf{q}',\mathbf{q},i\omega_n\right)$ is obtained by performing Matsubara summation followed by analytical continuation $i\omega_n \to \omega + i\delta$,
\begin{equation}
\Xi^R_{\nu'}\left(\mathbf{q}',\mathbf{q},\omega\right) 
= 
\sum_{n} \underset{\rm 1. BZ}{\int} \frac{\mathrm{d}^3 \mathbf{k}}{\left(2\pi\right)^3} \Bigl[ \frac{N_{+,n,\nu'} \left(\mathbf{q}',\mathbf{q},\mathbf{k}\right)}{\frac{\epsilon_{n,\mathbf{k} + \mathbf{q}'}}{\hbar} - \frac{\epsilon_{n,\mathbf{k}}}{\hbar} + \omega_{\nu',\mathbf{q}-\mathbf{q}'} - \omega - i\delta} - \frac{N_{-,n,\nu'} \left(\mathbf{q}',\mathbf{q},\mathbf{k}\right)}{\frac{\epsilon_{n,\mathbf{k} + \mathbf{q}'}}{\hbar} - \frac{\epsilon_{n,\mathbf{k}}}{\hbar} - \omega_{\nu',\mathbf{q}-\mathbf{q}'} - \omega - i \delta}\Bigr],
\label{eq:retarded_1-electron-2-phonon_convolution}
\end{equation}
with the shorthand occupation functions
\begin{equation}
\begin{split}
N_{+,n,\nu'} \left(\mathbf{q}',\mathbf{q},\mathbf{k}\right) =& \, 
n_{F} \left(\epsilon_{n,\mathbf{k}}\right) \Big[1-n_{F} \left( \epsilon_{n,\mathbf{k} + \mathbf{q}'} \right)\Big] 
+ 
\Big[ n_F \left(\epsilon_{n,\mathbf{k}}\right) - n_F \left(\epsilon_{n,\mathbf{k} + \mathbf{q}'}\right) \Big] n_B \left(\omega_{\nu',\mathbf{q} - \mathbf{q}'}\right) \\
N_{-,n,\nu'} \left(\mathbf{q}',\mathbf{q},\mathbf{k}\right) =& \, 
n_{F} \left(\epsilon_{n,\mathbf{k} + \mathbf{q}'}\right) \Big[1-n_{F} \left(\epsilon_{n,\mathbf{k}}\right)\Big] 
- 
\Big[ n_F \left(\epsilon_{n,\mathbf{k}}\right) - n_F \left(\epsilon_{n,\mathbf{k} + \mathbf{q}'}\right) \Big] n_B \left(\omega_{\nu',\mathbf{q} - \mathbf{q}'}\right). \label{eq:convolution_numerator_definition}
\end{split}
\end{equation}
If the $n$-th electron band is fully occupied, the Fermi-Dirac distributions are equal to unity, $n_{F} \left(\epsilon_{n,\mathbf{k}}\right) = 1$, such that the numerators vanish, $N_{\pm,n,\nu'} \left(\mathbf{q}',\mathbf{q},\mathbf{k}\right) = 0$. Consequently only partially filled electronic bands contribute to $\Xi^R_{\nu'}\left(\mathbf{q}',\mathbf{q},\omega\right)$. Using the convolution as defined in Eq.~\eqref{eq:convolution_Lindhard_phonon}, the one-electron-two-phonon self-energy can then be written as,
\begin{equation}
\Pi_{\nu\mu}^{\rm (1el-2ph)} \left(\mathbf{q}, \omega\right) 
= 
\frac{4 \Omega_0^2}{\hbar^2} 
\sum_{\nu'} \underset{\rm 1. BZ}{\int} \frac{\mathrm{d}^3 \mathbf{q}'}{\left(2\pi\right)^3} 
\Xi_{\nu'}^{\rm R} \left(\mathbf{q}',\mathbf{q},\omega\right) \sum_{\mathbf{G}} g_{\nu'\nu} \left(\mathbf{q},\mathbf{q}' + \mathbf{G}\right) g_{\nu'\mu}^* \left(\mathbf{q},\mathbf{q}' + \mathbf{G}\right).
\end{equation}
The function $\Xi_{\nu'}^{\rm R} \left(\mathbf{q}',\mathbf{q},\omega\right)$ depends on $\mathbf{q}'$ only through the electronic dispersion $\epsilon_{n,\mathbf{q}'+\mathbf{k}}$ and the phonon dispersion $\omega_{\nu',\mathbf{q}-\mathbf{q}'}$, both of which are periodic with respect to reciprocal lattice vectors $\mathbf{G}$. Therefore, it holds that $\Xi_{\nu'}^{\rm R} \left(\mathbf{q}',\mathbf{q},\omega\right) = \Xi_{\nu'}^{\rm R} \left(\mathbf{q}' + \mathbf{G},\mathbf{q},\omega\right)$, allowing us to write
\begin{equation}
\begin{split}
\Pi_{\nu\mu}^{\rm (1el-2ph)} \left(\mathbf{q}, \omega\right) 
= 
\frac{4 \Omega_0^2}{\hbar^2} 
\sum_{\nu'} 
\underset{\rm 1. BZ}{\int} \frac{\mathrm{d}^3 \mathbf{q}'}{\left(2\pi\right)^3} \sum_{\mathbf{G}} 
\Xi_{\nu'}^{\rm R} \left(\mathbf{q}' + \mathbf{G},\mathbf{q},\omega\right)  g_{\nu'\nu} \left(\mathbf{q},\mathbf{q}' + \mathbf{G}\right) g_{\nu'\mu}^* \left(\mathbf{q},\mathbf{q}' + \mathbf{G}\right).
\end{split}
\end{equation}
Introducing the momentum $\mathbf{Q} = \mathbf{q}' + \mathbf{G}$ allows to merge the integration over $\mathbf{q}'$ and summation over all reciprocal lattice vectors $\mathbf{G}$ into a single integral over $\mathbf{Q}$ spanning all of $\mathbb{R}^3$. This is possible because any vector $\mathbf{Q}$ in reciprocal space $\mathbb{R}^3$ can be uniquely decomposed as $\mathbf{Q} = \mathbf{q}' + \mathbf{G}$, with $\mathbf{q}'$ being restricted to the first Brillouin zone. In the long-range approximation, however, the dominant contributions to the integral arise from small values of $\mathbf{q}' + \mathbf{G}$, i.e. from momenta far away from the Brillouin zone boundary. In other words, the corresponding integrand effectively vanishes before reaching the boundary, allowing us to neglect terms with $\mathbf{G} \neq 0$, since they correspond to large $|\mathbf{Q}|$. Under this assumption, the integral over the first Brillouin zone can be extended to the entire reciprocal space without changing the result. Depending on numerical convenience, we may therefore perform the integration over either domain. This reasoning allows us to write
\begin{equation}
\begin{split}
\Pi_{\nu\mu}^{\rm (1el-2ph)} \left(\mathbf{q}, \omega\right) = \frac{4 \Omega_0^2}{\hbar^2} \sum_{\nu'} \underset{\mathbb{R}^3}{\int} \frac{\mathrm{d}^3 \mathbf{Q}}{\left(2\pi\right)^3} \Xi_{\nu'} \left(\mathbf{Q},\mathbf{q},\omega\right) g_{\nu'\nu} \left(\mathbf{q},\mathbf{Q}\right) g_{\nu'\mu}^* \left(\mathbf{q},\mathbf{Q}\right).
\end{split}
\end{equation}
Inserting the expression for the vertex factor $g_{\nu'\nu} \left(\mathbf{q},\mathbf{Q}\right)$ yields
\begin{equation}
\Pi_{\nu\mu}^{\rm (1el-2ph)} \left(\mathbf{q}, \omega\right) 
= 
\frac{e^4}{\hbar^2 \varepsilon_0^2} 
\sum_{\alpha,\beta} \sum_{\nu'} 
Y_{\nu'\nu,\alpha} \left(\mathbf{q}\right) Y_{\nu'\mu,\beta}^* \left(\mathbf{q}\right) \underset{\mathbb{R}^3}{\int} \frac{\mathrm{d}^3 \mathbf{Q}}{\left(2\pi\right)^3} \Xi_{\nu'} \left(\mathbf{Q},\mathbf{q},\omega\right)  \frac{Q_{\alpha} Q_{\beta}}{\left(\mathbf{Q}\cdot\boldsymbol{\varepsilon}_{\infty}\cdot \mathbf{Q}\right)^2}.
\label{eq:1-electron-2-phonon_self_energy_final_expression}
\end{equation}
\end{widetext}
This expression constitutes our main result regarding the nonlinear one-electron-two-phonon self-energy. It involves summations over the Cartesian components~$\alpha,\beta\in\lbrace x,y,z\rbrace$, as well as over the internal phonon scattering channels $\nu'$. The material-specific input enters through (i) the high-frequency dielectric constant $\boldsymbol{\varepsilon}_{\infty}$, (ii) the derivative of the dynamical matrix with respect to the electric field \textit{via} $Y_{\nu'\nu,\alpha} \left(\mathbf{q}\right)$, and (iii) the dependence of~$\Xi_{\nu'} \left(\mathbf{Q},\mathbf{q},\omega\right)$ on the phononic and electronic band structure. 

A useful way to read Eq.~\eqref{eq:1-electron-2-phonon_self_energy_final_expression} is to note that, unlike the electron-hole polarization bubble in Eq.~\eqref{eq:1-electron-1-phonon_self_energy_final}, the one-electron-two-phonon self-energy probes both the electronic and the phononic spectra simultaneously. The incoming phonon scatters into an electron-hole pair and an internal phonon, so the magnitude of $\Pi^{(\mathrm{1el-2ph})}(\mathbf{q},\omega)$ at a given $(\mathbf{q},\omega)$ reflects two distinct factors: how readily the conduction electrons can sustain a particle-hole excitation at the relevant energy, encoded in $\chi(\mathbf{q}',\omega')$, and how many phonon channels $\nu'$ are available to absorb the remaining momentum and frequency mismatch, encoded in $D^{(0)}_{\nu',\nu'}(\mathbf{q}-\mathbf{q}',\omega-\omega')$ together with the Bose-Einstein distribution factor $n_B(\omega_{\nu',\mathbf{q}-\mathbf{q}'})$. As a consequence, materials with many low-frequency phonon branches at thermally accessible energies open up additional intermediate channels and enhance $\Pi^{(\mathrm{1el-2ph})}$, showing that this process is particularly important in soft materials. For a fixed phonon spectrum, the same expression samples the electronic polarization through its convolution with the internal phonon propagator. The external phonon momentum can span the Brillouin zone even though the electronic momentum transfer entering the long-range vertex remains concentrated in the long-wavelength region. This dual sensitivity has no counterpart in the linear phonon self-energy and is the origin of the qualitative differences we report in the following.

In the next two paragraphs, we will show that the expression for $\Pi_{\nu\mu}^{\rm (1el-2ph)} \left(\mathbf{q}, \omega\right)$ in Eq.~\eqref{eq:1-electron-2-phonon_self_energy_final_expression} simplifies considerably in the long-range limit and under the assumption of isotropic materials, which facilitates its numerical evaluation. 

\paragraph{Long-range limit for $\Xi_{\nu'} \left(\mathbf{q}',\mathbf{q},\omega\right)$.} 

In the long-range limit, we approximate the phonon dispersion entering $\Xi_{\nu'} \left(\mathbf{q}',\mathbf{q},\omega\right)$ in Eq.~\eqref{eq:retarded_1-electron-2-phonon_convolution} as
\begin{equation}
\omega_{\nu,\mathbf{q}-\mathbf{q}'} \approx \omega_{\nu,\mathbf{q}} + \mathcal{O} \left(|\mathbf{q}'|\right), \label{eq:phonon_dispersion_long_range_limit}
\end{equation}
since the dominant contributions in the occupation functions $N_{\pm,n,\nu'} \left(\mathbf{q}',\mathbf{q},\mathbf{k}\right)$ arise from momenta $\mathbf{q}'$ far away from the edge of the Brillouin zone, as $k_{\rm F} \ll 1/a$, and so $q' \ll 1/a$. In other words, most contributions in the integral over the internal momentum $\mathbf{q}'$ will come from $\mathbf{q}'$ close to the $\Gamma$ point, while the external momentum $\mathbf{q}$ remains unrestricted. In contrast, when $\mathbf{q}'$ appears in the electronic dispersion, such as in $\epsilon_{n,\mathbf{k} + \mathbf{q}'}$, we can not approximate it with $\epsilon_{n,\mathbf{k}}$, as $q' \sim k_{\rm F}$. In our numerical implementation, we incorporate the long-range limit of $\Xi_{\nu'}^{\rm R} \left(\mathbf{q}',\mathbf{q},\omega\right)$ by applying the approximation from Eq.~\eqref{eq:phonon_dispersion_long_range_limit},~i.e.~$\Xi_{\nu'}^{\rm R} \left(\mathbf{q}',\mathbf{q},\omega\right) \to \Xi^{R,\left(\rm long\right)}_{\nu'}\left(\mathbf{q}',\mathbf{q},\omega\right)$. A practical advantage of this approach is that we need to evaluate $\omega_{\nu,\mathbf{q}}$ only once at the external phonon momentum $\mathbf{q}$, which can be done before solving the momentum space integral. As calling the Fourier interpolator is computationally expensive, the long-range limit of the phonon dispersion leads to a significant reduction in computational cost.

\paragraph{Isotropic Materials.} By isotropic material, we mean a material with a cubic point group whose electron band close to the band extremum is given by a quadratic dispersion
\begin{equation}
\epsilon_k = \frac{\hbar^2 |k|^2}{2m^*},
\label{eq:quadratic_electronic_dispersion}
\end{equation}
where $m^*$ is the band effective mass. In such materials, physical properties are the same in every direction because the system is invariant under rotations. As a result, tensor quantities like the dielectric response $\boldsymbol{\varepsilon}_{\infty}$ are diagonal, $\boldsymbol{\varepsilon}_{\infty} = \varepsilon_{\infty} \mathbf{1}$. For the one-electron-two-phonon diagram in Eq.~\eqref{eq:1-electron-2-phonon_self_energy_final_expression}, it turns out that the response function $\Xi_{\nu'}^{\rm R}\left(\mathbf{Q},\mathbf{q},\omega\right)$ depends only on the magnitude of the exchanged momentum, $\Xi_{\nu'}^{\rm R}\left(\mathbf{Q},\mathbf{q},\omega\right) = \Xi_{\nu'}^{\rm R}\left(Q,\mathbf{q},\omega\right)$, with $Q = |\mathbf{Q}|$. This allows us to split the integration over $\mathbf{Q}$ in Eq.~\eqref{eq:1-electron-2-phonon_self_energy_final_expression} into radial and angular contributions. The angular integral can be evaluated analytically and is proportional to $\delta_{\alpha\beta}$, see App.~\ref{sec:app_integral_identities}. The remaining radial integral can be expressed as a sum over the three Cartesian components,
\begin{equation}
\Pi^{(\text{1el-2ph})}_{\nu \mu} \left(\mathbf{q},\omega\right)
= \frac{1}{3} \sum_{\alpha} 
\Pi^{(\text{1el-2ph})}_{\nu \mu,\alpha} \left(\mathbf{q},\omega\right),
\label{eq:isotropic_1el_2ph_self_energy_split_into_components}
\end{equation}
with the component-wise expressions
\begin{equation}
\Pi^{(\text{1el-2ph})}_{\nu\mu,\alpha}(\mathbf{q},\omega) := \sum_{\nu'} 
Y_{\nu'\nu,\alpha} (\mathbf{q}) Y_{\nu'\mu,\alpha}^* (\mathbf{q}) \,
\xi^{\rm R} (\omega_{\nu',\mathbf{q}},\omega).
\end{equation}
Here we defined the shorthand function $\xi^{\rm R} (\omega_{\nu',\mathbf{q}},\omega)$ as
\begin{equation}
\xi^{\rm R} (\omega_{\nu',\mathbf{q}},\omega) 
:= \frac{1}{2\pi^2} 
\left( \frac{e^2}{\hbar\varepsilon_0 \varepsilon_\infty} \right)^{\!2}
\int_{0}^{\infty} 
\mathrm{d}Q
\,
\Xi_{\nu'}^{\rm R}(Q,\mathbf{q},\omega).
\end{equation}
For the numerical implementation of the response function $\Xi_{\nu'}^{\rm R}(Q,\mathbf{q},\omega)$ as written in Eq.~\eqref{eq:retarded_1-electron-2-phonon_convolution}, we align the phonon momentum $\mathbf{Q}$ along the $z$ axis without loss of generality, $\mathbf{Q} \to Q\mathbf{e}_z$, and apply the long-range approximation for the phonon frequencies according to Eq.~\eqref{eq:phonon_dispersion_long_range_limit}. We now relate the $x$ and $y$ components of $\Pi^{(\text{1el-2ph})}_{\nu\mu,\alpha}(\mathbf{q},\omega)$ to the $z$ component using the following rotation $\mathbf{R}$ from the cubic point group,
\begin{equation}
    \mathbf{R} = \left(\begin{matrix}
        0 & 0 & 1 \\
        1 & 0 & 0 \\
        0 & 1 & 0
    \end{matrix}\right). \label{eq:rotation_matrix_to_simplify_iso_calculation}
\end{equation}
In particular, $\mathbf{R}$ rotates the $z$ component of a vector to $x$, while the inverse $\mathbf{R}^{-1}$ rotates $y$ into $z$. If we rotate the wavevector $\mathbf{q}$ in the opposite direction to compensate, we can express the $x$- and $y$-components of the one-electron-two-phonon self-energy in terms of the $z$-component as follows,
\begin{equation}
\begin{split}
    \Pi^{(\text{1el-2ph})}_{\nu\mu,x}(\mathbf{q},\omega) =& \, \Pi^{(\text{1el-2ph})}_{\nu\mu,z}(\mathbf{R}^{-1}\cdot\mathbf{q},\omega), \\
    \Pi^{(\text{1el-2ph})}_{\nu\mu,y}(\mathbf{q},\omega) =& \, \Pi^{(\text{1el-2ph})}_{\nu\mu,z}(\mathbf{R}\cdot\mathbf{q},\omega). \label{eq:dia_iso_1el_2ph_self_energy_x_and_y_component}
\end{split}
\end{equation}
Here, we also use that $\xi^{\rm R} (\omega_{\nu',\mathbf{q}},\omega)$ only depends on the phonon momentum $\mathbf{q}$ and the band index $\nu'$ through the phonon frequencies $\omega_{\nu',\mathbf{q}}$. A detailed derivation that also includes the transformation properties of $Y_{\nu'\nu,\alpha} (\mathbf{q})$ is discussed in App.~\ref{sec:symmetry_transformations}. The result in Eq.~\eqref{eq:dia_iso_1el_2ph_self_energy_x_and_y_component} is particularly useful for numerical implementations in any cubic material, as we only need to evaluate $\Pi^{(\text{1el-2ph})}_{\nu\mu,z}(\mathbf{q},\omega)$ at symmetry-related, rotated $\mathbf{q}$ points. In particular, the self-energy as written in Eq.~\eqref{eq:isotropic_1el_2ph_self_energy_split_into_components} is invariant under point group operations $\mathbf{R}$, i.e.~$\Pi^{(\text{1el-2ph})}_{\nu \mu} \left(\mathbf{R}\cdot\mathbf{q},\omega\right) = \Pi^{(\text{1el-2ph})}_{\nu \mu} \left(\mathbf{q},\omega\right)$, which follows from the symmetry properties $\mathbf{R}^3 = 1$ and $\mathbf{R}^2 = \mathbf{R}^{-1}$. Therefore, the phonon self-energy inherits the same symmetries as the phonon frequencies, and it is sufficient to define it within the irreducible Brillouin zone. We want to emphasize that to derive the relations between the different Cartesian components of the one-electron-two-phonon self-energy for isotropic materials in Eq.~\eqref{eq:dia_iso_1el_2ph_self_energy_x_and_y_component}, we only used that the rotation $\mathbf{R}$ is in the point group of the material. Consequently, we can use Eq.~\eqref{eq:isotropic_1el_2ph_self_energy_split_into_components} and Eq.~\eqref{eq:dia_iso_1el_2ph_self_energy_x_and_y_component} to simplify the calculations for any cubic material.

\section{Linear and non-linear phonon renormalization for lithium fluoride and potassium tantalate}

\subsection{Physical properties of lithium fluoride}

Lithium fluoride, LiF, is a wide-gap ionic insulator with a band gap of $\Delta E = 8.9\,\text{eV}$~\cite{Sommer2012,Piacentini1976}. Its simple rock-salt structure and strong ionic bonding give rise to well-defined long-range electron-phonon interactions, making LiF an established model system for polaron physics~\cite{sioInitioTheoryPolarons2019,franchiniPolaronsMaterials2021}.

\begin{figure*}[!t]
    \centering
    \includegraphics[width=\linewidth]{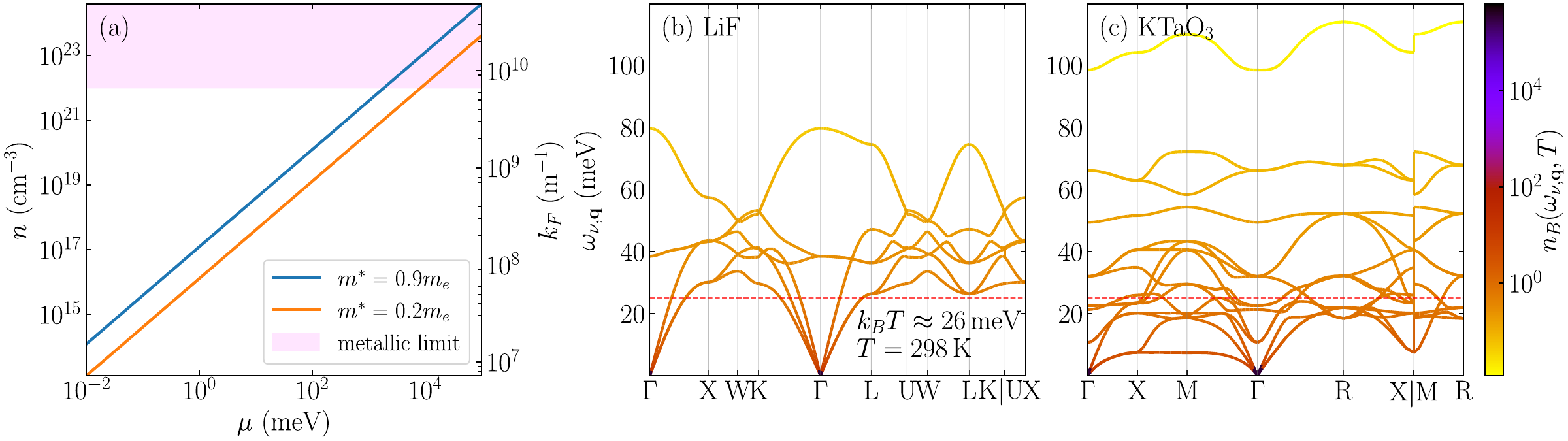}
    \vspace{-0.8cm}
    \caption{(a) Charge carrier density $n$ and Fermi wavevector $k_{\rm F}$ as a function of the chemical potential $\mu$ for a three-dimensional parabolic band, shown for the representative band effective masses of LiF~\cite{sioInitioTheoryPolarons2019} and CsPbI$_3$~\cite{filipPhononScreeningExcitons2021}. The shaded region highlights carrier densities typical of metals. For a fixed chemical potential, a larger band effective mass gives higher carrier densities and larger Fermi wavevectors. (b) and (c) Phonon band structures of LiF and KTaO$_3$ along a selected high-symmetry path in the Brillouin zone, cf.~\cite{houtputFirstprinciplesTheoryNonlinear2025a}. The color bar indicates the phonon occupation at room temperature according to the Bose-Einstein distribution. The dashed red line marks the energy scale for room temperature.}
    \label{fig:fig4}
\end{figure*}

The phonon spectrum of LiF extends up to energies of approximately $80\,\text{meV}$. In particular, at the $\Gamma$ point, the three optical branches display a pronounced LO-TO splitting, reflecting the polar nature of the material, while the transverse optical modes remain degenerate. At room temperature, thermal phonon occupation is dominated by the three acoustic branches, whereas the higher-energy optical modes are only weakly populated, as illustrated in Fig.~\ref{fig:fig4}(b).

  Most previous studies on electron-phonon physics in LiF have focused on the renormalization of electronic properties such as the band gap and the effective mass~\cite{antoniusDynamicalAnharmonicEffects2015,sioInitioTheoryPolarons2019,luoFirstprinciplesDiagrammaticMonte2025,houtputFirstprinciplesTheoryNonlinear2025a}, while the corresponding renormalization of phonon properties has received much less attention. In thermal equilibrium, the electronic contribution to the phonon self-energy is negligible, as the valence band is fully occupied and the conduction band is essentially empty. As a result, the density of intrinsic mobile carriers is exponentially small and therefore, any sizable renormalization of phonons by electrons requires an out-of-equilibrium electronic distribution.

To circumvent this constraint, we consider an \emph{effective} chemical potential inside the conduction band, which mimics a nonequilibrium carrier population, as it is for example realized by transient photodoping~\cite{crepaldiUltrafastPhotodopingEffective2012} or by hot-carrier injection at a metal/insulator interface~\cite{sunElectricalTuningEffect2021}, see the schematic in  Fig.~\ref{fig:fig1}(a). For a parabolic conduction band with band effective mass of $0.88 m_e$~\cite{sioInitioTheoryPolarons2019}, a representative choice of $\mu = 100~\mathrm{meV}$ corresponds to a dilute electron gas with a carrier density of around $10^{20}\,\text{cm}^{-3}$ and a Fermi momentum of $k_{\rm F} \sim 10^{9}~\mathrm{m}^{-1}$, cf. Fig.~\ref{fig:fig4}(a). Any electronic excitations are in the long-wavelength regime because the edge of the first Brillouin zone is at $2\pi / a \sim 10^{10}~\mathrm{m}^{-1}$ and therefore $k_{\rm F} < 2\pi / a$ ($a = 4\,$\AA~\cite{Dressler1987}). In this limit, the valence band remains fully occupied, while the conduction electrons provide a tunable screening channel for polar lattice vibrations. At the same time, the phonons of LiF are well in the harmonic regime, providing a reference background to compare the influence of the one-electron-one-phonon and one-electron-two-phonon contributions to the phonon self-energy. 

\subsection{Physical properties of potassium tantalate}

Potassium tantalate, KTaO$_3$, is a cubic quantum paraelectric in the perovskite structure featuring significant anharmonicity in  a soft polar mode~\cite{singhStabilityPhononsKTaO1996,ranalliTemperatureDependentAnharmonicPhonons2023,essweinFirstprinciplesCalculationElectronphonon2023}. The conduction band minimum is two-fold degenerate, which is a consequence of the large spin-orbit coupling of the heavy tantalum atom~\cite{kingSubbandStructureTwoDimensional2012,fan2DEGsPerovskiteInterfaces2014}. However, the theory of Sec.~\ref{subsubsec:one_electron_two_phonon_diagram}~c was derived assuming a single parabolic electron band, as given by Eq.~\eqref{eq:quadratic_electronic_dispersion}. In order to use this theory, we replace the band structure of KTaO$_3$ with a single parabolic band, and use the single-band effective mass~$m^* = 0.4 m_e$, as derived in Ref.~\cite{houtputFirstprinciplesTheoryNonlinear2025a} for the polaron ground state energy. We expect that this replacement does not change any of our qualitative results or conclusions.

In contrast to LiF, the band gap in KTaO$_3$ is smaller and lies between $3.6\,\mathrm{eV}$ for the indirect gap and $4.3\,\mathrm{eV}$ for the direct gap at the $\Gamma$ point~\cite{jellisonOpticalFunctionsKTaO2006,fan2DEGsPerovskiteInterfaces2014,javedEfficientFirstprinciplesEvaluation2025}. Therefore, the electronic contribution to the phonon self-energy from thermal electronic excitations will be larger than in LiF, but it remains small and similarly to before, we consider an \textit{effective} chemical potential inside the conduction band. 

The phonon band structure of KTaO$_3$ has 15 branches, with the largest phononic excitations around $115\,\mathrm{meV}$, see Fig.~\ref{fig:fig4}(c). Importantly, at room temperature, between 8 and 9 phonon bands are significantly populated, which is in contrast to LiF, where only around three acoustic branches are thermally activated. The phonon renormalization from the nonlinear one-electron-two-phonon interaction vertex in KTaO$_3$ should therefore be more pronounced than in LiF, as we will show below. \\

In the following, we compare the one-electron-one-phonon and one-electron-two-phonon contributions to the phonon self-energy as functions of the chemical potential, set by the doping density, and temperature for LiF and KTaO$_3$. In both materials, we find that the one-electron-one-phonon term mainly renormalizes phonons near the $\Gamma$ point, whereas the one-electron-two-phonon self-energy renormalizes the bands across the entire Brillouin zone due to relaxed kinematic constraints on the scattering process. Furthermore, the one-electron-two-phonon renormalization increases with temperature, as additional phonon branches become thermally accessible. In LiF, only the three acoustic branches are populated at room temperature, see~Fig.~\ref{fig:fig4}(b), resulting in a quantitatively small one-electron-two-phonon phonon renormalization. The phonon renormalization in KTaO$_3$ is about two to five times larger, which we attribute to the larger number of thermally available phononic scattering channels. This trend suggests that in materials with many low-frequency phonon modes, such as soft semiconductors, e.g.~lead halide perovskites~\cite{miyataLargePolaronsLead2017,martiradonnaRiddlesPerovskiteResearch2018,qianPhotocarrierinducedPersistentStructural2023} or Bi$_2$O$_2$Se~\cite{wangPhononSignaturesPolaron2022}, the nonlinear one-electron-two-phonon process becomes significantly more important.

\subsection{First-principles setup and numerical integration of the Feynman diagrams}

For the phonon self-energy calculations, we use the LiF and KTaO$_3$ phonon band structures and electron-phonon coupling vertices as reported in \cite{houtputFirstprinciplesTheoryNonlinear2025a}. For completeness, we reproduce the numerical parameters of this first-principles setup below: for more technical details, especially surrounding the calculation of the electric field derivative of the dynamical matrix $\partial \mathcal{D}_{\kappa \alpha, \kappa' \beta}(\mathbf{q})/\partial \boldsymbol{\mathcal{E}}$, we refer to Ref.~\cite{houtputFirstprinciplesTheoryNonlinear2025a}. The first-principles calculations were performed with the Vienna Ab Initio Simulation Package (VASP)~\cite{kresseInitioMolecularDynamics1993,kresseEfficientIterativeSchemes1996,kresseEfficiencyAbinitioTotal1996}, using the Perdew-Burke-Ernzerhof functional for solids (PBEsol)~\cite{perdewRestoringDensityGradientExpansion2008}. For all calculations, we use the default projector augmented wave (PAW) pseudopotentials from the potpaw.54 dataset included in VASP. This means the valence configurations are (1s$^2$ 2s$^2$) for LiF, (2s$^2$ 2p$^5$) for F, (3s$^2$ 3p$^6$ 4s$^1$) for K, (5p$^6$ 5d$^4$ 6s$^1$) for Ta, and (2s$^2$ 2p$^4$) for O. Supercells and displacements for VASP calculations have been generated with the PhonoPy package~\cite{togoFirstPrinciplesPhonon2015,togoImplementationStrategiesPhonopy2023,togoFirstprinciplesPhononCalculations2023}. For LiF, we perform phonon calculations on a $6 \times 6 \times 6$ supercell, and for KTaO$_3$ on a $4 \times 4 \times 4$ supercell. Unit cell calculations for LiF and KTaO$_3$ are performed with a $12 \times 12 \times 12$ $\Gamma$-centered $\mathbf{k}$-grid and an $8 \times 8 \times 8$ Monkhorst-Pack $\mathbf{k}$-grid, respectively; for the supercell calculations, we use an appropriately scaled $2 \times 2 \times 2$ $\mathbf{k}$-grid in both cases. We evaluate~$\partial \mathcal{D}_{\kappa \alpha, \kappa' \beta}(\mathbf{q})/\partial \boldsymbol{\mathcal{E}}$ by calculating the dynamical matrix at finite electric fields $\bm{\mathcal{E}} = \pm \mathcal{E} \mathbf{e}_z$, and then approximating the electric field derivative as a finite difference. For LiF we use $\mathcal{E} = 0.01 \,\text{V} / \text{\AA}$, and for KTaO$_3$ we use $\mathcal{E} = 0.005 \,\text{V} / \text{\AA}$. In both cases we set the energy convergence tolerance to a strict value of $10^{-10} \text{eV}$, to avoid numerical noise when calculating the finite difference derivative with respect to $\mathcal{E}$. Spin-orbit coupling is neglected in all phonon calculations; we only include it when calculating the electronic band structure of KTaO$_3$, to derive its effective single-band mass of $0.4 m_e$.

We focus in the following on the lowest order calculation of the one-electron-one-phonon and one-electron-two-phonon self-energy according to Eq.~\eqref{eq:1-electron-1-phonon_self_energy_final} and Eq.~\eqref{eq:isotropic_1el_2ph_self_energy_split_into_components}. To obtain the phonon renormalization, we diagonalize the rescaled inverse phonon propagator $(\boldsymbol{\tilde{D}}^{\rm R} (\mathbf{q},\omega))^{-1}$ evaluated at the unperturbed frequencies $\omega = \omega_{\nu,\mathbf{q}}$, see App.~\ref{subsec:diagonalizing_dyson_equation_with_off_diagonal_terms} as well as App.~\ref{subsec:app_numerical_integration} for details on the numerical computation of momentum-space integrals. Within this procedure, the complex self-energy gives rise to both the energy shift from its real part, $\Delta\omega_{\nu,\mathbf{q}}$, and the linewidth from its imaginary part, as summarized in Eq.~\eqref{eq:summary_diagonal_deg_approximate_renormalization} for the case that just the diagonal self-energy terms contribute to the renormalization. Below we focus on the energy renormalization as linewidths are typically dominated by other phonon scattering mechanisms.  

\subsection{The one-electron-one-phonon self-energy} 
\label{subsec:one_electron_one_phonon_self_energy}

The magnitude of the one-electron-one-phonon self-energy, calculated within the long-range approximation from Eq.~\eqref{eq:1-electron-1-phonon_self_energy_final}, is controlled by the long-range coupling vertex strength $g_{\nu} (\mathbf{q})$ and the retarded electron polarization bubble $\chi^{\rm R} (\mathbf{q},\omega)$, see Eq.~\eqref{eq:retarded_Lindhard_final}. The vertex sets the strength of the coupling, while the polarization bubble encodes the dynamical screening of polar lattice vibrations by electron-hole excitations near the Fermi level in the conduction band. The phonon is renormalized whenever its energy can be matched by an electron-hole excitation, i.e.~whenever the denominator of $\chi^{\rm R}$ becomes resonant, $\epsilon_{n,\mathbf{k}+\mathbf{q}} - \epsilon_{n,\mathbf{k}} \sim \hbar\omega$. For a parabolic conduction band with effective mass $m^*$ at zero temperature and small $q$, this resonance condition is satisfied within a continuum of particle-hole excitations bounded by $|\omega| \leq v_F\, q$, where $v_F = \hbar k_{\rm F} / m^*$ is the Fermi velocity. 

Thus, acoustic and optical phonons behave very differently with respect to this continuum. Acoustic phonons with sound velocities $v_s < v_F$ lie inside the particle-hole continuum and decay into electron-hole pairs (Landau damping), so they are simultaneously frequency-shifted and broadened by the real and imaginary part of $\chi^{\rm R}$. High-frequency optical phonons, by contrast, sit outside the continuum and are renormalized only through virtual electron-hole pairs, leaving their lifetime essentially unaffected within this channel. 

The same reasoning explains why the long-range approximation for the coupling vertex is self-consistent here. While the continuum of electron-hole excitations grows with $\mathbf{q}$, the phase-space overlap between occupied and empty states shrinks as governed by the Fermi-Dirac distributions, so $\chi^{\rm R} (\mathbf{q},\omega) \to 0$ for $q \gg k_{\rm F}$. For modest excitation densities, $k_{\rm F} \ll 2\pi / a$, contributions from large $\mathbf{q}$ are therefore suppressed by the decaying polarization bubble, and the renormalization is dominated by the small $\mathbf{q}$ region where the long-range vertex is accurate. Within this small-$\mathbf{q}$ window, a finer feature is the well-known Kohn anomaly~\cite{Kohn1959}. For a well-defined Fermi surface, $\mathrm{Re}\,\chi^{\rm R}(\mathbf{q},0)$ develops a nonanalyticity at wavevectors connecting points of the Fermi surface, which in an isotropic parabolic band occurs at $|\mathbf{q}|=2k_{\rm F}$. In three dimensions, the derivative of the susceptibility has a logarithmic singularity, which imprints sharp features on the renormalized phonon dispersion. 

\begin{figure}[t]
    \centering
    \includegraphics[width=\columnwidth]{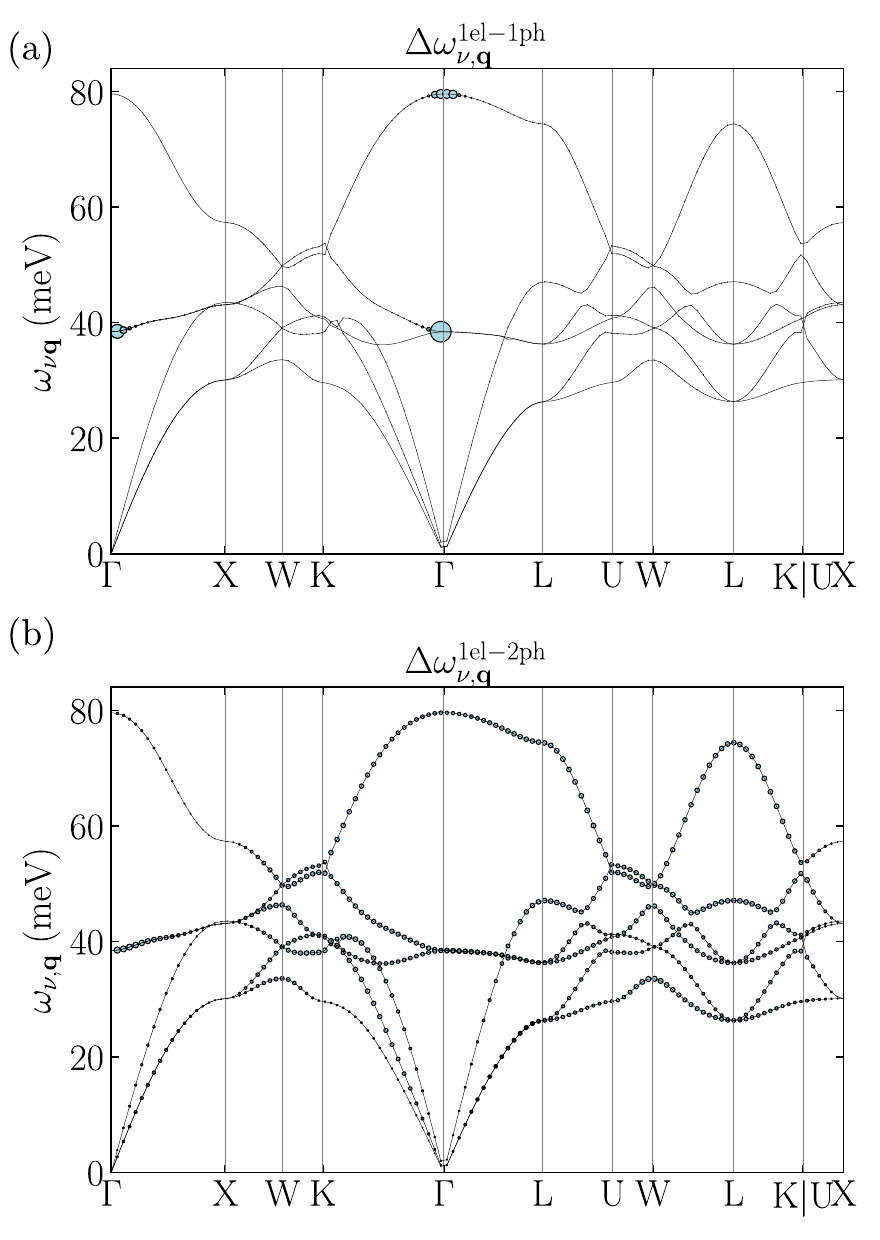}
    \vskip -0.5cm
    \caption{Phonon energy renormalization in LiF due to the one-electron-one-phonon self-energy in (a) and due to the one-electron-two-phonon self-energy in (b). The area of the circles in (a) is proportional to the correction $\Delta \omega_{\nu,\mathbf{q}}^{\rm 1el-1ph}$, where we introduced a cutoff for the maximum renormalization of~$40\,\mathrm{meV}$ due to the sharp renormalization features near $\Gamma$. Because the one-electron-one-phonon renormalization of the acoustic branches is smaller than for the optical branches, the circles on top of the acoustic branches are barely visible. The circle areas in (b) are proportional to the one-electron-two-phonon renormalization $\Delta \omega_{\nu,\mathbf{q}}^{\rm 1el-2ph}$, with the largest correction being~$\Delta \omega_{\nu,\mathbf{q}}^{\rm 1el-2ph} \approx 3.3\times 10^{-5}\,\mathrm{meV}$. This energy is significantly smaller than the shifts in (a), meaning the one-electron-two-phonon renormalization in LiF is essentially negligible. The chemical potential and temperature were set to $\mu = 100\,\mathrm{meV}$ and $348\,\mathrm{K}$ respectively.}
    \label{fig:fig5}
\end{figure}

We confirm this physical picture for LiF with our numerical calculation, see Fig.~\ref{fig:fig5}(a) and for complementary data App.~\ref{sec:app_lithium_fluoride}. The  phonon renormalization in KTaO$_3$ behaves qualitatively similar and is presented in Fig.~\ref{fig:KTaO3_phonon_self_energy_corrections_on_band_structure}(a) of App.~\ref{sec:app_potassium_tantalate}. In both materials, the one-electron-one-phonon renormalization is sharply localized near the $\Gamma$ point for all phonon branches, consistent with the $1/|\mathbf{q}|$ structure of the long-range polar vertex and with the rapid decay of $\chi^{\rm
R}(\mathbf{q},\omega)$ for $q \gg k_{\rm F}$. Within each phonon family, the
$\mathbf{q}$ dependence reflects the two physical regimes identified
above. The acoustic branches lie inside the particle-hole
continuum and are predominantly Landau-damped; their sub-meV modulation
along $L$-$U$-$W$-$L$-$K|U$ comes from the $\mathbf{q}$-dependent projection of
the mode polarity vector $\mathbf{p}_\nu(\mathbf{q})$ onto the long-range
polar direction, which varies smoothly along the path. On the other hand, the optical branches are renormalized through virtual electron-hole pairs, since they sit above the particle-hole continuum. The highest optical branch in particular displays a characteristic sign-changing oscillation around the $\Gamma$ point, see Fig.~\ref{fig:lithium_fluoride_energy_renormalization_1_electron_1_phonon_self_energy} of App.~\ref{sec:app_lithium_fluoride} for a more detailed view. This dispersive feature reflects the structure of the polarization bubble $\chi^{\rm R}(\mathbf{q},\omega)$, and is a well-documented signature of polar-phonon renormalization~\cite{bergesPhononSelfEnergyCorrections2023}.

\subsection{The one-electron-two-phonon self-energy}
\label{subsec:one_electron_two_phonon_self_energy}

The response function $\Xi^R_{\nu'}(\mathbf q',\mathbf q,\omega)$ defined in Eq.~\eqref{eq:retarded_1-electron-2-phonon_convolution} describes the combined electronic and phononic phase space of the nonlinear process. In contrast to the linear polarization bubble, where the external phonon momentum $\mathbf q$ is transferred directly to an electron-hole pair, the nonlinear process creates an electron-hole excitation carrying momentum $\mathbf q'$ together with an internal phonon carrying momentum $\mathbf q-\mathbf q'$. Energy conservation can therefore be satisfied through the sum or difference of the external and internal phonon energies, corresponding to phonon-emission and phonon-absorption channels. This additional degree of freedom relaxes the direct kinematic connection between the external phonon momentum and the electronic excitation, allowing the nonlinear self-energy to remain finite over a much larger portion of the Brillouin zone, cf.~Fig.~\ref{fig:fig5}(b). At the same time, Pauli blocking confines electronic excitations to the vicinity of the Fermi surface, whereas the additional phonon degree of freedom, weighted by Bose-Einstein occupation factors, expands the available scattering phase space and softens the sharp momentum selectivity of the electron-hole polarization bubble.

\begin{figure*}[!t]
    \centering
    \includegraphics[width=\textwidth]{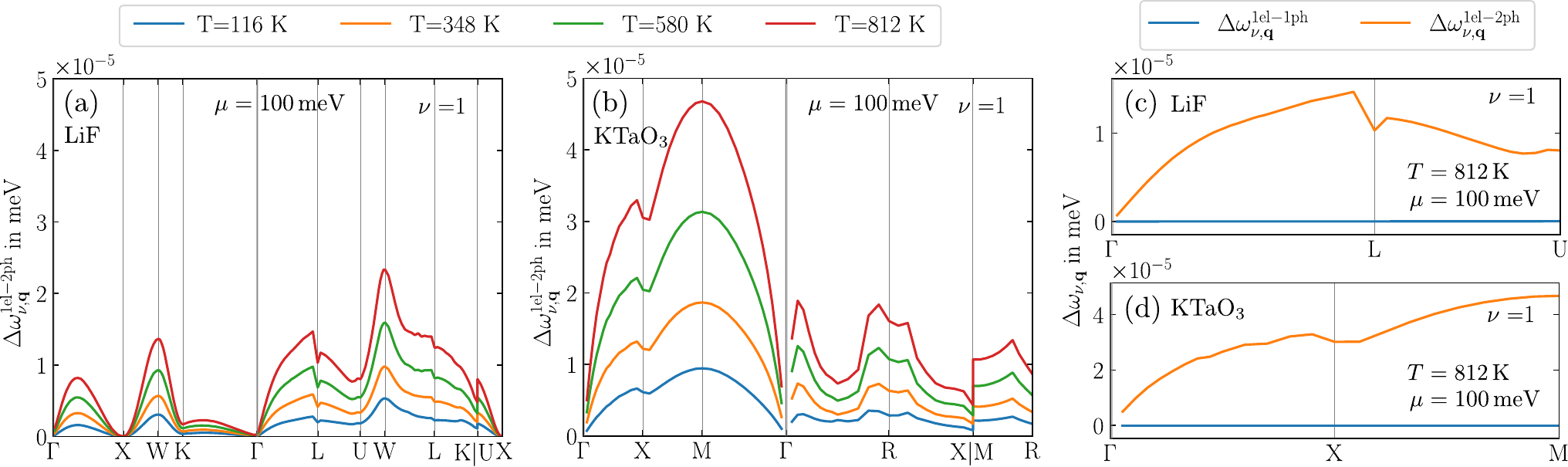}
    \vskip -0.2cm
    \caption{Panels (a) and (b) show the  temperature dependent renormalization of the first acoustic branch coming from the one-electron-two-phonon interaction for LiF and KTaO$_3$, respectively. The magnitude of the renormalization increases with temperature and varies across the Brillouin zone, reflecting the strong momentum dependent coupling to electronic excitations for this interaction process. Panels (c) and (d) show a direct comparison of the one-electron-one-phonon and one-electron-two-phonon contributions to the phonon self-energy for LiF and KTaO$_3$ respectively. For both materials, we show the respective renormalization for the first acoustic branch along path segments where the one-electron-two-phonon self-energy dominates the one-electron-one-phonon self-energy.}
    \label{fig:fig6}
\end{figure*}

Besides these distinct kinematic and phase space constraints, the one-electron-two-phonon self-energy has a qualitatively different temperature dependence. The internal phonon line carries a Bose-Einstein factor $n_B(\omega_{\nu',\mathbf{q}-\mathbf{q}'})$ that scales linearly with temperature at high $T$, $n_B\sim k_B T / (\hbar\omega)$, so every additional thermally accessible phonon mode contributes additively to the scattering. This is in sharp contrast to the linear self-energy, where temperature enters only through the Fermi-Dirac distribution, which broadens the occupation around the Fermi surface over an energy scale proportional to temperature but does not grow in magnitude. We expect, and observe in Figs.~\ref{fig:fig6}(a) and~\ref{fig:fig6}(b), that the nonlinear renormalization grows substantially for both LiF and KTaO$_3$ as soft phonon modes become thermally populated. 

As a result of both relaxed kinematic and phase space constraints, the two-phonon contribution is less sharply peaked in momentum space and exhibits a comparatively uniform renormalization across the Brillouin zone, albeit with a smaller overall magnitude due to the reduced probability of these higher-order scattering processes, cf.~\cite{houtputFirstprinciplesTheoryNonlinear2025a}.  

Regarding the overall magnitude, we find that the real parts of the one-electron-two-phonon self-energy are of order $10^{-5}\,\text{meV}$ in both materials studied here, as shown in Figs.~\ref {fig:fig6}(a) and~\ref {fig:fig6}(b). The correction in KTaO$_3$ is nevertheless two to three times larger than in LiF. This enhancement partly follows from the larger number of thermally occupied phonon branches in KTaO$_3$, which provide additional intermediate scattering channels. At room temperature, approximately eight to nine branches are appreciably populated in KTaO$_3$, whereas thermal occupation in LiF is largely restricted to its three acoustic branches, cf.~Figs.~\ref{fig:fig4}(b) and~\ref{fig:fig4}(c). However, in both cases the Bose-Einstein factor in the internal phonon line remains comparatively small, and because the scattering amplitude~$Y_{\nu'\nu,\alpha} (\mathbf{q}) Y_{\nu'\mu,\alpha}^* (\mathbf{q})$ is of the order of $10^{-6}\,\text{\AA}^{2}$ in both of these materials~\cite{houtputFirstprinciplesTheoryNonlinear2025a}, the impact of the nonlinear channel remains very small compared to the linear channel. We also note that the frequency renormalization extends almost uniformly across the entire Brillouin zone. However, this behavior is generally subject to symmetry constraints of the coupling vertex $Y_{\nu'\nu,\alpha} (\mathbf{q})$. In particular, the derivative of the dynamical matrix with respect to the electric field vanishes at the $X$ point in the rock salt structure, which suppresses the renormalization at this high-symmetry point in LiF, see Fig.~\ref{fig:fig6}(a).

The qualitative difference between the linear and nonlinear self-energy contributions is illustrated in Fig.~\ref{fig:fig6}(c) and Fig.~\ref{fig:fig6}(d), which show the phonon renormalization of the first acoustic branch along the high-symmetry path $\Gamma$-$L$-$U$ for LiF and $\Gamma$-$X$-$M$ for KTaO$_3$, respectively.

It turns out that for the acoustic branch, the linear contribution is suppressed by symmetry essentially everywhere along the path. The Born effective charge tensors and the mode polarity vector $\mathbf{p}_\nu(\mathbf{q})$ project the long-range vertex $g_\nu
(\mathbf{q})$ onto a vanishing component, leaving no scattering channel
through the polarization bubble.

In contrast, the nonlinear vertex ${Y}_{\nu'\nu,\alpha}(\mathbf{q})$ has a different symmetry structure, and the corresponding self-energy remains finite in these regions. The same behavior holds at any momentum where the linear vertex is symmetry suppressed, so the renormalized phonon dispersion along such a path can isolate the nonlinear contribution. In materials where the absolute magnitude of the renormalization is large enough to resolve experimentally, this provides a route to distinguish the two interaction channels by inelastic x-ray or neutron scattering and Terahertz spectroscopy. Natural candidates are for example soft polar semiconductors such as the lead halide perovskites or Bi$_2$O$_2$Se~\cite{miyataLargePolaronsLead2017,martiradonnaRiddlesPerovskiteResearch2018,qianPhotocarrierinducedPersistentStructural2023,wangPhononSignaturesPolaron2022}. 

Beyond distinguishing the two interaction channels, the structure of the nonlinear self-energy suggests a more direct application. In the one-electron-two-phonon self-energy as written in Eq.~\eqref{eq:1-electron-2-phonon_self_energy_final_expression}, the
polarization bubble~$\chi^{\rm R}(\mathbf{q}',\omega')$ enters as an
internal factor convolved with the phonon propagator, integrated
over the internal momentum $\mathbf{q}'$. The renormalized phonon
dispersion therefore carries direct information about the electronic
susceptibility of the medium across the Brillouin zone, with the
nonlinear vertex ${Y}_{\nu'\nu,\alpha}(\mathbf{q})$ acting as the coupling that probes it. This kind of indirect access to a medium response function through an interacting impurity is a general feature of quasiparticle self-energies~\cite{alhyder2022, Alhyder2024,alhyder2025}. In materials where the nonlinear renormalization is large enough to resolve, the measured phonon dispersion would carry information about $\chi^{\rm R}(\mathbf{q},\omega)$ at finite momentum, which is otherwise difficult to access by standard optical or
photoemission probes. 


\subsection{Corrections to the heat capacity}
\label{subsec:heat_capacity_corrections}

So far, we have characterized the electron-phonon-induced renormalization of individual phonon modes as a function of momentum and temperature. To assess whether these mode-resolved shifts, that are relevant for spectroscopy, accumulate to give macroscopic thermodynamic effects, we subsequently consider how the energy shifts affect the phonon heat capacity. This quantity sums over the entire phonon spectrum while weighting each mode according to its thermal occupation. It therefore emphasizes phonons with energies comparable to or smaller than $k_BT$ and indicates how strongly the thermally populated part of the spectrum is modified. Moreover, the heat capacity depends on the temperature derivative of the renormalized frequencies and is consequently sensitive not only to the magnitude of a phonon shift, but also to how that shift evolves with temperature~\cite{Allen1980,fultzVibrationalThermodynamicsMaterials2010,allenAnharmonicPhononQuasiparticle2015}.

For each prescribed effective chemical potential, we treat the electronic distribution as being in quasi-equilibrium and evaluate the temperature derivative at fixed volume and chemical potential. At constant volume, the lattice heat capacity can be obtained from $C_{\mu,V}^{\mathrm{ph}}(T) = T \left(\partial S_{\mathrm{ph}} / \partial T\right)_{\mu,V}$. Following the phonon-quasiparticle treatment of Refs.~\cite{Allen1980,allenAnharmonicPhononQuasiparticle2015}, we assume that the electron-phonon renormalization is sufficiently weak for the phonons to remain well-defined excitations, see also~Fig.~\ref{fig:fig6} that substantiates this assumption. The phonon entropy can then be written in terms of the renormalized frequencies $\widetilde\omega_{\nu, \mathbf{q}} (T,\mu)$,
\begin{equation}
S_{\mathrm{ph}}(T)
=
\sum_{\nu, \mathbf{q}}
k_B\left[
\frac{x_{\nu,\mathbf{q}}}{e^{x_{\nu,\mathbf{q}}}-1}
-
\ln\!\left(1-e^{-x_{\nu,\mathbf{q}}}\right)
\right],
\label{eq:Sph_Tdep}
\end{equation}
with the shorthand $x_{\nu, \mathbf{q}}(T,\mu)=\hbar\widetilde\omega_{\nu,\mathbf{q}}(T,\mu) / (k_B T)$. If the phonon frequencies are temperature independent, differentiation of Eq.~\eqref{eq:Sph_Tdep} gives the standard harmonic result
$C_V^{\mathrm{ph}}(T) = \sum_{\nu,\mathbf{q}} C_{\nu,\mathbf{q}}\!\left(T\right)$~\cite{Mahan2000}, with the contribution per mode and phonon momentum being
\begin{equation}
C_{\nu, \mathbf{q}}(T)
=
k_B\,x_{\nu, \mathbf{q}}^2\, n_B(x_{\nu, \mathbf{q}})\bigl(n_B(x_{\nu, \mathbf{q}})+1\bigr).
\label{eq:heat_capacity_harmonic_factor}
\end{equation}
For temperature-dependent quasiparticle frequencies, the derivative of Eq.~\eqref{eq:Sph_Tdep} also acts on the renormalized frequencies $\widetilde\omega_{\nu,\mathbf{q}}(T,\mu)$, which gives~\cite{Allen1980,allenAnharmonicPhononQuasiparticle2015}
\begin{equation}
\tilde{C}_V^{\mathrm{ph}}(T)
=
\sum_{\nu,\mathbf{q}}
C_{\nu,\mathbf{q}}\!\left(T\right)
\left[
1-\Bigr(\frac{\partial \ln\tilde{\omega}_{\nu,\mathbf{q}} (T,\mu)}{\partial \ln T}\Bigl)_{\mu, V}
\right],
\label{eq:Cv_ph_entropy}
\end{equation}
where $C_{\nu,\mathbf{q}}(T,\mu)$ is evaluated using the renormalized frequency. A detailed derivation of this result is given in App.~\ref{sec:app_heat_capacity}.

The second term in Eq.~\eqref{eq:Cv_ph_entropy} accounts for the explicit temperature dependence of the phonon spectrum. A mode that hardens with increasing temperature has~$\partial\widetilde\omega_{\nu,\mathbf{q}}/\partial T>0$ and therefore reduces the heat capacity, while a mode that softens has the opposite effect. The correction is consequently also governed by the temperature derivative of the phonon renormalization rather than by the frequency shift alone. A phonon may, for example, remain softened relative to its unperturbed frequency while its heat-capacity contribution is suppressed if its renormalized frequency increases with temperature.

\begin{figure}[h]
    \centering
    \includegraphics[width=\columnwidth]
    {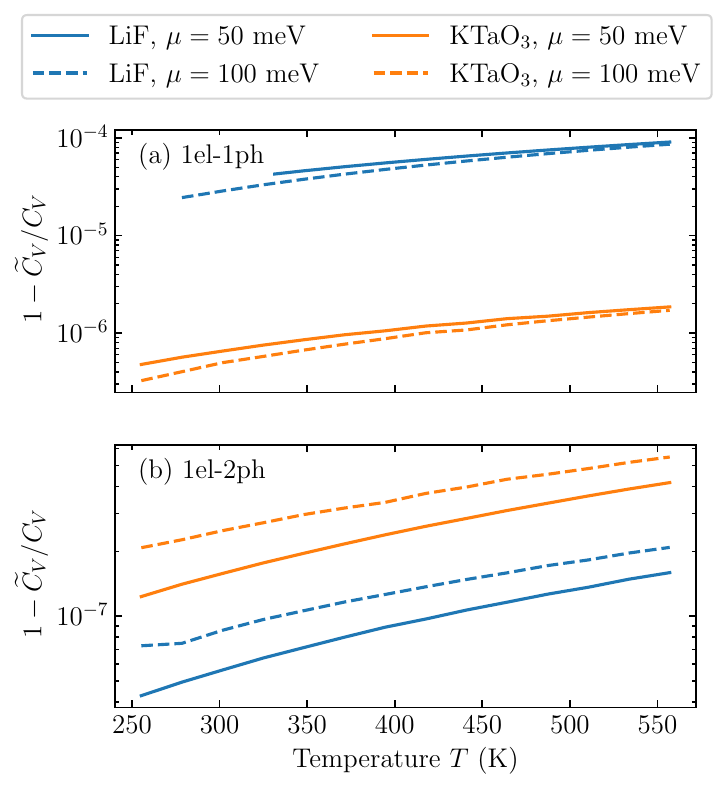}
    \vskip -0.5cm
    \caption{Relative suppression of the phonon heat capacity, $1-\widetilde C_{V,\mu}^{\mathrm{ph}} / C_{V,0}^{\mathrm{ph}}$, arising from (a) the one-electron-one-phonon interaction and (b) only the one-electron-two-phonon interaction. All values are positive, which indicates that the electron-phonon-induced renormalization lowers the phonon heat capacity. Blue and orange curves show the results for LiF and KTaO$_3$, respectively. Solid and dashed curves correspond to $\mu=50$ and $100\,\mathrm{meV}$.}
    \label{fig:fig7}
\end{figure}

We may now calculate two different values for the heat capacity in the materials LiF and KTaO3. The expression in Eq.~\eqref{eq:heat_capacity_harmonic_factor} yields the unperturbed heat capacity $C_V^{\mathrm{ph}}(T)$, calculated with the unperturbed, temperature-independent phonon spectrum. Additionally, Eq.~\eqref{eq:Cv_ph_entropy} yields the corrected heat capacity $\tilde{C}_V^{\mathrm{ph}}(T)$, which takes the temperature-dependent phonon quasiparticle energies into account. In Fig.~\ref{fig:fig7}, we plot the relative difference between these two, as given by $1-\widetilde C_{V}^{\mathrm{ph}} / C_{V}^{\mathrm{ph}}$, for both LiF and KTaO$_3$. Because this relative correction is positive throughout the temperature range considered, the electron-phonon induced renormalization decreases the phonon heat capacity in both materials. Furthermore, this suppression increases with temperature, which is consistent with the increasing hardening of the renormalized phonon frequencies observed in Fig.~\ref{fig:fig6}.

Interestingly, the two interaction channels show different chemical-potential dependences. For the one-electron-one-phonon contribution, increasing $\mu$ reduces the heat-capacity correction and the curves for different chemical potentials approach one another as the temperature increases, see Fig.~\ref{fig:fig7}(a). This behavior reflects the saturation of the Lindhard polarization bubble in the limit $\mu \gg k_B T$, see the following discussion and Fig.~\ref{fig:lindhard_polarization_bubble_saturation} of App.~\ref{sec:app_lithium_fluoride}. At larger chemical potentials, the degeneracy of the electronic system increases and its excitations become less sensitive to temperature changes which are smaller compared to the Fermi-energy scale. Away from particle-hole excitation thresholds, the standard Sommerfeld expansion gives leading thermal corrections to a smooth electronic response of order $(k_BT/\mu)^2$~\cite{maldagueManybodyCorrectionsPolarizability1978}. Increasing $\mu$ can therefore enhance the phonon renormalization while reducing its temperature derivative and, consequently, its heat-capacity correction.

For the one-electron-two-phonon contribution, on the other hand, the heat capacity correction increases with temperature and $\mu$, and the separation between the two chemical potentials in Fig.~\ref{fig:fig7}(b) remains visible throughout the investigated temperature range. Within this interaction channel, the electronic response is sampled together with the thermally occupied internal phonon spectrum. More specifically, the temperature dependence of the intermediate phonon occupations outweighs the reduced thermal sensitivity of the electronic polarization bubble at larger $\mu$. Therefore, this balance accounts for the increase with chemical potential and temperature observed in Fig.~\ref{fig:fig7}(b).

When comparing the two materials studied here, the one-electron--one-phonon correction is substantially larger in LiF than in KTaO$_3$, whereas the one-electron--two-phonon correction is larger in KTaO$_3$. This reversal can be understood by comparing the electronic and phononic phase spaces that control both interaction channels. 

For the one-electron-one-phonon contribution, the magnitude of the long-wavelength polarization bubble is governed by the electronic density of states near the Fermi level, while its momentum range is set by $k_{F}$. Within the three-dimensional parabolic-band model and at fixed $\mu$ relative to the conduction-band minimum, the Fermi momentum becomes $k_{\mathrm F}\propto\sqrt{m^*\mu}$ and the density of states is proportional to $(m^*)^{3/2}\sqrt{\mu}$. Since the effective mass is smaller in KTaO$_3$ than in LiF, both the magnitude and the absolute momentum range of the polarization bubble are reduced, resulting in a smaller one-electron-one-phonon correction for KTaO$_3$. The precise ratio between the two materials, however, also depends on band degeneracies and the material-specific long-range coupling vertices.

Conversely, the one-electron-two-phonon contribution is also governed by the internal phonon spectrum and extends over a larger fraction of the Brillouin zone. At room temperature, approximately eight to nine branches are appreciably occupied in KTaO$_3$, compared with the three populated acoustic branches in LiF, as shown in~Fig.~\ref{fig:fig4}. The larger number of thermally accessible intermediate scattering channels in KTaO$_3$ consequently produces a stronger temperature dependence and allows the nonlinear correction to accumulate over more phonon branches and momenta. Since the heat capacity sums over the full phonon spectrum, this broadly distributed contribution is larger in KTaO$_3$ despite its weaker linear polarization response.

The largest relative heat capacity correction, attributed to the one-electron-one-phonon interaction in LiF, reaches approximately $10^{-4}$, whereas the corresponding relative correction in KTaO$_3$ and the one-electron-two-phonon corrections remain between $10^{-6}$ and $10^{-7}$. Resolving effects of this magnitude calorimetrically requires high-precision differential measurements and careful separation from electronic and anharmonic backgrounds. Nevertheless, the opposite carrier-density dependences of the two interaction channels provide a differential signature through which linear and nonlinear electron-phonon interactions could be distinguished experimentally. Additionally, we note that the effect of linear and nonlinear electron-phonon interactions are strongly material dependent. They are small in LiF and KTaO3, but they are expected to be stronger in other materials, such as the soft halide perovskites and Bi$_2$O$_2$Se~\cite{miyataLargePolaronsLead2017,martiradonnaRiddlesPerovskiteResearch2018,qianPhotocarrierinducedPersistentStructural2023,wangPhononSignaturesPolaron2022,Houtput2026}.

\section{Conclusions and Outlook}

In this article, we derived the contribution to the phonon self-energy arising from the long-range one-electron-two-phonon interaction, using the first-principles vertex recently obtained in Ref.~\cite{houtputFirstprinciplesTheoryNonlinear2025a}. We find three qualitative features that distinguish this contribution from the standard electron-hole polarization bubble. First, the one-electron-two-phonon process renormalizes the phonon band structure across the entire Brillouin zone, in contrast to the linear contribution which is sharply concentrated near the $\Gamma$ point. Second, the Bose-Einstein factor associated with the internal phonon mode generates a pronounced temperature dependence that is absent at the same order in the linear self-energy. Third, the heat capacity correction due the one-electron-two-phonon interaction increases as a function of the carrier density, while the correction due to the one-electron-one-phonon interaction decreases.

To benchmark the theory, we computed both self-energies as a function of chemical potential and temperature for the electron-doped materials LiF and KTaO$_3$. The qualitative differences between the two contributions are most clearly seen along high-symmetry paths in momentum space. At momenta where the symmetry of the Born effective charges and mode polarities suppresses the linear vertex, the nonlinear contribution can remain finite. The renormalized phonon dispersion in such regions therefore isolates the nonlinear coupling and provides a direct experimental target. The same self-energy structure also carries information about the electronic susceptibility of the medium across the Brillouin zone, with the nonlinear vertex acting as a coupling that exposes finite-momentum features of~$\chi^{\rm R}(\mathbf{q},\omega)$ that are otherwise difficult to access experimentally.

The nonlinear phonon renormalization is small in LiF because only the three acoustic branches are appreciably populated at room temperature, suppressing contributions from the internal phonon line through the Bose-Einstein factor. By contrast, the greater number of thermally populated low-energy phonon modes in KTaO$_3$ enhances the nonlinear renormalization by up to a factor of five. Following this trend, we expect the effect to be substantially larger in materials with many low-energy phonon modes, where the corresponding occupation factors are of order unity. Natural candidates are the lead halide perovskites such as CsPbI$_3$ and MAPbI$_3$, and the layered semiconductor Bi$_2$O$_2$Se~\cite{miyataLargePolaronsLead2017,martiradonnaRiddlesPerovskiteResearch2018,qianPhotocarrierinducedPersistentStructural2023,wangPhononSignaturesPolaron2022}. The same mechanism has recently been shown to produce a strong additional temperature dependence in the electron mobility of soft polar semiconductors~\cite{Houtput2026}, supporting the expectation that phonon properties in these materials will display correspondingly larger renormalizations coming from electron-phonon interactions. The general qualitative conclusions that we found here for LiF and KTaO$_3$ are still expected to hold in other materials, despite the fact that the nonlinear phonon renormalization is small in both LiF and KTaO$_3$. As such, these conclusions will help to understand and design both experiments and calculations on materials where the nonlinear electron-phonon interaction is stronger.

Several extensions of this work follow naturally. From an experimental perspective, the electron-phonon induced lattice hardening may leave signatures in observables sensitive to the phonon spectrum, including elastic constants, sound velocities, lattice heat capacity, thermal conductivity, and the infrared dielectric response, thereby providing complementary probes of the nonlinear phonon renormalization. A particularly interesting extension concerns excitons and their exciton-phonon coupling~\cite{antoniusTheoryExcitonphononCoupling2022}, where the long-range linear coupling vanishes for a charge-neutral composite particle at $\mathbf{q} \to 0$~\cite{daiTheoryExcitonicPolarons2024}, while the nonlinear vertex couples through the polarizability response and remains finite~\cite{cavalcanteStarkShiftExcitons2018}. Consequently, the two-phonon channel may dominate the exciton-phonon renormalization in regimes where the linear contribution is locally suppressed. Finally, soft semiconductors such as lead-halide perovskites exhibit strong intrinsic phonon-phonon interactions that substantially renormalize and broaden their phonon spectra~\cite{whalleyPhononAnharmonicityLifetimes2016,tadanoFirstPrinciplesPhononQuasiparticle2022,zachariasAnharmonicLatticeDynamics2023}. A complete treatment of their lattice dynamics should therefore include intrinsic anharmonic and carrier-induced phonon self-energies on equal footing. The diagrammatic framework developed here is well suited to such an extension, since the nonlinear electron-phonon vertex and the three-phonon anharmonic vertex enter the phonon self-energy through structurally similar diagrams. Yet another direction is to include the short-range contribution to the nonlinear one-electron-two-phonon vertex, which requires evaluating the second-order variation of the self-consistent electronic potential with respect to pairs of atomic displacements.

\section*{Acknowledgments}

We thank Zhanybek Alpichshev, Artem Volosniev, Volker Karle and Georgios Koutentakis for stimulating discussions along the way. M.H. acknowledges the Research Foundation Flanders (FWO), file numbers 1224724N and V472923N for their funding of this research. C.F. and J.T., acknowledge support from the joint Austrian Science Fund (FWF) - FWO project 10.55776/PIN5456724. J.T. acknowledges financial support by the Research Foundation Flanders (FWO), Projects No. G0AIY25N, No. G0A9F25N, and No. G060820N. R. A. received funding from the Austrian Academy of Sciences ÖAW grant No. PR1029OEAW03.

\bibliographystyle{apsrev4-2}
\bibliography{methodology}

\clearpage
\newpage

\appendix 

\section{Properties of the electron-phonon Hamiltonian} \label{sec:properties_electron_phonon_hamiltonian}

\subsection{Electron and phonon Green's functions}
\label{subsec:electron_and_phonon_greens_functions}

The bare electron and phonon propagators are defined as time-ordered correlators between creating an electron or phonon at time zero and annihilating it at time $t$, $G_{nm} (\mathbf{k},t) = - \langle \mathcal{T}_t \, \hat c_{n,\mathbf{k}} (t) \hat c^\dagger_{m,\mathbf{k}}(0) \rangle_0$, and $D^{(0)}_{\nu\mu}(\mathbf{q},t) = - \langle \mathcal{T}_t \, \hat A_{\nu,\mathbf q}(t) \hat A_{\mu,-\mathbf q}(0) \rangle_0$, where $\mathcal{T}_t$ denotes the time-ordering operator and $\langle\cdots\rangle_0$ the thermal average over the noninteracting electronic and harmonic phonon systems. The symmetric phonon operator is defined as $\hat{A}_{\nu,\mathbf{q}} = \hat{a}_{\nu,\mathbf{q}} + \hat{a}_{\nu,-\mathbf{q}}^{\dagger}$. In the electronic band basis and the harmonic phonon normal-mode basis, the bare propagators are diagonal and take the frequency-space forms
\begin{equation}
G_{nm} (\mathbf{k},i\Omega_p)
=
\delta_{nm}
\frac{1}{i\Omega_p - \frac{\epsilon_{n,\mathbf{k}}}{\hbar}},
\label{eq:electron_matsubara_green_function}
\end{equation}
for electrons, and 
\begin{equation}
D^{(0)}_{\nu\mu}(\mathbf{q},i\omega_n) 
= 
\delta_{\nu\mu} \left( \frac{1}{i\omega_n - \omega_{\nu,\mathbf{q}}} - \frac{1}{i\omega_n + \omega_{\nu,\mathbf{q}}} \right). \label{eq:phonon_matsubara_green_function}
\end{equation}
for phonons. Here, $\epsilon_{n,\mathbf{k}}=\epsilon_{n,\mathbf{k}}-\mu$ is the electronic energy measured relative to the chemical potential, while $\hbar\Omega_p = \pi (2p+1) / \beta$ and $\hbar\omega_n = 2\pi n / \beta$, with $n,p\in \mathbb{Z}$, define the fermionic and bosonic Matsubara frequencies. The inverse temperature is $\beta=(k_BT)^{-1}$. The two poles of the phonon propagator at positive and negative phonon energies arise because $\hat{A}_{\nu,\mathbf q}$ contains both phonon annihilation and creation operators.

The bare electron and phonon propagators in Eq.~\eqref{eq:electron_matsubara_green_function} and Eq.~\eqref{eq:phonon_matsubara_green_function} form the internal lines of the self-energy diagrams considered in this work, cf.~Fig.~\ref{fig:self_energy_feynman_diagrams}. Electron-phonon interactions dress the phonon propagator as described within the Dyson equation, and the corresponding retarded quantities are obtained by the analytic continuation, $i\omega_n\rightarrow\omega+i0^+$. In the numerical calculations, the infinitesimal $0^+$ is replaced by the finite broadening $\delta$ that is also discussed in App.~\ref{subsec:app_numerical_integration}.

\subsection{Long-range approximation} \label{subsec:long_range_approximation}

In both vertices, when explicitly working in the long-range approximation and only considering non-degenerate electronic bands, the matrix elements of the Bloch states reduce to a Kronecker delta~\cite{verdiFrohlichElectronPhononVertex2015},
\begin{equation}
    \lim_{\mathbf{q}+\mathbf{G} \to 0} \langle \psi_{\mathbf{k}+\mathbf{q},m} | e^{i\left(\mathbf{q} + \mathbf{G}\right)\cdot \left(\mathbf{r}-\boldsymbol{\tau}_{\kappa}\right)} | \psi_{\mathbf{k},n} \rangle = \delta_{mn} \delta_{\mathbf{G},0}.
\end{equation}
This reduction is useful as it simplifies the calculation of the self-energies. \\

\subsection{Summation over Bloch states} \label{subsec:summation_over_bloch_states}

In the derivation of the diagrammatic expressions for the one-electron-one-phonon and one-electron-two-phonon self-energies with coupling vertices in the long-range approximation, we have used that we can rewrite a sum over two electronic band indices into a sum over one electronic band index and reciprocal lattice vectors. We start by proving the identity
\begin{widetext}
\begin{equation}
I := \sum_{m,n} |g_{mn\nu}^{\rm (long)} \left(\mathbf{k},\mathbf{q}\right)|^2 f\left(\epsilon_{\mathbf{k},n},\epsilon_{\mathbf{q}+\mathbf{k},m}\right) = \sum_{\mathbf{G}} |\mathbf{V}_{\nu} \left(\mathbf{q} + \mathbf{G}\right) \cdot \mathbf{p}_{\nu} \left(\mathbf{q}\right)|^2 \sum_{n} f\left(\epsilon_{\mathbf{k},n},\epsilon_{\mathbf{q}+\mathbf{k},n}\right), \label{eq:app_bloch_band_sum_1el-1ph_interaction}
\end{equation}
which was used to rewrite the definition of the one-electron-one-phonon self-energy from Eq.~\eqref{eq:def_1_electron_1_phonon_self_energy} into Eq.~\eqref{eq:1_electron_1_phonon_self_energy_after_summation_over_to_G_conversion}. Here, the function $f\left(\epsilon_{\mathbf{k},n},\epsilon_{\mathbf{q}+\mathbf{k},m}\right)$ depends on the electronic band indices $n$ and $m$ only through the electronic disperion relations $\epsilon_{\mathbf{k},n}$ and $\epsilon_{\mathbf{q}+\mathbf{k},m}$. We can prove the  identity in Eq.~\eqref{eq:app_bloch_band_sum_1el-1ph_interaction} by noting that the long-range coupling vertex contains a sum over reciprocal lattice vectors, and depends on the electronic band indices $m$ and $n$ just through the overlap matrix element, see Eq.~\eqref{eq:simple_1el-1ph_vertex_long_range}. Then 
\begin{equation}
\begin{split}
I = \sum_{m,n} f\left(\epsilon_{\mathbf{k},n},\epsilon_{\mathbf{q}+\mathbf{k},m}\right) \sum_{\mathbf{G}_1\neq -\mathbf{q},\mathbf{G}_2\neq -\mathbf{q}} &\left(\mathbf{V}_{\nu} \left(\mathbf{q}+\mathbf{G}_1\right) \cdot \mathbf{p}_{\nu} \left(\mathbf{q}\right)\right) \left(\mathbf{V}_{\nu} \left(\mathbf{q}+\mathbf{G}_2\right) \cdot \mathbf{p}_{\nu} \left(\mathbf{q}\right)\right)^* \\
\times & \left\langle \psi_{\mathbf{k}, n} \left| e^{-i (\mathbf{q} + \mathbf{G}_2) \cdot (\hat{\mathbf{r}} - \boldsymbol{\tau}_\kappa)} \right| \psi_{\mathbf{k} + \mathbf{q}, m} \right\rangle \left\langle \psi_{\mathbf{k} + \mathbf{q}, m} \left| 
e^{i (\mathbf{q} + \mathbf{G}_1) \cdot (\hat{\mathbf{r}} - \boldsymbol{\tau}_\kappa)} 
\right| \psi_{\mathbf{k}, n} \right\rangle
\end{split}
\end{equation}
Under the long-range assumption, we can drop all terms with $\mathbf{G}_1 \neq \mathbf{G}_2$, and only keep the diagonal contributions such that $\sum_{\mathbf{G}_1,\mathbf{G}_2} = \sum_{\mathbf{G}}$. Furthermore $\langle \psi_{\mathbf{k}+\mathbf{q},m} | e^{i\left(\mathbf{q}+\mathbf{G}_1\right)\cdot \mathbf{r}} | \psi_{\mathbf{k},n} \rangle \approx 0$ unless $\epsilon_{\mathbf{k},n} \approx \epsilon_{\mathbf{k},m}$ and $\epsilon_{\mathbf{q}+\mathbf{k},n} \approx \epsilon_{\mathbf{q}+\mathbf{k},m}$ because $\mathbf{q} \approx 0$. Here we note that in the literature, for non-degenerate bands, it is a common assumption that the long-range limit of $\langle \psi_{\mathbf{k} + \mathbf{q}, m} | 
e^{i (\mathbf{q} + \mathbf{G}) \cdot (\hat{\mathbf{r}} - \boldsymbol{\tau}_\kappa)} 
| \psi_{\mathbf{k}, n}\rangle$ is $\delta_{m,n}$, see Sec.~\ref{subsec:long_range_approximation}. Therefore, we can replace $f\left(\epsilon_{\mathbf{k},n};\epsilon_{\mathbf{q}+\mathbf{k},m}\right) \approx f\left(\epsilon_{\mathbf{k},n};\epsilon_{\mathbf{q}+\mathbf{k},n}\right)$ and reorder the sum over electronic band indices $\sum_m$, which gives
\begin{equation}
I = \sum_{n} f\left(\epsilon_{\mathbf{k},n},\epsilon_{\mathbf{q}+\mathbf{k},n}\right) \sum_{\mathbf{G}\neq -\mathbf{q}} \left|\mathbf{V}_{\nu} \left(\mathbf{q}+\mathbf{G}\right) \cdot \mathbf{p}_{\nu} \left(\mathbf{q}\right)\right|^2 \langle \psi_{\mathbf{k}, n} | e^{-i (\mathbf{q} + \mathbf{G}) \cdot (\hat{\mathbf{r}} - \boldsymbol{\tau}_\kappa)} \sum_m | \psi_{\mathbf{k} + \mathbf{q}, m} \rangle \langle \psi_{\mathbf{k} + \mathbf{q}, m} | 
e^{i (\mathbf{q} + \mathbf{G}) \cdot (\hat{\mathbf{r}} - \boldsymbol{\tau}_\kappa)} 
| \psi_{\mathbf{k}, n} \rangle
\end{equation}
Since Bloch states form a complete basis, $\sum_m | \psi_{\mathbf{k}+\mathbf{q},m} \rangle \langle \psi_{\mathbf{k}+\mathbf{q},m} | = \hat{1}$, we can further simplify to
\begin{equation}
I = \sum_{n} f\left(\epsilon_{\mathbf{k},n},\epsilon_{\mathbf{q}+\mathbf{k},n}\right) \sum_{\mathbf{G}\neq -\mathbf{q}} \left|\mathbf{V}_{\nu} \left(\mathbf{q}+\mathbf{G}\right) \cdot \mathbf{p}_{\nu} \left(\mathbf{q}\right)\right|^2 \langle \psi_{\mathbf{k}, n} | \psi_{\mathbf{k}, n} \rangle
\end{equation}
and due to the normalization of Bloch states, $\langle \psi_{\mathbf{k},n} | \psi_{\mathbf{k},n} \rangle = 1$,
\begin{equation}
I = \sum_{n} f\left(\epsilon_{\mathbf{k},n},\epsilon_{\mathbf{q}+\mathbf{k},n}\right) \sum_{\mathbf{G}\neq -\mathbf{q}} \left|\mathbf{V}_{\nu} \left(\mathbf{q}+\mathbf{G}\right) \cdot \mathbf{p}_{\nu} \left(\mathbf{q}\right)\right|^2.
\end{equation}
\end{widetext}
This completes the proof of the identity in Eq.~\eqref{eq:app_bloch_band_sum_1el-1ph_interaction}. Next we prove the analogous identity for the one-electron-two-phonon self-energy with the long-range interaction vertex, which was used in the derivation in the main text when going from the definition in Eq.~\eqref{eq:1electron_2phonon_self_energy_first_expr} to Eq.~\eqref{eq:1el-2ph_expression_for_self_energy_after_sum_rewriting}. Explicitly, we want to show that 
\begin{widetext}
\begin{equation}
\begin{split}
I :=& 
\sum_{m,n} 
f\left(\epsilon_{\mathbf{k},n},\epsilon_{\mathbf{k}+\mathbf{Q},m}\right) \,
g_{mn\nu_1\nu_2}^* \left(\mathbf{k},-\mathbf{q},\mathbf{q}+\mathbf{Q}\right) 
g_{mn\nu_3\nu_2} \left(\mathbf{k},-\mathbf{q},\mathbf{q}+\mathbf{Q}\right)  \\
=& 
\sum_{\mathbf{G}} g_{\nu_1\nu_2}^* \left(\mathbf{q},\mathbf{Q} + \mathbf{G}\right) g_{\nu_3\nu_2} \left(\mathbf{q},\mathbf{Q} + \mathbf{G}\right) \sum_n f\left(\epsilon_{\mathbf{k},n},\epsilon_{\mathbf{k}+\mathbf{Q},n}\right) \label{eq:app_sum_identity_second_order}
\end{split}
\end{equation}
holds. By using the definition of the long-range vertex $g_{mn\nu_1\nu_2}^{\rm (long)} \left(\mathbf{k},-\mathbf{q},\mathbf{q}+\mathbf{Q}\right)$ from Eq.~\eqref{eq:1-el-2-ph_coupling_element}, we can write 
\begin{equation}
\begin{split}
I = \sum_{m,n} \sum_{\mathbf{G}_1,\mathbf{G}_2} & f\left(\epsilon_{\mathbf{k},n};\epsilon_{\mathbf{k}+\mathbf{Q},m}\right) g_{\nu_3,\nu_2} \left(\mathbf{q},\mathbf{Q}+\mathbf{G}_1\right) g_{\nu_1,\nu_2}^* \left(\mathbf{q},\mathbf{Q}+\mathbf{G}_2\right) \\
\times &\langle \psi_{\mathbf{k},n} | e^{-i\left(\mathbf{Q}+\mathbf{G}_2\right)\cdot \mathbf{r}} | \psi_{\mathbf{k}+\mathbf{Q},m} \rangle \langle \psi_{\mathbf{k}+\mathbf{Q},m} | e^{i\left(\mathbf{Q}+\mathbf{G}_1\right)\cdot \mathbf{r}} | \psi_{\mathbf{k},n} \rangle
\end{split}
\end{equation}
Again we drop all the terms with $\mathbf{G}_1 \neq \mathbf{G}_2$, setting $\mathbf{G} := \mathbf{G}_1 = \mathbf{G}_2$ and keeping only the diagonal sum over reciprocal lattice vectors, $\sum_{\mathbf{G}_1,\mathbf{G}_2} = \sum_{\mathbf{G}}$. Furthermore $\langle \psi_{\mathbf{k}+\mathbf{Q},m} | e^{i\left(\mathbf{Q}+\mathbf{G}_1\right)\cdot \mathbf{r}} | \psi_{\mathbf{k},n} \rangle \approx 0$ unless $\epsilon_{\mathbf{k},n} \approx \epsilon_{\mathbf{k},m}$ and $\epsilon_{\mathbf{k}+\mathbf{Q},n} \approx \epsilon_{\mathbf{k}+\mathbf{Q},m}$ because $\mathbf{Q} \approx 0$. Consequently, we can replace $f\left(\epsilon_{\mathbf{k},n};\epsilon_{\mathbf{k}+\mathbf{Q},m}\right) \approx f\left(\epsilon_{\mathbf{k},n};\epsilon_{\mathbf{k}+\mathbf{Q},n}\right)$ and reorder the sum over electronic band indices $\sum_m$,
\begin{equation}
\begin{split}
I = \sum_{n} \sum_{\mathbf{G}} & f\left(\epsilon_{\mathbf{k},n};\epsilon_{\mathbf{k}+\mathbf{Q},n}\right) g_{\nu_3,\nu_2} \left(\mathbf{q},\mathbf{Q}+\mathbf{G}\right) g_{\nu_1,\nu_2}^* \left(\mathbf{q},\mathbf{Q}+\mathbf{G}\right) \\
\times &\langle \psi_{\mathbf{k},n} | e^{-i\left(\mathbf{Q}+\mathbf{G}\right)\cdot \mathbf{r}} \sum_m | \psi_{\mathbf{k}+\mathbf{Q},m} \rangle \langle \psi_{\mathbf{k}+\mathbf{Q},m} | e^{i\left(\mathbf{Q}+\mathbf{G}\right)\cdot \mathbf{r}} | \psi_{\mathbf{k},n} \rangle
\end{split}
\end{equation}
Again, as Bloch states form a complete and normalized basis, we can simplify the sum to
\begin{equation}
I = \sum_{n} \sum_{\mathbf{G}} f\left(\epsilon_{\mathbf{k},n};\epsilon_{\mathbf{k}+\mathbf{Q},n}\right) g_{\nu_1,\nu_2}^* \left(\mathbf{q},\mathbf{Q}+\mathbf{G}\right) g_{\nu_3,\nu_2} \left(\mathbf{q},\mathbf{Q}+\mathbf{G}\right),
\end{equation}
\end{widetext}
which proves the identity in Eq.~\eqref{eq:app_sum_identity_second_order}. We note that both relations in Eq.~\eqref{eq:app_bloch_band_sum_1el-1ph_interaction} and Eq.~\eqref{eq:app_sum_identity_second_order} are independent of the crystal structure and only hold when the short-range interaction can be ignored, which is material dependent.

\subsection{Feynman rules} \label{appsec:feynman_rules}

When writing down the mathematical expressions for the diagrams in Fig.~\ref{fig:self_energy_feynman_diagrams}, we obey the following Feynman rules. At every 1-electron-1-phonon and 1-electron-2-phonon vertex, we multiply by $\sqrt{\Omega_0 / \Omega} ( 1 / \hbar ) g_{mn\nu}\left(\mathbf{k}, \mathbf{q}\right)$ and $( \Omega_0 / \Omega ) ( 2 / \hbar ) g_{mn\nu_1\nu_2}(\mathbf{k}, \mathbf{q}_1, \mathbf{q}_2)$ respectively. Electron and phonon lines correspond to the non-interacting electron and phonon Green's function from Eq.~\eqref{eq:electron_matsubara_green_function} and Eq.~\eqref{eq:phonon_matsubara_green_function}. Integration over internal frequencies and momenta is done with the standard prescriptions
\begin{equation}
\frac{1}{\beta} \sum_{i\omega_n} 
\text{ and } 
\sum_{\mathbf{k}} = \frac{\Omega}{(2\pi)^3} \underset{\text{1.BZ}}{\int} \mathrm{d}^3 \mathbf{k}.
\end{equation}
Finally, the whole diagram is multiplied with the combinatorial factor obtained in general from Wick expansion, 
\begin{equation}
\frac{1}{\mathrm{\#}G} \left(-1\right)^F \left(2S+1\right)^F, \label{eq:factor_feynman_rules}
\end{equation}
where $S = 1 / 2$ is the electron spin, $F$ denotes the number of closed fermion loops, and $\mathrm{\#}G$ represents the total number of symmetries of the diagram~\cite{Goldberg1985}. These are the Feynman rules when working with Matsubara frequencies. When working with real frequencies, we additionally need to multiply with a factor $i^m$, where $m$ is the number of free four-momenta. 

\section{Symmetry transformations} \label{sec:symmetry_transformations}

In this section, we establish the transformation properties needed to exploit symmetry in the calculation of the one-electron-two-phonon self-energy. In Sec.~\ref{subsec:app_gauge_covariance_under_unitary_transformations_of_phonon_eigenvectors}, we show that the gauge freedom of the phonon eigenvectors induces covariant transformations of both the one-electron-two-phonon vertex and the self-energy under unitary rotations within degenerate phonon subspaces, without affecting physical observables. We then derive the point-group transformation of the vertex in Sec.~\ref{subsec:symmetry_transformation_one_electron_two_phonon_vertex} and use it in Sec.~\ref{subsec:symmetry_transformations_one_electron_two_phonon_self_energy} to relate the Cartesian self-energy components of isotropic cubic materials at symmetry-equivalent wavevectors, which is the main derivation for Eq.~\eqref{eq:dia_iso_1el_2ph_self_energy_x_and_y_component} of the main text.

\subsection{Gauge covariance under unitary transformations of the phonon eigenvectors} 
\label{subsec:app_gauge_covariance_under_unitary_transformations_of_phonon_eigenvectors}

\subsubsection{Gauge freedom in the choice of phonon eigenvectors}

The phonon eigenvectors obtained by diagonalizing the dynamical matrix are not uniquely defined. For a nondegenerate phonon mode, the corresponding eigenvector is defined only up to a phase factor, whereas within a $n_{\rm deg}$-fold degenerate subspace, any two
orthonormal eigenvector bases are related by a unitary transformation
in~$\mathrm{U}(n_{\rm deg})$. For example, for a set of orthonormal eigenvectors $\mathbf{e}_{\kappa\alpha,\nu} \left(\mathbf{q}\right)$, we can generate another orthonormal basis by applying a unitary transform, 
\begin{equation}
\tilde{\mathbf{e}}_{\kappa\alpha,\nu} \left(\mathbf{q}\right) = \sum_{\nu'} \mathbf{e}_{\kappa\alpha,\nu'} \left(\mathbf{q}\right) U_{\nu'\nu} \left(\mathbf{q}\right),
\label{eq:phonon_eigenvectors_unique_definition}
\end{equation}
or equivalently in matrix form
\begin{equation}
    \tilde{\mathbf{e}}_{\kappa\alpha} \left(\mathbf{q}\right) = \mathbf{e}_{\kappa\alpha} \left(\mathbf{q}\right) \cdot \mathbf{U} \left(\mathbf{q}\right).
\end{equation}
Here, $U_{\nu' \nu} \left(\mathbf{q}\right)$ is a block-diagonal matrix, i.e. $U_{\nu' \nu}(\mathbf{q}) = 0$ unless $\omega_{\nu,\mathbf{q}} = \omega_{\nu',\mathbf{q}}$, and a unitary matrix:
\begin{equation}
\sum_{\nu''} U_{\nu\nu''} \left(\mathbf{q}\right) U_{\nu'\nu''}^* \left(\mathbf{q}\right) = \delta_{\nu\nu'}. \label{eq:unitary_condition_eigenvectors}
\end{equation}
The sizes of the blocks are equal to the dimensions $n_{\rm deg}$ of the degenerate subspaces. We note that an alternative choice would be to choose~$\tilde{\mathbf{U}}$ which obeys $\mathbf{U} = \tilde{\mathbf{U}}^{\rm T}$ and multiply $\tilde{\mathbf{U}}$ from the left, $\tilde{\mathbf{e}}_{\kappa\alpha} \left(\mathbf{q}\right) = \tilde{\mathbf{U}} \left(\mathbf{q}\right) \cdot \mathbf{e}_{\kappa\alpha} \left(\mathbf{q}\right)$. In this manuscript, all the proofs rely on the choice of $\mathbf{U}$, although some proofs could be simplified by using $\tilde{\mathbf{U}}$. Generally speaking, this freedom in the choice of the phonon eigenvectors may be regarded as a gauge freedom, as it reduces to a $\mathrm{U}(1)$ phase freedom for an isolated mode and becomes a generally non-Abelian $\mathrm{U}(n_{\rm deg})$ freedom within a $n_{\rm deg}$-fold degenerate subspace. Subsequently, in Sec.~\ref{subsubsec:unitary_transformation_of_vertex_factor} and Sec.~\ref{subsubsec:unitary_transformation_of_one_electron_two_phonon_self_energy}, we will discuss the consequences of the unitary freedom when choosing phonon eigenvectors on $Y_{\nu_1\nu_2,\alpha} \left(\mathbf{q}\right)$ and the one-electron-two-phonon self-energy.

\subsubsection{Gauge covariance of the one-electron-two-phonon vertex} 
\label{subsubsec:unitary_transformation_of_vertex_factor}

Since the definition of $Y_{\nu_1\nu_2,\alpha} \left(\mathbf{q}\right)$ in Eq.~\eqref{eq:def_Y_nu1nu2_tensor} involves the phonon eigenvectors, it is not uniquely defined due to the transformation property of $\mathbf{e}_{\kappa\alpha} \left(\mathbf{q}\right)$ as written in Eq.~\eqref{eq:phonon_eigenvectors_unique_definition}. Suppose that we have found one set of~$Y_{\nu_1\nu_2,\alpha} \left(\mathbf{q}\right)$ numerically. Then, all possible values of~$\tilde{Y}_{\nu_1\nu_2,\alpha} \left(\mathbf{q}\right)$ can be written as 
\begin{equation}
\tilde{Y}_{\nu_1\nu_2,\alpha} \left(\mathbf{q}\right) 
= 
\sum_{\nu_1',\nu_2'} 
U_{\nu_1'\nu_1}^* \left(\mathbf{q}\right) 
Y_{\nu_1'\nu_2',\alpha} \left(\mathbf{q}\right) 
U_{\nu_2'\nu_2} \left(\mathbf{q}\right), \label{eq:gauge_invariance_vertex_factor}
\end{equation}
where $U_{\nu'\nu} \left(\mathbf{q}\right)$ is a unitary matrix fulfilling Eq.~\eqref{eq:unitary_condition_eigenvectors}. We can prove Eq.~\eqref{eq:gauge_invariance_vertex_factor} by inserting the transformation of the eigenvectors from Eq.~\eqref{eq:phonon_eigenvectors_unique_definition} into the definition of the vertex factor $\mathbf{Y}_{\nu_1'\nu_2',\alpha} \left(\mathbf{q}\right)$ from Eq.~\eqref{eq:def_Y_nu1nu2_tensor}, which gives
\begin{widetext}
\begin{equation}
\begin{aligned}
\tilde{\mathbf{Y}}_{\nu_{1},\nu_{2}}(\mathbf{q})
&= \frac{1}{i e}
\sqrt{\frac{\hbar}{2 \omega_{\nu_{1},\mathbf{q}}}
      \frac{\hbar}{2 \omega_{\nu_{2},\mathbf{q}}}}
\sum_{\kappa \alpha} \sum_{\kappa' \beta}
\tilde{\mathbf{e}}_{\kappa \alpha,\nu_{1}}(-\mathbf{q}) \,
\frac{\partial \mathcal{D}_{\kappa \alpha,\kappa' \beta}(\mathbf{q})}{\partial \boldsymbol{\mathcal{E}}} \,
\tilde{\mathbf{e}}_{\kappa' \beta,\nu_{2}}(\mathbf{q}) \\
&= \sum_{\nu_1' \nu_2'} 
U^{*}_{\nu_1',\nu_{1}}(\mathbf{q}) \,
U_{\nu_2',\nu_{2}}(\mathbf{q}) \,
\frac{1}{i e}
\sqrt{\frac{\hbar}{2 \omega_{\nu_{1},\mathbf{q}}}
      \frac{\hbar}{2 \omega_{\nu_{2},\mathbf{q}}}}
\sum_{\kappa \alpha} \sum_{\kappa' \beta}
\mathbf{e}_{\kappa \alpha,\nu_1'}(-\mathbf{q}) \,
\frac{\partial \mathcal{D}_{\kappa \alpha,\kappa' \beta}(\mathbf{q})}{\partial \boldsymbol{\mathcal{E}}} \,
\mathbf{e}_{\kappa' \beta,\nu_2'}(\mathbf{q}) .
\end{aligned}
\end{equation}
Because the matrix $U_{\nu'\nu} \left(\mathbf{q}\right)$ is block diagonal, we can replace the phonon frequencies $\omega_{\nu_1,\mathbf{q}}$ and $\omega_{\nu_2,\mathbf{q}}$ with $\omega_{\nu_1',\mathbf{q}}$ and $\omega_{\nu_2',\mathbf{q}}$. All terms in the sum where $\omega_{\nu_1,\mathbf{q}} \neq \omega_{\nu_1',\mathbf{q}}$ would be zero because of the block diagonal structure of $U_{\nu'\nu} \left(\mathbf{q}\right)$. As a consequence, we can write 
\begin{equation}
    \tilde{Y}_{\nu_{1},\nu_{2}}(\mathbf{q}) = \sum_{\nu_1' \nu_2'} 
U^{*}_{\nu_1',\nu_{1}}(\mathbf{q}) \,
U_{\nu_2',\nu_{2}}(\mathbf{q}) \,
\frac{1}{i e}
\sqrt{\frac{\hbar}{2 \omega_{\nu_{1}',\mathbf{q}}}
      \frac{\hbar}{2 \omega_{\nu_{2}',\mathbf{q}}}}
\sum_{\kappa \alpha} \sum_{\kappa' \beta}
\mathbf{e}_{\kappa \alpha,\nu_1'}(-\mathbf{q}) \,
\frac{\partial \mathcal{D}_{\kappa \alpha,\kappa' \beta}(\mathbf{q})}{\partial \boldsymbol{\mathcal{E}}} \,
\mathbf{e}_{\kappa' \beta,\nu_2'}(\mathbf{q})
\end{equation}
\end{widetext}
In this expression, we recognize the definition of $\mathbf{Y}_{\nu_{1}',\nu_{2}'}(\mathbf{q})$ in terms of the original eigenvectors $\mathbf{e}_{\kappa \alpha,\nu_1'}(-\mathbf{q})$, allowing us to write
\begin{equation*}
    \tilde{\mathbf{Y}}_{\nu_1\nu_2,\alpha} \left(\mathbf{q}\right) = \sum_{\nu_1',\nu_2'} U_{\nu_1'\nu_1}^* \left(\mathbf{q}\right) \mathbf{Y}_{\nu_1'\nu_2',\alpha} \left(\mathbf{q}\right) U_{\nu_2'\nu_2} \left(\mathbf{q}\right),
\end{equation*}
which completes the proof Eq.~\eqref{eq:gauge_invariance_vertex_factor}. In matrix form, we can write the transformation in Eq.~\eqref{eq:gauge_invariance_vertex_factor} equivalently as 
\begin{equation}
    \tilde{\mathbf{Y}}_{\alpha} \left(\mathbf{q}\right) = \mathbf{U}^{\dagger} (\mathbf{q}) \cdot \mathbf{Y}_{\alpha} (\mathbf{q}) \cdot \mathbf{U} (\mathbf{q}).
\end{equation}
The proofs presented in this work can be sometimes simplified when writing the involved objects in this matrix form.

\subsubsection{Gauge covariance of the one-electron-two-phonon self-energy} 
\label{subsubsec:unitary_transformation_of_one_electron_two_phonon_self_energy}

The Cartesian $\alpha$ and $\beta$ components of the one-electron-two-phonon self-energy from Eq.~\eqref{eq:1-electron-2-phonon_self_energy_final_expression} can be compactly written as
\begin{equation}
\Pi_{\nu\mu,\alpha\beta}^{\rm (1el-2ph)} \left(\mathbf{q}, \omega\right) 
=  
\sum_{\nu'} 
Y_{\nu'\nu,\alpha} \left(\mathbf{q}\right) 
Y_{\nu'\mu,\beta}^* \left(\mathbf{q}\right) 
f(\dots),
\label{eq:1_electron_2_phonon_self_energy_compact_writing_for_symmetry_transformation}
\end{equation}
where $f(\dots)$ is an appropriately defined function that does not depend on the phonon eigenvectors. According to the transformation of the vertex factor from Eq.~\eqref{eq:gauge_invariance_vertex_factor}, we can make the substitution $Y_{\nu_1\nu_2,\alpha} \left(\mathbf{q}\right) \to \tilde{Y}_{\nu_1\nu_2,\alpha} \left(\mathbf{q}\right)$ and plug it into $\Pi_{\nu\mu,\alpha\beta}^{\rm (1el-2ph)} \left(\mathbf{q}, \omega\right)$, which gives, 
\begin{widetext}
\begin{equation}
\begin{split}
\Pi_{\nu\mu,\alpha\beta}^{\rm (1el-2ph)} \left(\mathbf{q}, \omega\right) 
=& 
\sum_{\nu'} \tilde{\mathbf{Y}}_{\nu'\nu,\alpha} \left(\mathbf{q}\right) \tilde{\mathbf{Y}}_{\nu'\mu,\beta}^* \left(\mathbf{q}\right) f\left(\dots\right) \\
=& \sum_{\nu'} \sum_{\nu_1',\nu_2'} \sum_{\nu_1'',\nu_2''} U_{\nu_1'\nu'}^* \left(\mathbf{q}\right) \mathbf{Y}_{\nu_1'\nu_2',\alpha} \left(\mathbf{q}\right) U_{\nu_2'\nu} \left(\mathbf{q}\right) U_{\nu_1''\nu'} \left(\mathbf{q}\right) \mathbf{Y}_{\nu_1''\nu_2'',\beta}^* \left(\mathbf{q}\right) U_{\nu_2''\mu}^* \left(\mathbf{q}\right) f\left(\dots\right) \\
=& \sum_{\nu_1',\nu_2'} \sum_{\nu_1'',\nu_2''} U_{\nu_2'\nu} \left(\mathbf{q}\right) U_{\nu_2''\mu}^* \left(\mathbf{q}\right) \left(\sum_{\nu'} U_{\nu_1'\nu'}^* \left(\mathbf{q}\right) U_{\nu_1''\nu'} \left(\mathbf{q}\right) \right) \mathbf{Y}_{\nu_1'\nu_2',\alpha} \left(\mathbf{q}\right) \mathbf{Y}_{\nu_1''\nu_2'',\beta}^* \left(\mathbf{q}\right) f\left(\dots\right) \\
=& \sum_{\nu_1',\nu_2'} \sum_{\nu_1'',\nu_2''} U_{\nu_2'\nu} \left(\mathbf{q}\right) U_{\nu_2''\mu}^* \left(\mathbf{q}\right) \delta_{\nu_1'\nu_1''} \mathbf{Y}_{\nu_1'\nu_2',\alpha} \left(\mathbf{q}\right) \mathbf{Y}_{\nu_1''\nu_2'',\beta}^* \left(\mathbf{q}\right) f\left(\dots\right) \\
=& \sum_{\nu_1',\nu_2'} \sum_{\nu_2''} U_{\nu_2'\nu} \left(\mathbf{q}\right) U_{\nu_2''\mu}^* \left(\mathbf{q}\right) \mathbf{Y}_{\nu_1'\nu_2',\alpha} \left(\mathbf{q}\right) \mathbf{Y}_{\nu_1'\nu_2'',\beta}^* \left(\mathbf{q}\right) f\left(\dots\right) \\
=& 
\sum_{\nu_2'',\nu_2'} 
U_{\nu_2''\mu}^* \left(\mathbf{q}\right)
\sum_{\nu_1'}
\mathbf{Y}_{\nu_1'\nu_2'',\beta}^* \left(\mathbf{q}\right)
f\left(\dots\right)
\mathbf{Y}_{\nu_1'\nu_2',\alpha} \left(\mathbf{q}\right)
U_{\nu_2'\nu} \left(\mathbf{q}\right) 
\end{split}
\end{equation}
\end{widetext}
From the second to the third line, we used that the transformation is unitary, and the rest are algebraic rearrangements. We can now employ Eq.~\eqref{eq:1_electron_2_phonon_self_energy_compact_writing_for_symmetry_transformation} to write 
\begin{equation}
\Pi_{\nu\mu,\alpha\beta}^{\rm (1el-2ph)} \left(\mathbf{q}, \omega\right) 
=
\sum_{\nu_2'',\nu_2'} 
U_{\nu_2''\mu}^* \left(\mathbf{q}\right)
\Pi_{\nu_2'\nu_2'',\alpha\beta}^{\rm (1el-2ph)} \left(\mathbf{q}, \omega\right) 
U_{\nu_2'\nu} \left(\mathbf{q}\right).
\label{eq:unitary_transformation_1_electron_2_phonon_self_energy}
\end{equation}
From Eq.~\eqref{eq:unitary_transformation_1_electron_2_phonon_self_energy}, we see that any sum of the type $\Pi_{\nu\mu,\alpha\beta}^{\rm (1el-2ph)} \left(\mathbf{q}, \omega\right)$ transforms under the unitary transformation of the phonon eigenvectors. However, as the secular equation is invariant under unitary transformations, this does not affect physical observables such as phonon frequencies. We will need the transformation written in~Eq.~\eqref{eq:unitary_transformation_1_electron_2_phonon_self_energy} later when relating the different Cartesian components of the one-electron-two-phonon self-energy for isotropic materials to one another, see in particular the derivation in Eq.~\eqref{eq:derivation_for_the_y_component}.

\subsection{Symmetry transformations of the one-electron-two-phonon vertex}
\label{subsec:symmetry_transformation_one_electron_two_phonon_vertex}

Crystalline materials possess space group symmetries that transform the lattice into itself. In general they are a combination of a rotational part with matrix $\mathbf{R}$, and a translation part with translation vector $\boldsymbol{\tau}$. The rotation can also be rotation inversion, though we will keep the term rotation matrix for brevity. The rotation transformation can take an atom $\kappa$ to a different position occupied by an atom of the same species. We denote this different position as $R (\kappa)$, which is defined as 
\begin{equation}
\mathbf{R}\cdot\boldsymbol{\tau}_{\kappa} + \boldsymbol{\tau} \equiv \boldsymbol{\tau}_{R(\kappa)}, \qquad \text{up to a lattice vector},
\end{equation}
where $\boldsymbol{\tau}_{\kappa}$ is the position of atom $\kappa$. The space group of the material consists of the set of all these symmetries. The set of rotation matrices $\mathbf{R}$ also forms a group, known as the point group of the material. Here, we only need the point group. All rotation matrices are orthogonal and satisfy $\mathbf{R}^{\rm T} = \mathbf{R}^{-1}$. In the following, we show that if $\mathbf{R}$ is an element of the point group of the material, then $Y_{\nu_1\nu_2,\alpha} (\mathbf{q})$ satisfies,
\begin{widetext}
\begin{equation}
Y_{\nu_1 \nu_2, \alpha}(\mathbf{q})
= 
\sum_{\nu_1' \nu_2'} 
U_{\nu_1' \nu_1}^* (\mathbf{q}, \mathbf{R})
\left( \sum_{\beta} \mathbf{R}_{\beta \alpha} \,
Y_{\nu_1' \nu_2', \beta}(\mathbf{R} \cdot \mathbf{q}) \right)
U_{\nu_2' \nu_2}(\mathbf{q}, \mathbf{R}), \label{eq:transformation_of_Ys_with_rotation}
\end{equation}
where $U_{\nu' \nu}(\mathbf{q}, \mathbf{R})$ is a block diagonal unitary matrix with $U_{\nu\nu'} \left(\mathbf{q}\right) = 0$, unless $\omega_{\nu,\mathbf{q}} = \omega_{\nu',\mathbf{q}}$, and $\sum_{\nu''} U_{\nu\nu''} \left(\mathbf{q}\right) U_{\nu'\nu''}^* \left(\mathbf{q}\right) = \delta_{\nu\nu'}$. For the proof of Eq.~\eqref{eq:transformation_of_Ys_with_rotation}, we need to know how the dynamical matrix, phonon eigenvectors and phonon frequencies transform under the point group operation $\mathbf{R}$. Each of these quantities depends on the phonon wavevector $\mathbf{q}$ and on the external electric field $\boldsymbol{\mathcal{E}}$. When the dependence on $\boldsymbol{\mathcal{E}}$ is not written, the quantity is evaluated at $\boldsymbol{\mathcal{E}} = 0$. It is straightforward to derive the following relations, where we use the c-type convention for the phonon eigenvectors and dynamical matrix:
\begin{equation}
\begin{split}
\mathcal{D}_{\kappa\kappa'}(\mathbf{q}, \boldsymbol{\mathcal{E}})
=&
\,
\mathbf{R}^{-1}\cdot
\mathcal{D}_{R(\kappa)R(\kappa')}(\mathbf{R}\cdot\mathbf{q}, \mathbf{R}\cdot\boldsymbol{\mathcal{E}})
\cdot
\mathbf{R}, \\
\mathbf{e}_{\kappa,\nu}(\mathbf{q};\boldsymbol{\mathcal{E}})
=&
\sum_{\nu'}
\mathbf{R}^{-1}\cdot
\mathbf{e}_{R(\kappa),\nu'}(\mathbf{R}\cdot\mathbf{q};\mathbf{R}\cdot\boldsymbol{\mathcal{E}})
U_{\nu',\nu}(\mathbf{q};\boldsymbol{\mathcal{E}};\mathbf{R}), \\
\omega_{\nu,\mathbf{q}}^2 
=&
\,
\omega_{\nu,\mathbf{R}\cdot \mathbf{q}}^2
\end{split}
\label{eq:point_group_symmetry_transformations_of_dynamical_matrix_evecs_and_evals_symmary}
\end{equation}
As a first derivation step, we insert the transformation of the phonon eigenvectors into the general definition of $\mathbf{Y}_{\nu_1 \nu_2}(\mathbf{q})$ as defined in Eq.~\eqref{eq:def_Y_nu1nu2_tensor}. This gives
\begin{equation}
\begin{split}
\mathbf{Y}_{\nu_1 \nu_2}(\mathbf{q}) 
=& 
\frac{1}{i e} 
\sqrt{ \frac{\hbar}{2 \omega_{\mathbf{q}, \nu_1}} } 
\sqrt{ \frac{\hbar}{2 \omega_{\mathbf{q}, \nu_2}} } 
\sum_{\kappa \alpha} \sum_{\kappa' \beta} 
\mathbf{e}^*_{\kappa \alpha, \nu_1}(\mathbf{q}) 
\frac{\partial \mathcal{D}_{\kappa \alpha, \kappa' \beta}(\mathbf{q})}{\partial \boldsymbol{\mathcal{E}}} 
\mathbf{e}_{\kappa' \beta, \nu_2}(\mathbf{q})
\\[0.5em]
=&
\frac{1}{i e} 
\sqrt{ \frac{\hbar}{2 \omega_{\mathbf{R}\cdot\mathbf{q}, \nu_1}} } 
\sqrt{ \frac{\hbar}{2 \omega_{\mathbf{R}\cdot\mathbf{q}, \nu_2}} } 
\sum_{\kappa \alpha} \sum_{\kappa' \beta} 
\left[
\sum_{\nu_1'}
U^{*}_{\nu_1',\nu_1}(\mathbf{q};\mathbf{R})
\left(
\mathbf{R}^{-1}\cdot
\mathbf{e}_{R(\kappa),\nu_1'}(\mathbf{R}\cdot\mathbf{q})
\right)^*_{\alpha}
\right]
\frac{\partial \mathcal{D}_{\kappa \alpha, \kappa' \beta}(\mathbf{q})}{\partial \boldsymbol{\mathcal{E}}}\\
&\hspace{7em}\times
\left[
\sum_{\nu_2'}
\left(
\mathbf{R}^{-1}\cdot
\mathbf{e}_{R(\kappa'),\nu_2'}(\mathbf{R}\cdot\mathbf{q})
\right)_{\beta}
U_{\nu_2',\nu_2}(\mathbf{q};\mathbf{R})
\right]
\end{split}
\end{equation}
We can rearrange the sum over the primed indices $\nu_1'$ and $\nu_2'$ together with the two unitary matrices,
\begin{equation}
\begin{split}
\mathbf{Y}_{\nu_1 \nu_2}(\mathbf{q})  
=&
\sum_{\nu_1'\nu_2'}
U^{*}_{\nu_1',\nu_1}(\mathbf{q};\mathbf{R})
\Bigg[
\frac{1}{i e} 
\sqrt{ \frac{\hbar}{2 \omega_{\mathbf{R}\cdot\mathbf{q}, \nu_1'}} } 
\sqrt{ \frac{\hbar}{2 \omega_{\mathbf{R}\cdot\mathbf{q}, \nu_2'}} } 
\sum_{\kappa \alpha} \sum_{\kappa' \beta}
\left(
\mathbf{R}^{-1}\cdot
\mathbf{e}_{R(\kappa),\nu_1'}(\mathbf{R}\cdot\mathbf{q})
\right)^*_{\alpha}
\frac{\partial \mathcal{D}_{\kappa \alpha, \kappa' \beta}(\mathbf{q})}
{\partial \boldsymbol{\mathcal{E}}}
\\
&\hspace{11em}\times
\left(
\mathbf{R}^{-1}\cdot
\mathbf{e}_{R(\kappa'),\nu_2'}(\mathbf{R}\cdot\mathbf{q})
\right)_{\beta}
\Bigg]
U_{\nu_2',\nu_2}(\mathbf{q};\mathbf{R}),
\end{split}
\end{equation}
and insert the matrix multiplication $\left(
\mathbf{R}^{-1}\cdot
\mathbf{e}_{R(\kappa'),\nu_2'}(\mathbf{R}\cdot\mathbf{q})
\right)_{\beta} = \sum_{\beta'} R_{\beta'\beta}
e_{R(\kappa')\beta',\nu_2'}(\mathbf{R}\cdot\mathbf{q})$, which gives
\begin{equation}
\begin{split}
\mathbf{Y}_{\nu_1 \nu_2}(\mathbf{q})
=&
\sum_{\nu_1'\nu_2'}
U^{*}_{\nu_1',\nu_1}(\mathbf{q};\mathbf{R})
\Bigg[
\frac{1}{i e} 
\sqrt{ \frac{\hbar}{2 \omega_{\mathbf{R}\cdot\mathbf{q}, \nu_1'}} } 
\sqrt{ \frac{\hbar}{2 \omega_{\mathbf{R}\cdot\mathbf{q}, \nu_2'}} } 
\sum_{\kappa \kappa'}
\sum_{\alpha\beta}
\sum_{\alpha'\beta'}
e^{*}_{R(\kappa)\alpha',\nu_1'}(\mathbf{R}\cdot\mathbf{q})
R_{\alpha'\alpha}
\\
&\hspace{5em}\times
\frac{\partial \mathcal{D}_{\kappa \alpha, \kappa' \beta}(\mathbf{q})}
{\partial \boldsymbol{\mathcal{E}}}
R_{\beta'\beta}
e_{R(\kappa')\beta',\nu_2'}(\mathbf{R}\cdot\mathbf{q})
\Bigg]
U_{\nu_2',\nu_2}(\mathbf{q};\mathbf{R}).
\end{split}
\label{eq:Y_point_group_transformation_step_three}
\end{equation}
To further manipulate this expression, we need to consider how the derivative of the dynamical matrix with respect to the electric field transforms under $\mathbf{R}$. We already know that the dynamical matrix transforms under $\mathbf{R}$ according to~Eq.~\eqref{eq:point_group_symmetry_transformations_of_dynamical_matrix_evecs_and_evals_symmary}. Now we show how its derivative with respect to the electric field transforms. Under a point group operation $\mathbf{R}$, the electric field transforms as an ordinary polar vector, $\boldsymbol{\mathcal{E}} \to \boldsymbol{\mathcal{E}}' = \mathbf{R} \cdot \boldsymbol{\mathcal{E}}$. Applying the chain rule gives
\begin{equation}
\frac{\partial}{\partial \mathcal{E}_{\eta}}
=
\sum_{\eta'}
\frac{\partial (\mathbf{R} \cdot \boldsymbol{\mathcal{E}})_{\eta'}}
{\partial \mathcal{E}_{\eta}}
\frac{\partial}{\partial (\mathbf{R} \cdot \boldsymbol{\mathcal{E}})_{\eta'}}
=
\sum_{\eta'}
\frac{\partial}{\partial \mathcal{E}_{\eta}}
\left(
\sum_{\eta''} R_{\eta'\eta''} \mathcal{E}_{\eta''}
\right)
\frac{\partial}{\partial (\mathbf{R} \cdot \boldsymbol{\mathcal{E}})_{\eta'}}
=
\sum_{\eta'} R_{\eta'\eta}
\frac{\partial}{\partial \mathcal{E}'_{\eta'}} .
\label{eq:electric_field_chain_rule}
\end{equation}
Considering the transformation of the dynamical matrix under the point group operation, and the orthogonality of $\mathbf{R}$, we can write 
\begin{equation}
\begin{split}
\mathcal{D}_{\kappa\alpha,\kappa'\beta}
(\mathbf{q},\boldsymbol{\mathcal{E}})
=&
\sum_{\alpha'\beta'}
(R^{-1})_{\alpha\alpha'}
\,
\mathcal{D}_{R(\kappa)\alpha',R(\kappa')\beta'}
(\mathbf{R}\cdot \mathbf{q},\mathbf{R} \cdot \boldsymbol{\mathcal{E}})
\,
R_{\beta'\beta} \\
=&
\sum_{\alpha'\beta'}
R_{\alpha'\alpha}
\,
\mathcal{D}_{R(\kappa)\alpha',R(\kappa')\beta'}
(\mathbf{R}\cdot \mathbf{q},\mathbf{R} \cdot \boldsymbol{\mathcal{E}})
\,
R_{\beta'\beta}.
\end{split}
\label{eq:dynamical_matrix_symmetry_components}
\end{equation}
and therefore,
\begin{equation}
\frac{\partial \mathcal{D}_{\kappa\alpha,\kappa'\beta}(\mathbf{q})}
{\partial \mathcal{E}_{\eta}}
=
\sum_{\tilde{\alpha}\tilde{\beta}\eta'}
R_{\tilde{\alpha}\alpha}
R_{\tilde{\beta}\beta}
R_{\eta'\eta}
\frac{\partial 
\mathcal{D}_{R(\kappa)\tilde{\alpha},R(\kappa')\tilde{\beta}}
(\mathbf{R}\cdot\mathbf{q})}
{\partial \mathcal{E}_{\eta'}}.
\end{equation}
In the intermediate step in Eq.~\eqref{eq:Y_point_group_transformation_step_three}, we can now use the orthogonality relation $\sum_{\alpha}R_{\alpha'\alpha}R_{\tilde{\alpha}\alpha}
=
\delta_{\alpha'\tilde{\alpha}}$, to evaluate the sum over $\alpha$ and $\beta$. This gives
\begin{equation}
\begin{split}
\sum_{\alpha\beta}
R_{\alpha'\alpha}
\frac{\partial \mathcal{D}_{\kappa\alpha,\kappa'\beta}(\mathbf{q})}
{\partial \mathcal{E}_{\eta}}
R_{\beta'\beta}
&=
\sum_{\alpha\beta}
\sum_{\tilde{\alpha}\tilde{\beta}\eta'}
R_{\alpha'\alpha}
R_{\tilde{\alpha}\alpha}
R_{\tilde{\beta}\beta}
R_{\beta'\beta}
R_{\eta'\eta}
\frac{\partial 
\mathcal{D}_{R(\kappa)\tilde{\alpha},R(\kappa')\tilde{\beta}}
(\mathbf{R}\cdot\mathbf{q})}
{\partial \mathcal{E}_{\eta'}}
\\
&=
\sum_{\eta'}
R_{\eta'\eta}
\frac{\partial 
\mathcal{D}_{R(\kappa)\alpha',R(\kappa')\beta'}
(\mathbf{R}\cdot\mathbf{q})}
{\partial \mathcal{E}_{\eta'}} \\
&=
\left[
\mathbf{R}^{-1}\cdot
\frac{\partial 
\mathcal{D}_{R(\kappa)\alpha',R(\kappa')\beta'}
(\mathbf{R}\cdot\mathbf{q})}
{\partial \boldsymbol{\mathcal{E}}}
\right]_{\eta},
\end{split}
\end{equation}
As a last step, we simplify Eq.~\eqref{eq:Y_point_group_transformation_step_three} and relabel the indices. This gives
\begin{equation}
\begin{split}
\mathbf{Y}_{\nu_1 \nu_2}(\mathbf{q}) =&
\sum_{\nu_1'\nu_2'}
U^{*}_{\nu_1',\nu_1}(\mathbf{q};\mathbf{R})
\Bigg[
\frac{1}{i e} 
\sqrt{ \frac{\hbar}{2 \omega_{\mathbf{R}\cdot\mathbf{q}, \nu_1'}} } 
\sqrt{ \frac{\hbar}{2 \omega_{\mathbf{R}\cdot\mathbf{q}, \nu_2'}} } 
\sum_{\kappa \kappa'}
\sum_{\alpha'\beta'}
e^{*}_{R(\kappa)\alpha',\nu_1'}(\mathbf{R}\cdot\mathbf{q})
\\
&\hspace{7em}\times
\left[
\mathbf{R}^{-1}\cdot
\frac{\partial 
\mathcal{D}_{R(\kappa)\alpha',R(\kappa')\beta'}
(\mathbf{R}\cdot\mathbf{q})}
{\partial \boldsymbol{\mathcal{E}}}
\right]
e_{R(\kappa')\beta',\nu_2'}(\mathbf{R}\cdot\mathbf{q})
\Bigg]
U_{\nu_2',\nu_2}(\mathbf{q};\mathbf{R})
\\[0.5em]
=&
\sum_{\nu_1'\nu_2'}
U^{*}_{\nu_1',\nu_1}(\mathbf{q};\mathbf{R})
\mathbf{R}^{-1}\cdot
\Bigg[
\frac{1}{i e} 
\sqrt{ \frac{\hbar}{2 \omega_{\mathbf{R}\cdot\mathbf{q}, \nu_1'}} } 
\sqrt{ \frac{\hbar}{2 \omega_{\mathbf{R}\cdot\mathbf{q}, \nu_2'}} } 
\sum_{\kappa'' \alpha'} \sum_{\kappa''' \beta'}
e^{*}_{\kappa''\alpha',\nu_1'}(\mathbf{R}\cdot\mathbf{q})
\\
&\hspace{9em}\times
\frac{\partial 
\mathcal{D}_{\kappa''\alpha',\kappa'''\beta'}
(\mathbf{R}\cdot\mathbf{q})}
{\partial \boldsymbol{\mathcal{E}}}
e_{\kappa'''\beta',\nu_2'}(\mathbf{R}\cdot\mathbf{q})
\Bigg]
U_{\nu_2',\nu_2}(\mathbf{q};\mathbf{R})
\\[0.5em]
=&
\sum_{\nu_1'\nu_2'}
U_{\nu_1',\nu_1}^{*}(\mathbf{q};\mathbf{R})
\mathbf{R}^{-1}\cdot
\mathbf{Y}_{\nu_1',\nu_2'}(\mathbf{R}\cdot\mathbf{q})
U_{\nu_2',\nu_2}(\mathbf{q};\mathbf{R}) .
\end{split}
\label{eq:Y_symmetry_transformation}
\end{equation}
Writing the last line in index form, we retain Eq.~\eqref{eq:transformation_of_Ys_with_rotation}, which completes the proof. From this derivation, it is clear that the additional rotation matrix comes from the transformation of the electric field.
\end{widetext}

\subsection{Symmetry transformation of the one-electron-two-phonon self-energy}
\label{subsec:symmetry_transformations_one_electron_two_phonon_self_energy}

In the following, we show how the self-energy for isotropic materials, as written in Eq.~\eqref{eq:isotropic_1el_2ph_self_energy_split_into_components}, transforms under rotation of $Y_{\nu'\nu,\alpha} (\mathbf{q})$, which we derived in Eq.~\eqref{eq:transformation_of_Ys_with_rotation}. We follow the idea that a rotation matrix $\mathbf{R}_{\beta \alpha}$ of the point group scrambles the components of $Y_{\nu_1 \nu_2, \alpha}(\mathbf{q})$, which we can use to write the $\alpha = x$ and $\alpha = y$ components of the one-electron-two-phonon self-energy in terms of the $\alpha = z$ component. First, we show how the one-electron-two-phonon self-energy transforms under point group operations in general, and then we restrict to a particular example to proof Eq.~\eqref{eq:dia_iso_1el_2ph_self_energy_x_and_y_component} from the main text. Plugging Eq.~\eqref{eq:transformation_of_Ys_with_rotation} into Eq.~\eqref{eq:isotropic_1el_2ph_self_energy_split_into_components} and rearranging the sums gives 
\begin{widetext}
\begin{equation}
\begin{split}
\Pi^{(\text{1el-2ph})}_{\nu\mu,\alpha}(\mathbf{q},\omega) 
&= \sum_{\nu'} 
\Bigg(
\sum_{\nu_1', \nu_2'} U_{\nu_1' \nu'}^* (\mathbf{q}, \mathbf{R})
\sum_{\beta} \mathbf{R}_{\beta \alpha} \, Y_{\nu_1' \nu_2', \beta}(\mathbf{R}\cdot \mathbf{q}) \,
U_{\nu_2' \nu}(\mathbf{q}, \mathbf{R})
\Bigg) \\
&\qquad\times
\Bigg(
\sum_{\tilde\nu_1', \tilde\nu_2'} U_{\tilde\nu_1' \nu'} (\mathbf{q}, \mathbf{R})
\sum_{\beta'} \mathbf{R}_{\beta' \alpha}^* \, Y^*_{\tilde\nu_1' \tilde\nu_2', \beta'}(\mathbf{R}\cdot \mathbf{q}) \,
U^*_{\tilde\nu_2' \mu}(\mathbf{q}, \mathbf{R})
\Bigg)\,
\xi(\omega_{\nu'\mathbf{q}},\omega) \\
&= 
\sum_{\nu_1', \nu_2'} 
\sum_{\tilde\nu_1', \tilde\nu_2'} 
\Big(\sum_{\nu'}
U_{\nu_1' \nu'}^* (\mathbf{q}, \mathbf{R}) 
U_{\tilde\nu_1' \nu'} (\mathbf{q}, \mathbf{R})\Big) 
U_{\nu_2' \nu}(\mathbf{q}, \mathbf{R}) 
U^*_{\tilde\nu_2' \mu}(\mathbf{q}, \mathbf{R})\\
&\qquad \times 
\sum_{\beta,\beta'}
\mathbf{R}_{\beta \alpha} 
\mathbf{R}_{\beta' \alpha}^*
Y_{\nu_1' \nu_2', \beta}(\mathbf{R}\cdot \mathbf{q}) \,
\, Y^*_{\tilde\nu_1' \tilde\nu_2', \beta'}(\mathbf{R}\cdot \mathbf{q}) \,
\,\xi(\omega_{\tilde{\nu}_1'\mathbf{q}},\omega) 
\end{split} \label{eq:Pi_plugged}
\end{equation}
Here, we used the block-diagonal property of $U_{\tilde{\nu}_1'\nu'}$ to replace $\omega_{\nu'\mathbf{q}} \to \omega_{\tilde{\nu}_1'\mathbf{q}}$. Now we can use the unitarity of $U$ to perform the sum over $\nu'$, as $\sum_{\nu'} U_{\nu_1' \nu'}^* (\mathbf{q}, \mathbf{R})\,
U_{\tilde\nu_1' \nu'} (\mathbf{q}, \mathbf{R})
= \delta_{\nu_1',\tilde\nu_1'}$, and the fact that rotation maps the branch labels so that $\xi(\omega_{\nu'\mathbf{q}},\omega) \to \xi(\omega_{\tilde{\nu}_1',\mathbf{R}\cdot\mathbf{q}},\omega)$ along the same unitary change of basis. In other words, the set of eigenfrequencies at $\mathbf q$ equals that at $\mathbf R\cdot \mathbf q$, and $U(\mathbf q,\mathbf R)$ is the rotation-induced basis change among the degenerate/near-degenerate modes. Consequently, Eq.~\eqref{eq:Pi_plugged} becomes
\begin{equation}
\begin{split}
\Pi^{(\text{1el-2ph})}_{\nu\mu,\alpha}(\mathbf{q},\omega)
=\sum_{\nu_2',\tilde\nu_2'} 
U_{\nu_2'\nu}(\mathbf{q}, \mathbf{R}) \Big[
\sum_{\nu_1'}
\Big(
\sum_{\beta} 
\mathbf{R}_{\beta \alpha}\,
Y_{\nu_1'\nu_2',\beta}(\mathbf{R}\cdot\mathbf{q})
\Big)
\xi(\omega_{\nu_1',\mathbf{R}\cdot\mathbf{q}},\omega)
\Big(
\sum_{\beta'}
\mathbf{R}_{\beta' \alpha}^*
Y^*_{\nu_1'\tilde\nu_2',\beta'}(\mathbf{R}\cdot\mathbf{q})\,
\Big) \Big] U^*_{\tilde\nu_2'\mu}(\mathbf{q}, \mathbf{R}). 
\end{split} \label{eq:transformation_under_rotation_and_unitary}
\end{equation}
In matrix form, this transformation can be written as
\begin{equation}
\Pi^{(\text{1el-2ph})}_{\nu\mu,\alpha}(\mathbf{q},\omega) \to 
\mathbf{U}^{\rm T} (\mathbf{q}, \mathbf{R}) 
\Big(\mathbf{R}_{\beta\alpha} \mathbf{Y}_{\beta}^{\rm T} (\mathbf{R}\cdot \mathbf{q})\Big)
\boldsymbol{\xi} (\boldsymbol{\omega}_{\mathbf{\mathbf{R}\cdot\mathbf{q}}},\omega)
\Big(\mathbf{R}_{\beta'\alpha}^* \mathbf{Y}_{\beta'}^* (\mathbf{R}\cdot \mathbf{q})\Big)
\mathbf{U}^* (\mathbf{q}, \mathbf{R}), \label{eq:transformation_under_rotation_and_unitary_matrix_form}
\end{equation}
where we introduced the vector notation for $\xi$, i.e. $(\boldsymbol{\xi}  (\boldsymbol{\omega}_{\mathbf{\mathbf{R}\cdot\mathbf{q}}},\omega))_{\nu} = \xi(\omega_{\nu,\mathbf{R}\cdot\mathbf{q}},\omega)$. From noting that the inverse of $\mathbf{R}$ is also in the group, and that point group transformations are represented by real orthogonal matrices, it follows that 
\begin{equation}
\Pi^{(\text{1el-2ph})}_{\nu\mu,\alpha}(\mathbf{q},\omega) \to 
\mathbf{U}^{\rm T} (\mathbf{q}, \mathbf{R}^{-1}) 
\Big(
\mathbf{R}_{\alpha\beta} \mathbf{Y}_{\beta}^{\rm T} (\mathbf{R}^{-1}\cdot \mathbf{q})
\Big)
\boldsymbol{\xi} (\boldsymbol{\omega}_{\mathbf{\mathbf{R}^{-1}\cdot\mathbf{q}}},\omega)
\Big(
\mathbf{R}_{\alpha\beta'}^* \mathbf{Y}_{\beta'}^* (\mathbf{R}^{-1}\cdot \mathbf{q})
\Big)
\mathbf{U}^* (\mathbf{q}, \mathbf{R}^{-1}). \label{eq:transformation_under_rotation_and_unitary_matrix_form_inverse_matrix}
\end{equation}
We will use these transformations to express the $\alpha = x,y$ components of the self-energy in terms of the $z$ component. Every cubic point group contains the following element, 
\begin{equation}
    \mathbf{R} = \left(\begin{matrix}
        0 & 0 & 1 \\
        1 & 0 & 0 \\
        0 & 1 & 0
    \end{matrix}\right). \label{eq:rotation_matrix_to_simplify_iso_calculation}
\end{equation}
Fixing $\alpha = y$, we see that $\mathbf{R}_{\beta y} = \delta_{\beta,z}$. Plugging this into Eq.~\eqref{eq:transformation_under_rotation_and_unitary} results in
\begin{equation}
\begin{split}
&\Pi^{(\text{1el-2ph})}_{\nu\mu,y}(\mathbf{q},\omega)
= \\
=&\sum_{\nu_2',\tilde\nu_2'} 
U_{\nu_2'\nu}(\mathbf{q}, \mathbf{R}) \Big[
\sum_{\nu_1'}
\Big(
\sum_{\beta} 
\delta_{\beta,z}\,
Y_{\nu_1'\nu_2',\beta}(\mathbf{R}\cdot\mathbf{q})
\Big)
\xi(\omega_{\nu_1',\mathbf{R}\cdot\mathbf{q}},\omega)
\Big(
\sum_{\beta'}
\delta_{\beta',z}^*
Y^*_{\nu_1'\tilde\nu_2',\beta'}(\mathbf{R}\cdot\mathbf{q})\,
\Big) \Big] U^*_{\tilde\nu_2'\mu}(\mathbf{q}, \mathbf{R}) \\
=&\sum_{\nu_2',\tilde\nu_2'} 
U_{\nu_2'\nu}(\mathbf{q}, \mathbf{R}) \Big[
\sum_{\nu_1'}
Y_{\nu_1'\nu_2',z}(\mathbf{R}\cdot\mathbf{q})
\xi(\omega_{\nu_1',\mathbf{R}\cdot\mathbf{q}},\omega)
Y^*_{\nu_1'\tilde\nu_2',z}(\mathbf{R}\cdot\mathbf{q})\,
\Big] U^*_{\tilde\nu_2'\mu}(\mathbf{q}, \mathbf{R}) \\
=&\sum_{\nu_2',\tilde\nu_2'} 
U_{\nu_2'\nu}(\mathbf{q}, \mathbf{R}) \Big[
\Pi^{(\text{1el-2ph})}_{\nu_2'\tilde{\nu}_2',z}(\mathbf{R}\cdot\mathbf{q},\omega)
\Big] U^*_{\tilde\nu_2'\mu}(\mathbf{q}, \mathbf{R}) \\
=& \Pi^{(\text{1el-2ph})}_{\nu\mu,z}(\mathbf{R}\cdot\mathbf{q},\omega),
\label{eq:derivation_for_the_y_component}
\end{split}
\end{equation}
where, in the last two lines, we used the transformation property of the one-electron-two-phonon self-energy under the unitary freedom of the choice of the phonon eigenvectors as derived in Eq.~\eqref{eq:unitary_transformation_1_electron_2_phonon_self_energy}. Similarly, we see that $\mathbf{R}_{x\beta} = \delta_{\beta,z}$, which we plug into Eq.~\eqref{eq:transformation_under_rotation_and_unitary_matrix_form_inverse_matrix} to write 
\begin{equation}
\begin{split}
&\Pi^{(\text{1el-2ph})}_{\nu\mu,x}(\mathbf{q},\omega)
= \\
=&\sum_{\nu_2',\tilde\nu_2'} 
U_{\nu_2'\nu}(\mathbf{q}, \mathbf{R}^{-1}) \Big[
\sum_{\nu_1'}
\Big(
\sum_{\beta} 
\delta_{\beta,z}\,
Y_{\nu_1'\nu_2',\beta}(\mathbf{R}^{-1}\cdot\mathbf{q})
\Big)
\xi(\omega_{\nu_1',\mathbf{R}^{-1}\cdot\mathbf{q}},\omega)
\Big(
\sum_{\beta'}
\delta_{\beta',z}^*
Y^*_{\nu_1'\tilde\nu_2',\beta'}(\mathbf{R}^{-1}\cdot\mathbf{q})\,
\Big) \Big] U^*_{\tilde\nu_2'\mu}(\mathbf{q}, \mathbf{R}^{-1}) \\
=&\sum_{\nu_2',\tilde\nu_2'} 
U_{\nu_2'\nu}(\mathbf{q}, \mathbf{R}^{-1}) \Big[
\sum_{\nu_1'}
Y_{\nu_1'\nu_2',z}(\mathbf{R}^{-1}\cdot\mathbf{q})
\xi(\omega_{\nu_1',\mathbf{R}^{-1}\cdot\mathbf{q}},\omega)
Y^*_{\nu_1'\tilde\nu_2',z}(\mathbf{R}^{-1}\cdot\mathbf{q})\,
\Big] U^*_{\tilde\nu_2'\mu}(\mathbf{q}, \mathbf{R}^{-1}) \\
=& \Pi^{(\text{1el-2ph})}_{\nu\mu,z}(\mathbf{R}^{-1}\cdot\mathbf{q},\omega).
\end{split}
\end{equation}
Consequently, we can express the sum of the one-electron-two-phonon self-energy components as 
\begin{equation}
\Pi^{(\text{1el-2ph})}_{\nu\mu,x}(\mathbf{q},\omega) 
+ \Pi^{(\text{1el-2ph})}_{\nu\mu,y}(\mathbf{q},\omega)
+ \Pi^{(\text{1el-2ph})}_{\nu\mu,z}(\mathbf{q},\omega) 
= 
\Pi^{(\text{1el-2ph})}_{\nu\mu,z}(\mathbf{R}^{-1}\cdot\mathbf{q},\omega)
+ \Pi^{(\text{1el-2ph})}_{\nu\mu,z}(\mathbf{R}\cdot\mathbf{q},\omega)
+ \Pi^{(\text{1el-2ph})}_{\nu\mu,z}(\mathbf{q},\omega),\label{eq:app_summation_over_rotated_Brillouin_zone_for_isotropic_self_energy}
\end{equation}
which is the expression that we wanted to show. We note that it also holds for the off-diagonal self-energy components for the isotropic one-electron-two-phonon self-energy. 
\end{widetext}

\section{Phonon Renormalization from the Dyson Equation} \label{sec:app_dyson_equation}

\subsection{Diagonalization of the Dyson Equation with Off-Diagonal Self-Energies} \label{subsec:diagonalizing_dyson_equation_with_off_diagonal_terms}

When physical processes give rise to significant off-diagonal phonon self-energies, it is convenient to define a rescaled inverse phonon propagator, 
\begin{equation}
\Big(\tilde{\boldsymbol{D}}^{\rm R}  \left(\mathbf{q},\omega\right) \Big) ^{-1} \equiv S \left(\boldsymbol{D}^{\rm R} \left(\mathbf{q},\omega\right) \right)^{-1} S,
\end{equation}
with the diagonal transformation matrix $S = \mathrm{diag} (\sqrt{2 \omega_{1,\mathbf{q}}}, \dots, \sqrt{2 \omega_{3N,\mathbf{q}}})$, following~\cite{Hermes2013,Hermes20132}. The eigenvalue problem is then recast into
\begin{equation}
(\tilde{\boldsymbol{D}}^{\rm R}  \left(\mathbf{q},\omega\right) )^{-1} = \omega^2 1 - \boldsymbol{\Omega}^2 (\mathbf{q}) - \boldsymbol{\mathcal{V}} (\mathbf{q},\omega),
\end{equation}
with diagonal matrix $\boldsymbol{\Omega}^2 (\mathbf{q}) = \mathrm{diag} (\omega_{1,\mathbf{q}}^2, \omega_{2,\mathbf{q}}^2, \dots, \omega_{3N,\mathbf{q}}^2)$ and \textit{perturbation} matrix
\begin{widetext}
\begin{equation}
    \boldsymbol{\mathcal{V}} (\mathbf{q},\omega) = 2
    \begin{pmatrix}
         \omega_{1,\mathbf{q}} \Pi^{\rm R}_{11} \left(\mathbf{q},\omega\right) 
        & \sqrt{ \omega_{1,\mathbf{q}} \omega_{2,\mathbf{q}} } \, \Pi^{\rm R}_{12} \left(\mathbf{q},\omega\right) 
        & \cdots 
        & \sqrt{ \omega_{1,\mathbf{q}} \omega_{3N,\mathbf{q}} } \,\Pi^{\rm R}_{1,3N} \left(\mathbf{q},\omega\right) \\
        \sqrt{ \omega_{2,\mathbf{q}}  \omega_{1,\mathbf{q}} } \, \Pi^{\rm R}_{21} \left(\mathbf{q},\omega\right) 
        & \omega_{2,\mathbf{q}} \Pi^{\rm R}_{22} \left(\mathbf{q},\omega\right)
        & \cdots 
        & \sqrt{ \omega_{2,\mathbf{q}}  \omega_{3N,\mathbf{q}} } \, \Pi^{\rm R}_{2,3N} \left(\mathbf{q},\omega\right) \\
        \vdots 
        & \vdots 
        & \ddots 
        & \vdots \\
        \sqrt{ \omega_{3N,\mathbf{q}} \omega_{1,\mathbf{q}} } \, \Pi^{\rm R}_{3N,1} \left(\mathbf{q},\omega\right)
        & \sqrt{ \omega_{3N,\mathbf{q}}  \omega_{2,\mathbf{q}} } \, \Pi^{\rm R}_{3N,2} \left(\mathbf{q},\omega\right) 
        & \cdots 
        & \omega_{3N,\mathbf{q}} \Pi^{\rm R}_{3N,3N} \left(\mathbf{q},\omega\right).
    \end{pmatrix} \label{eq:perturbation_def_of_interacting_phonon_propagator}
\end{equation}
\end{widetext}
Defining the auxiliary matrix $\boldsymbol{\mathcal{A}} (\mathbf{q},\omega)= \boldsymbol{\Omega}^2 (\mathbf{q}) + \boldsymbol{\mathcal{V}} (\mathbf{q},\omega)$, solving the secular equation $\det (\tilde{\boldsymbol{D}}^{\rm R}  \left(\mathbf{q},\omega\right) )^{-1} = 0$ is equivalent to solving the nonlinear eigenvalue problem $\boldsymbol{\mathcal{A}} (\mathbf{q},\omega) |\phi\rangle = \omega^2 |\phi\rangle$. In case the self-energies are zero, we recover the non-interacting phonon band structure, $\boldsymbol{\Omega}^2 (\mathbf{q}) |\mathbf{e}_{\nu,\mathbf{q}}^{(0)}\rangle = (\omega_{\nu,\mathbf{q}}^{(0)})^2 | \mathbf{e}_{\nu,\mathbf{q}}^{(0)}\rangle$, with eigenvectors in the standard basis, and $(\omega_{\nu,\mathbf{q}}^{(0)})^2 \equiv \omega_{\nu,\mathbf{q}}^2$. In general, however, the eigenvectors $|\phi\rangle$ are related to the eigenvectors in the normal mode basis via some normalization coming from $S$.  For non-zero self-energies, we obtain the correction of the eigenvalue $(\omega_{\nu,\mathbf{q}}^{(0)})^2$ by evaluating $\boldsymbol{\mathcal{A}}$ at this frequency, and solve $\boldsymbol{\mathcal{A}} (\mathbf{q},\omega_{\nu,\mathbf{q}}^{(0)}) |\phi_{\nu,n}\rangle = (\omega_{\nu,n}^{(1)})^2 |\phi_{\nu,n}\rangle$, where $n \in \lbrace 1,\dots, 3N\rbrace$ labels the eigenvalues. The accepted eigenvalue with index $n$ is then the one where the corresponding eigenvector maximizes the overlap to the unperturbed eigenvector, $\underset{n}{\mathrm{max}} \langle \mathbf{e}_{\nu,\mathbf{q}}^{(0)} | \phi_{\nu,n}\rangle$. These diagonalization steps are repeated iteratively until convergence.

\section{Integration details} \label{sec:app_integral_identities}

\subsection{Surface integral} 

In the derivation of Eq.~\eqref{eq:isotropic_1el_2ph_self_energy_split_into_components} from Eq.~\eqref{eq:1-electron-2-phonon_self_energy_final_expression}, we rescale the momenta, $\mathbf{Q}_i = Q n_i^{\mathbf{Q}}$, where $n_i^{\mathbf{Q}}$ is the $i$-th component of the unit vector in $\mathbf{Q}$ direction, with the integration measure change $\mathrm{d}^3 \mathbf{Q} = \mathrm{d} Q\, Q^2 \mathrm{d}^2 \mathbf{n}$. The magnitude $Q$ in the numerator and denominator cancels, and we use the surface integral identity 
\begin{equation}
\underset{S^2}{\int} \mathrm{d}^2 \mathbf{n} \, n_{\alpha}^{\mathbf{Q}} n_{\beta}^{\mathbf{Q}} = \frac{4\pi}{3} \delta_{\alpha\beta}, \label{eq:app_surface_integral_identity}
\end{equation}
to rewrite the three-dimensional integral into an integral over the radial part.

\subsection{Numerical Integration in Momentum Space} 
\label{subsec:app_numerical_integration}

For our calculations, we set the imaginary part to a value of $\delta = 5\,\text{meV}$. This broadening regularizes the resonant electron-hole contributions and was chosen such that the momentum-space integrals remain numerically well converged while retaining their characteristic momentum dependence.

All momentum-space integrals were evaluated using the adaptive stratified-sampling algorithm Divonne from the Cuba library through its PyCuba interface~\cite{hahnCubaLibraryMultidimensional2005}. The real and imaginary parts were integrated simultaneously as two components of the same integrand. Internal momenta were parameterized by fractional reciprocal coordinates in the interval $[0,1)$ and transformed to Cartesian coordinates using the reciprocal-lattice vectors. The Lindhard polarization entering the one-electron-one-phonon self-energy requires a three-dimensional integration over the internal electronic momentum $\mathbf{k}$. For the one-electron-two-phonon contribution, the isotropic reduction introduced in Sec.~\ref{subsubsec:one_electron_two_phonon_diagram} allows the angular integral over the internal phonon momentum to be evaluated analytically, reducing the original six-dimensional integral over $(\mathbf{Q},\mathbf{k})$ to a four-dimensional integral over $(Q,k_x,k_y,k_z)$. The dimensionless radial coordinate was integrated up to a cutoff radius, beyond which contributions are neglected within the long-range approximation. 

Throughout the calculation, we requested a relative integration accuracy of $\varepsilon_{\mathrm{rel}}=10^{-6}$ and allowed at most $N_{\mathrm{eval}}=10^6$ integrand evaluations for each external momentum, phonon branch, and frequency.  The Divonne sampling parameters were set to $({\tt key1},{\tt key2},{\tt key3})=(300,1,1)$ and ${\tt maxpass}=5$, ${\tt border}=0$, ${\tt maxchisq}=10$, and ${\tt mindeviation}=0.25$, while retaining the default seed of zero. The integration errors returned by Divonne were recorded separately for the real and imaginary parts.

For the LiF calculation, the external high-symmetry path was sampled with 16 points per ordinary path segment. Segments adjacent to $\Gamma$ were refined using 30 points over the fraction $0.6$ closest to $\Gamma$ and six points over the remainder of the segment. For the KTaO$_3$ calculation, on the other hand, ordinary high-symmetry path segments were sampled with 10 points, while segments adjacent to $\Gamma$ used 10 points over the half closest to $\Gamma$ and six points over the remaining half. This choice was made because the computations for KTaO$_3$ are computationally more demanding. 

\section{Derivation of the heat capacity correction} \label{sec:app_heat_capacity}

In this section, we derive the heat-capacity expression used in Sec.~\ref{subsec:heat_capacity_corrections} for weakly temperature-dependent phonon frequencies, as given by the electron-phonon interactions on the phonon self-energy.

For a prescribed effective chemical potential $\mu$ in the conduction band, we consider temperature dependent renormalized phonon frequencies $\omega_{\nu,\mathbf q}(T,\mu)$ and suppress the explicit $\mu$ dependence below for compactness. Following Refs.~\cite{Allen1980,allenAnharmonicPhononQuasiparticle2015}, we retain terms through second order in the ionic displacement amplitudes within an adiabatic Rayleigh-Schr{\" o}dinger framework. The electron-phonon-induced shifts are assumed to be sufficiently weak so that the phonons remain well-defined bosonic quasiparticles characterized by $\omega_{\nu,\mathbf{q}}(T)$ and Bose-Einstein occupations at each temperature. As can be seen in Fig.~\ref{fig:fig6}, this is an excellent approximation in the context of this manuscript.

To isolate the contribution calculated in this work, we retain only the temperature dependence generated by the electronic occupations and neglect intrinsic anharmonic phonon-phonon interactions. The latter provides an additional contribution and need not be small, particularly in KTaO$_3$. Under the quasiparticle approximation, the phonon entropy retains the noninteracting Bose form after replacing the bare frequencies by their temperature-dependent quasiparticle values~\cite{Allen1980,allenAnharmonicPhononQuasiparticle2015}. The resulting entropy is given in Eq.~\eqref{eq:Sph_Tdep}.

\begin{widetext}
Let us first differentiate $S$ with respect to $T$ at fixed $V$ by applying the chain rule, which gives
\begin{equation}
\left(\frac{\partial S_{\mathrm{ph}}}{\partial T}\right)_V
=
\sum_{\mathbf q\nu} k_B\,
\frac{\partial}{\partial x}\left[\frac{x}{e^{x}-1}-\ln\!\left(1-e^{-x}\right)\right]_{x=x_{\mathbf q\nu}}
\left(\frac{\partial x_{\nu,\mathbf{q}}}{\partial T}\right)_V .
\label{eq:dSdT_chainrule}
\end{equation}
The derivative with respect to $x$ is
\begin{equation}
\frac{\partial}{\partial x}\left[\frac{x}{e^{x}-1}-\ln\!\left(1-e^{-x}\right)\right]
=
x\,\frac{\partial n_B}{\partial x}
=
- x \frac{e^{x}}{(e^{x}-1)^2}
=
- x \,n_B(x)\bigl[n_B(x)+1\bigr],
\label{eq:dfdx_final}
\end{equation}
while the derivative of $x_{\mathbf q\nu}$ with respect to $T$, computed with the chain rule, is
\begin{equation}
\left(\frac{\partial x_{\nu,\mathbf{q}}}{\partial T}\right)_V
=
\frac{\partial}{\partial T}\left(\frac{\hbar\omega_{\nu,\mathbf{q}}(T)}{k_B T}\right)_V
=
\frac{\hbar}{k_B}\left(\frac{1}{T}\frac{\partial\omega_{\nu,\mathbf{q}}}{\partial T}
-\frac{\omega_{\nu,\mathbf{q}}}{T^2}\right)_V 
=
-\frac{x_{\nu,\mathbf{q}}}{T}\left[1-\left(\frac{\partial\ln\omega_{\nu,\mathbf{q}}}{\partial\ln T}\right)_V\right].
\label{eq:dxdT}
\end{equation}
Inserting Eqs.~\eqref{eq:dfdx_final} and \eqref{eq:dxdT} back into Eq.~\eqref{eq:dSdT_chainrule} produces,
\begin{equation}
C_V^{\mathrm{ph}}(T)
=
T\left(\frac{\partial S_{\mathrm{ph}}}{\partial T}\right)_V
=
\sum_{\mathbf q\nu}
k_B\,x_{\nu,\mathbf{q}}^2\, n_B(x_{\nu,\mathbf{q}})\bigl(n_B(x_{\nu,\mathbf{q}})+1\bigr)
\left[1-\left(\frac{\partial\ln\omega_{\nu,\mathbf{q}}}{\partial\ln T}\right)_V\right].
\end{equation}
Defining the usual harmonic mode heat capacity as
\begin{equation}
C_{\nu,\mathbf{q}}(T)
=
k_B\,x_{\nu,\mathbf{q}}^2\, n_B(x_{\nu,\mathbf{q}})\bigl(n_B(x_{\nu,\mathbf{q}})+1\bigr)
=
k_B\left(\frac{\hbar\omega_{\nu,\mathbf{q}}(T)}{k_B T}\right)^2
n_B(\omega_{\nu,\mathbf{q}} (T),T)\bigl(n_B(\omega_{\nu,\mathbf{q}} (T),T)+1\bigr),
\end{equation}
we obtain the compact final result,
\begin{equation}
C_V^{\mathrm{ph}}(T)
=
\sum_{\mathbf q\nu}
C_{\nu,\mathbf{q}}\!\left(T\right)
\left[
1-\frac{\partial \ln\omega_{\nu,\mathbf{q}} (T)}{\partial \ln T}
\right],
\end{equation}
as it is also shown in Eq.~\eqref{eq:Cv_ph_entropy}. The factor in square brackets arises from the implicit temperature dependence of the renormalized phonon frequencies. It modifies the rate at which $x_{\nu,\mathbf q}=\hbar\omega_{\nu,\mathbf q}/(k_BT)$ changes with temperature. Phonon softening, $\partial\omega_{\nu,\mathbf q}/\partial T<0$, makes this factor larger than unity and enhances the corresponding heat-capacity contribution relative to the result for temperature independent phonon frequency. On the other hand, if the phonons harden, the heat capacity gets suppressed.
\end{widetext}

\section{Supplementary data for lithium fluoride} \label{sec:app_lithium_fluoride}

In this section, we present additional results that complement our analysis of the one-electron-one-phonon and one-electron-two-phonon renormalization in LiF as discussed in Sec.~\ref{subsec:one_electron_one_phonon_self_energy} and Sec.~\ref{subsec:one_electron_two_phonon_self_energy} of the main text. 

\begin{figure}[t]
    \centering
    \includegraphics[width=0.8\columnwidth]{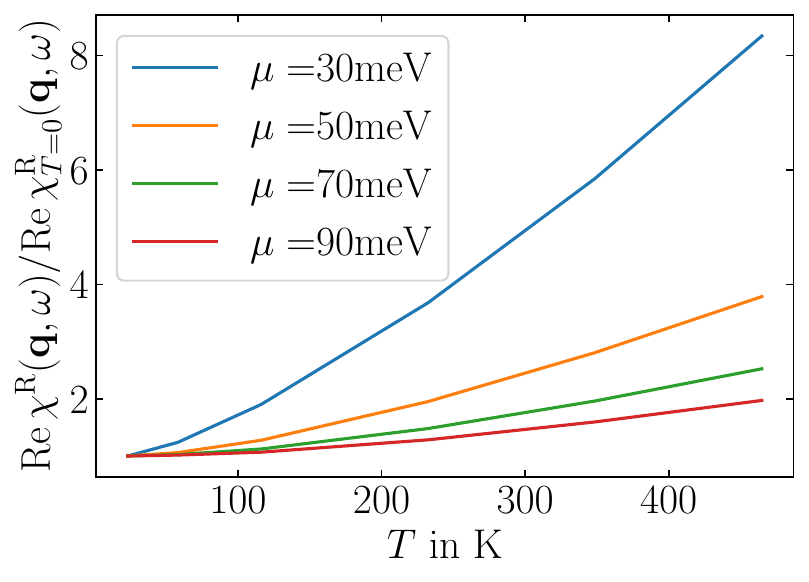}
    \vskip -0.3cm
    \caption{Temperature dependence of the real part of the Lindhard polarization bubble, normalized by its zero-temperature value at the same chemical potential. The response is evaluated at the representative momentum $\mathbf q=(0.2,0,0)$ in reciprocal-lattice units and at the phonon frequency $\omega=50\,\mathrm{meV}$. Increasing $\mu$ reduces the relative thermal variation of the polarization bubble.}
    \label{fig:lindhard_polarization_bubble_saturation}
\end{figure}

We first examine how electronic degeneracy affects the temperature dependence of the Lindhard polarization bubble. Figure~\ref{fig:lindhard_polarization_bubble_saturation} shows the real part of the polarization bubble $\chi^{\mathrm R}(\mathbf q,\omega)$ at a representative phonon momentum and frequency, normalized by its zero-temperature value at the same chemical potential. The normalized response varies most strongly at small $\mu$, whereas its temperature dependence becomes weaker as $\mu$ increases. This behavior reflects the crossover toward the degenerate regime, $\mu\gg k_BT$, in which the thermal broadening of the electronic occupations becomes small relative to the Fermi-energy scale. The results therefore support the reduced relative thermal sensitivity of the electronic polarization at larger chemical potentials.

\begin{figure}[h!]
    \centering
    \includegraphics[width=\columnwidth]{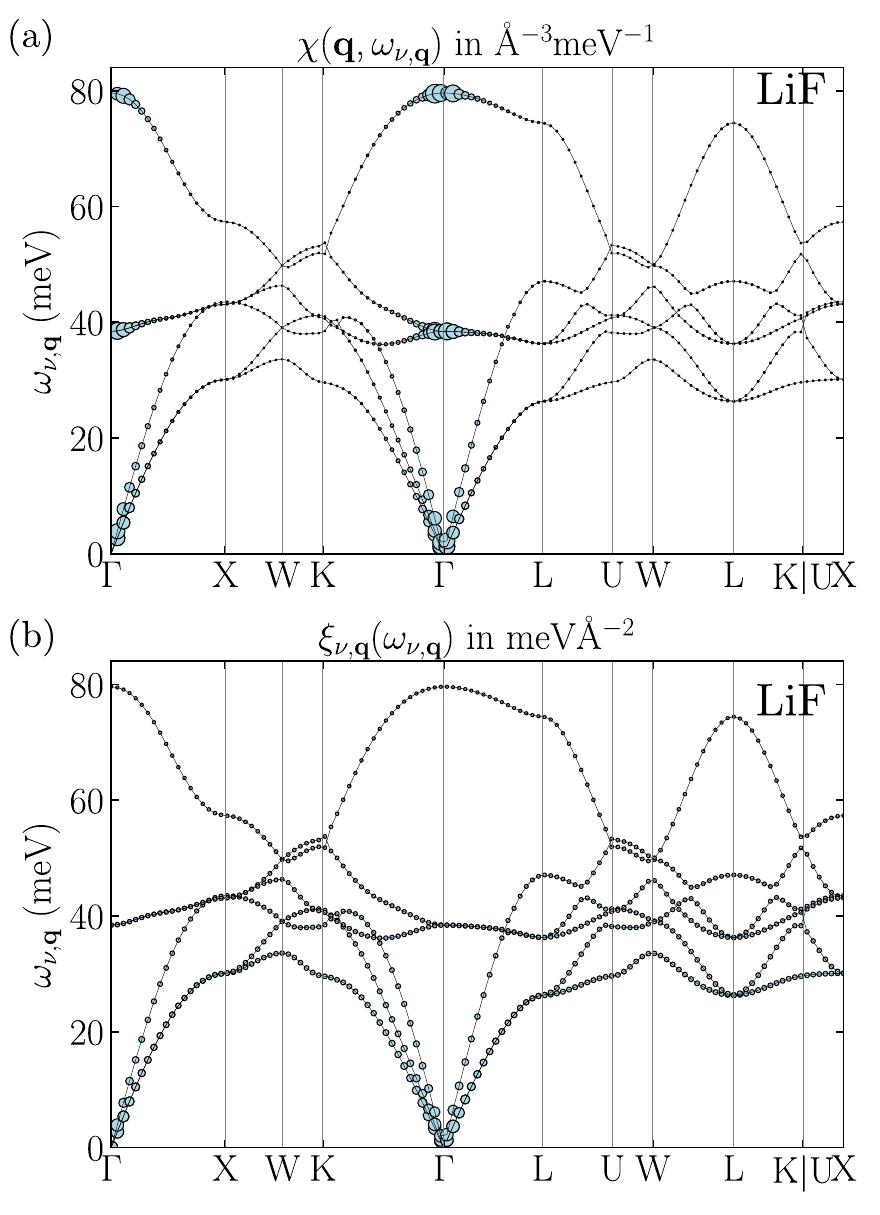}
    \vskip -0.5cm
    \caption{Real parts of the response functions entering the (a) one-electron-one-phonon and (b) one-electron-two-phonon self-energies of LiF, evaluated at $\omega=\omega_{\nu,\mathbf q}$ and superimposed on the phonon dispersion. The circle sizes represent their magnitudes, with the largest circles corresponding to $|\operatorname{Re}\chi^{\rm R}|\approx 3.7\times10^{-6}\,\mathrm{\AA}^{-3}\mathrm{meV}^{-1}$ in panel (a) and $|\operatorname{Re}\xi^{\rm R}_{\nu\mathbf q}|\approx 6.6\times10^{-7}\,\mathrm{meV}\,\mathrm{\AA}^{-2}$ in panel (b). The calculations were performed at $\mu=100\,\mathrm{meV}$ and $T=348\,\mathrm{K}$.}
    \label{fig:LiF_chi_and_xi_comparison}
\end{figure}

To further illustrate the distinct momentum dependences of the linear and nonlinear renormalization channel, as discussed in the main text, Fig.~\ref{fig:LiF_chi_and_xi_comparison} shows $\chi^{\rm R}(\mathbf q,\omega_{\nu,\mathbf q})$ and $\xi^{\rm R}_{\nu,\mathbf q}(\omega_{\nu,\mathbf q})$ along a high-symmetry path through the Brillouin zone of LiF. The Lindhard polarization is concentrated near $\Gamma$, whereas the nonlinear susceptibility remains finite throughout the path. This directly corroborates the observation that in the linear channel, the external phonon momentum directly fixes the momentum transferred to the electron-hole pair, while in the nonlinear channel, this transfer is shared between the external and internal phonons, and the sum over the internal momentum distributes the response over a larger portion of the Brillouin zone.

\begin{figure*}[t]
    \centering
    \includegraphics[width=\textwidth]{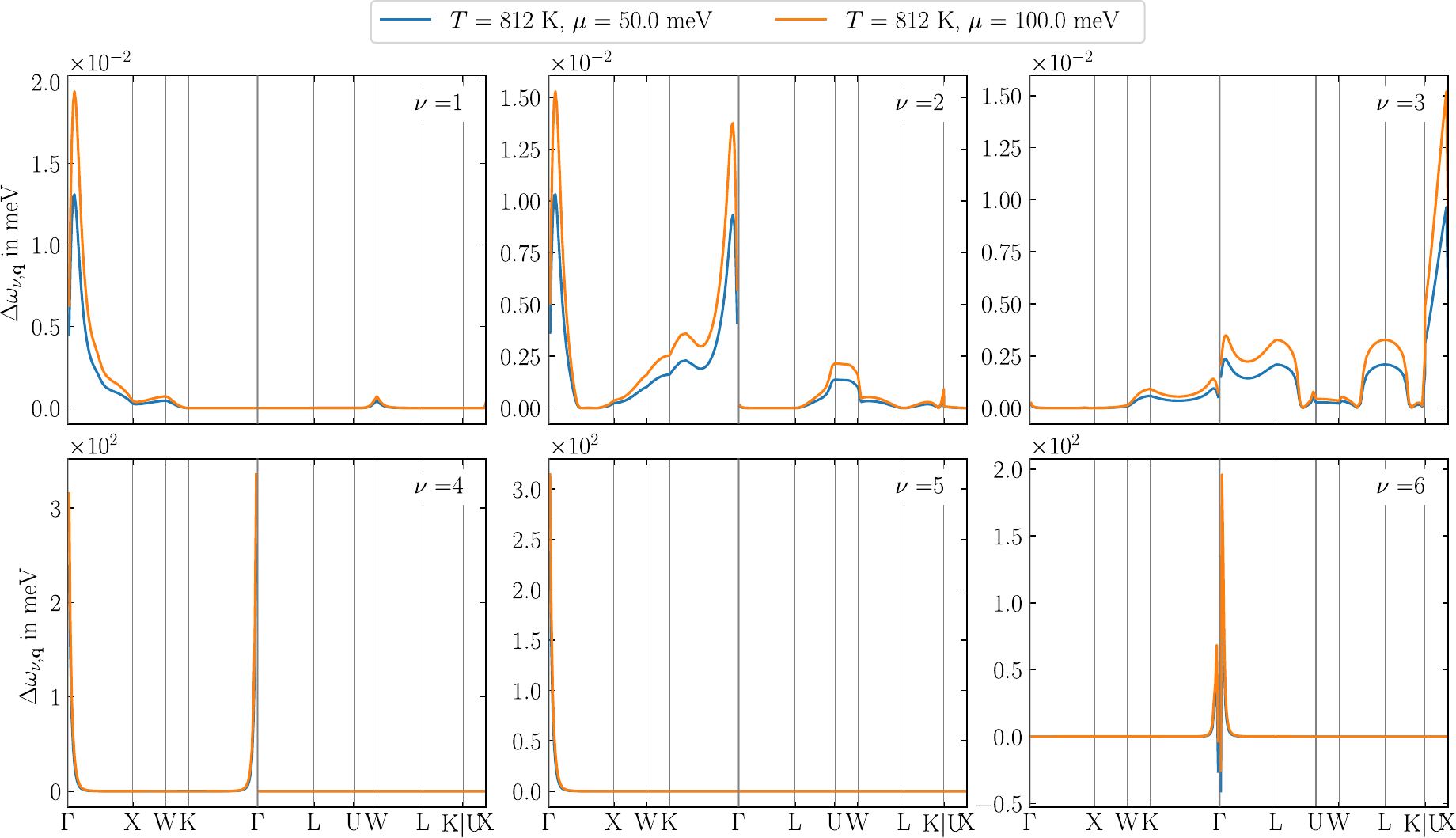}
    \caption{Chemical-potential dependence of the phonon energy renormalization in LiF arising from the one-electron-one-phonon interaction. The correction $\Delta\omega_{\nu,\mathbf{q}}$ is shown along high-symmetry lines for the six phonon branches at $T=812\,\mathrm{K}$ and chemical potentials $\mu=50$ and $100\,\mathrm{meV}$. The branches are indexed in order of increasing energy near $\Gamma$, with $\nu=1,2,3$ denoting the acoustic branches and $\nu=4,5,6$ the optical branches. Note the different vertical scales used for the acoustic and optical modes.}
    \label{fig:lithium_fluoride_energy_renormalization_1_electron_1_phonon_self_energy}
\end{figure*}

Having established the restricted momentum range of the Lindhard polarization, we next resolve the linear phonon renormalization branch by branch to expose the role of the mode-dependent coupling vertex and to complement Fig.~\ref{fig:fig5}(a) of the main text. Figure~\ref{fig:lithium_fluoride_energy_renormalization_1_electron_1_phonon_self_energy} shows the correction at fixed temperature for two chemical potentials, with the branches indexed in order of increasing energy near $\Gamma$. The branch-resolved results make the pronounced difference between the acoustic and optical corrections in Fig.~\ref{fig:fig5}(a) more apparent. The long-range coupling to the acoustic branches is suppressed near $\Gamma$ because their nearly rigid sublattice motion carries a small mode effective charge. By contrast, the relative sublattice motion of the polar optical modes produces a finite mode effective charge and a strongly enhanced long-range coupling near $\Gamma$.

The acoustic corrections in the upper panels increase with $\mu$, consistent with the larger electronic phase space available at higher carrier densities. Within the parabolic-band picture used here, the characteristic Fermi momentum scales as $k_{\rm F}\propto\sqrt{\mu}$ at fixed effective mass, thereby extending the range of electron-hole excitations contributing to the acoustic renormalization. The branch-resolved representation also makes explicit the sign-changing structure of the highest optical branch near $\Gamma$ discussed in the main text.

\begin{figure*}
    \centering
    \includegraphics[width=\textwidth]{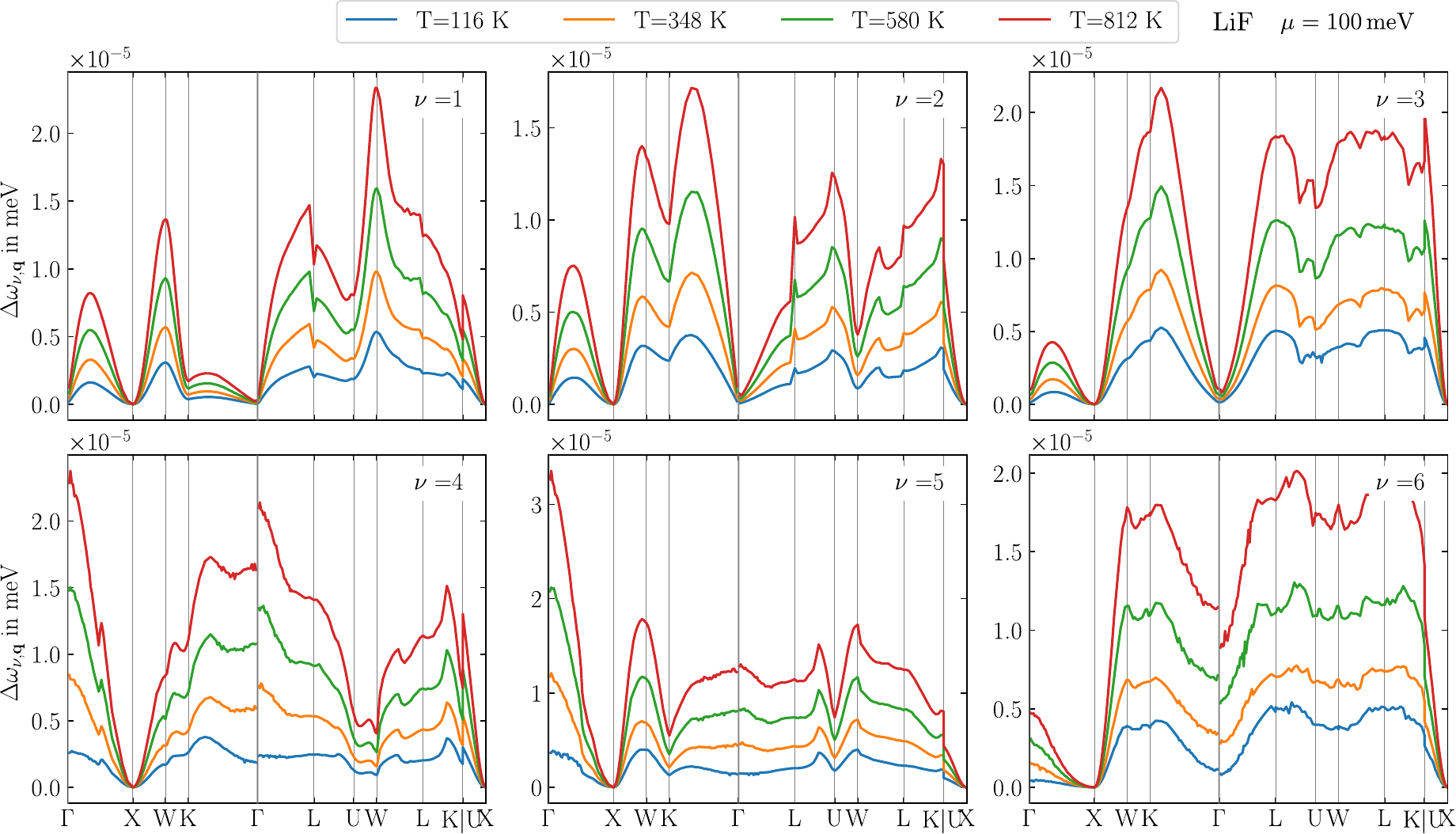}
    \caption{Mode-resolved temperature dependence of the phonon energy renormalization in LiF arising from the one-electron-two-phonon interaction. The correction $\Delta\omega_{\mathbf q\nu}$ is shown along high-symmetry lines for the six phonon branches at fixed chemical potential $\mu=100\,\mathrm{meV}$ and temperatures $T=116$, $348$, $580$, and $812\,\mathrm{K}$. Similar to Fig.~\ref{fig:lithium_fluoride_energy_renormalization_1_electron_1_phonon_self_energy}, the branches are indexed in order of increasing energy near $\Gamma$, with $\nu=1,2,3$ denoting the acoustic branches and $\nu=4,5,6$ the optical branches. The correction increases with temperature across the entire phonon spectrum, while its momentum dependence remains approximately unchanged.}
    \label{fig:lithium_fluoride_energy_renormalization_1_electron_2_phonon_self_energy_temperature_dependence}
\end{figure*}

\begin{figure*}
    \centering
    \includegraphics[width=\textwidth]{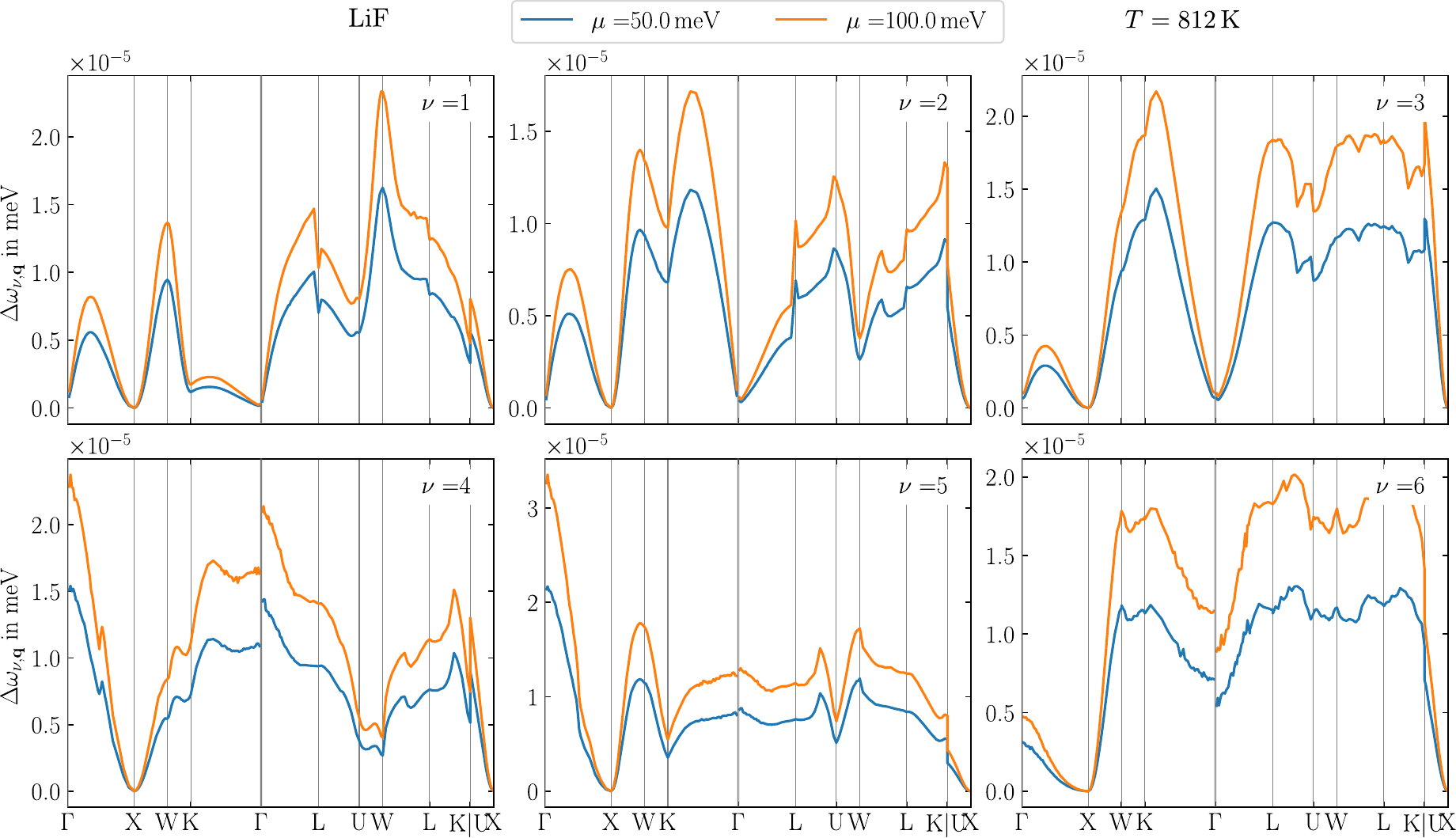}
    \caption{Mode-resolved chemical-potential dependence of the phonon energy renormalization in LiF arising from the one-electron-two-phonon interaction. The correction $\Delta\omega_{\mathbf q\nu}$ is shown along the same high-symmetry path for $\mu=50$ and $100\,\mathrm{meV}$ at fixed temperature $T=812\,\mathrm{K}$. Increasing $\mu$ enhances the correction for all six branches without substantially changing its momentum dependence.}
    \label{fig:lithium_fluoride_energy_renormalization_1_electron_2_phonon_self_energy_chemical_potential_dependence}
\end{figure*}

Finally, we complement the results for the first acoustic branch as shown in Fig.~\ref{fig:fig6}(a) of the main text with a mode-resolved analysis of the nonlinear correction for all six LiF phonon branches. For this purpose, Fig.~\ref{fig:lithium_fluoride_energy_renormalization_1_electron_2_phonon_self_energy_temperature_dependence} shows the temperature dependence of the one-electron-two-phonon self energy renormalization at fixed chemical potential, while Fig.~\ref{fig:lithium_fluoride_energy_renormalization_1_electron_2_phonon_self_energy_chemical_potential_dependence} compares two chemical potentials at fixed temperature. Across all branches, the nonlinear interaction produces a hardening of order $10^{-5}\,\mathrm{meV}$ that increases with both temperature and $\mu$. The temperature dependence follows from the Bose-Einstein occupation of the internal phonon line, whereas the chemical-potential dependence reflects the increasing electronic phase space. In both cases, the overall momentum dependence remains largely unchanged, showing that temperature and chemical potential primarily control the magnitude of the correction, while its momentum structure is already set by the nonlinear coupling vertex. In particular, the correction vanishes at $X$ for every branch, consistent with the symmetry constraint discussed of $\boldsymbol{Y}_{\nu_1\nu_2} (\mathbf{q})$ in the rock salt structure, as discussed in the main text.

We observe that the discontinuities in the corrections for the optical branches around the $\Gamma$ point, when approaching it from the $K$ or $L$ point come from discontinuities when evaluating $Y_{\nu'\nu,\alpha} (\mathbf{q}) Y_{\nu'\mu,\alpha}^* (\mathbf{q})$ and $\xi^{\rm R} (\omega_{\nu',\mathbf{q}}, \omega)$ along rotated paths in the Brillouin zone. By contrast, the response function $\xi^{\rm R}(\omega_{\nu',\mathbf q},\omega)$ remains smooth, as shown in Fig.~\ref{fig:LiF_chi_and_xi_comparison}(b), indicating that the discontinuities are primarily governed by the symmetry and mode dependence of the nonlinear coupling vertex rather than by a singularity of the coupled electron-phonon response.

\section{Supplementary data for potassium tantalate} \label{sec:app_potassium_tantalate}

\begin{figure}[h]
    \centering
    \includegraphics[width=\columnwidth]{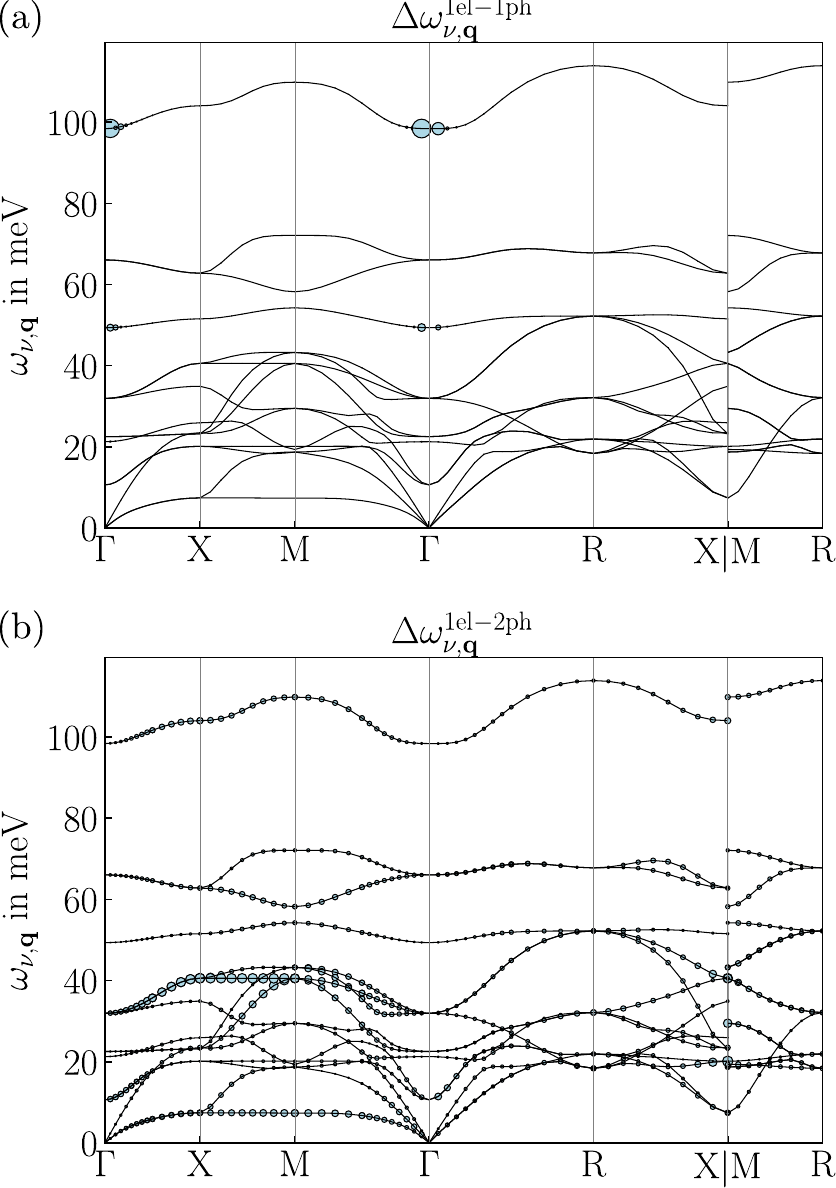}
    \vskip -0.5cm
    \caption{Phonon energy renormalization in KTaO$_3$ due to the one-electron-one-phonon self-energy in (a) and due to the one-electron-two-phonon self-energy in (b). The area of the circles in (a) is proportional to the correction $\Delta \omega_{\nu,\mathbf{q}}^{\rm 1el-1ph}$, with the largest circle corresponding to a renormalization of $\Delta \omega_{\nu,\mathbf{q}}^{\rm 1el-1ph} \approx 20\,\mathrm{meV}$, which is so large due to the coupling vertex divergence when approaching the $\Gamma$ point. Similar to the LiF data in Fig.~\ref{fig:fig5}, the one-electron-one-phonon renormalization of the acoustic branches is smaller than for the optical branches, and therefore, the circles on top of the acoustic branches are barely visible. The circle areas in (b) are proportional to the one-electron-two-phonon renormalization $\Delta \omega_{\nu,\mathbf{q}}^{\rm 1el-2ph}$, with the largest correction being~$\Delta \omega_{\nu,\mathbf{q}}^{\rm 1el-2ph} \approx 1.3\times 10^{-4}\,\mathrm{meV}$. The chemical potential and temperature were set to $\mu = 100\,\mathrm{meV}$ and $812\,\mathrm{K}$ respectively.}
    \label{fig:KTaO3_phonon_self_energy_corrections_on_band_structure}
\end{figure}

To establish that the distinct momentum structures of the two interaction channels are not specific to LiF, we perform the same analysis for KTaO$_3$. The plots in Fig.~\ref{fig:KTaO3_phonon_self_energy_corrections_on_band_structure} of App.~\ref{sec:app_potassium_tantalate} show the corresponding phonon energy shifts superimposed on the phonon band structure. The one-electron-one-phonon correction in panel (a) is again sharply concentrated near the $\Gamma$ point and is strongest for the polar optical branches, reflecting the combined momentum dependence of the long-range electron-phonon vertex and the Lindhard polarization bubble. By contrast, the one-electron-two-phonon correction in panel (b) remains within the same order of magnitude along the entire high-symmetry path and is distributed over many phonon branches. Although the nonlinear correction remains much smaller than the linear contribution, it is larger than in LiF, consistent with the greater number of thermally populated low-energy modes in KTaO$_3$, cf.~Fig.~\ref{fig:fig4}. We omit separate branch-by-branch plots analogous to those presented for LiF in App.~\ref{sec:app_lithium_fluoride}, since displaying all 15 phonon branches would obscure rather than clarify the relevant trends. Accordingly, the comparison between LiF and KTaO$_3$ in Fig.~\ref{fig:fig6} uses the first acoustic branch as a representative example, whereas the relative changes in the lattice heat capacity shown in Fig.~\ref{fig:fig7} includes the renormalization of the entire phonon spectrum.

\end{document}